\documentclass[letterpaper,12pt,titlepage,oneside,final]{book}
 
\newcommand{\href}[1]{#1} 

\newcommand{\thetitle}{A Holographic Approach to Spacetime Entanglement}

\usepackage{ifthen}
\newboolean{PrintVersion}
\setboolean{PrintVersion}{false} 

\usepackage{amsmath,amssymb,amstext} 
\usepackage[pdftex]{graphicx} 
\usepackage{subfig}
\usepackage{cancel}
\usepackage{mathtools}
\usepackage{cite}

\usepackage[pdftex,letterpaper=true,pagebackref=false]{hyperref} 
\hypersetup{
    plainpages=false,       
    pdfpagelabels=true,     
    bookmarks=true,         
    unicode=false,          
    pdftoolbar=true,        
    pdfmenubar=true,        
    pdffitwindow=false,     
    pdfstartview={FitH},    
    pdftitle={\thetitle},    
    pdfauthor={Jason Wien},    
    pdfsubject={Physics},  
    pdfkeywords={holography} {entropy} {entanglement}{differential entropy}, 
    pdfnewwindow=true,      
    colorlinks=true,        
    linkcolor=blue,         
    citecolor=green,        
    filecolor=magenta,      
    urlcolor=cyan           
}
\ifthenelse{\boolean{PrintVersion}}{   
\hypersetup{	
    citecolor=black,%
    filecolor=black,%
    linkcolor=black,%
    urlcolor=black}
}{} 

\let\origdoublepage\cleardoublepage
\newcommand{\clearemptydoublepage}{%
  \clearpage{\pagestyle{empty}\origdoublepage}}
\let\cleardoublepage\clearemptydoublepage

\def\({\left(}
\def\){\right)}
\def\[{\left[}
\def\]{\right]}

\newcommand{\labell}[1]{\qquad\mt{#1}\label{#1}}

\newcommand{\reef}[1]{(\ref{#1})}

\newcommand{\mt}[1]{\textrm{\tiny #1}}

\newcommand{\pd}[2]{\frac{\partial#1}{\partial#2}}
\newcommand{\deriv}[2]{\frac{d#1}{d#2}}

\newcommand{\ban}[1]{\begin{align}#1\end{align}}
\newcommand{\eg}{{\it e.g.}\ }
\newcommand{\coords}[1]{\left\{#1 \right\}}
\newcommand{\la}{\lambda}
\newcommand{\ga}{\gamma}
\newcommand{\G}{\Gamma}
\newcommand{\ie}{{\it i.e.}\ }
\newcommand{\cL}{\mathcal L}
\newcommand{\be}{\begin{equation}}
\newcommand{\ee}{\end{equation}}
\newcommand{\bulkc}{\gamma}

\begin{document}


\pagestyle{empty}
\pagenumbering{roman}

\begin{titlepage}
        \begin{center}
        \vspace*{1.0cm}

        \Huge
        {\bf \thetitle }

        \vspace*{1.0cm}

        \normalsize
        by \\

        \vspace*{1.0cm}

        \Large
        Jason Wien \\

        \vspace*{1.5cm}
        \normalsize Supervised by: \emph{Robert C. Myers}
        \vspace*{1cm}

        \normalsize
        An essay \\
        presented to the Perimeter Institute \\ 
        for the completion of \\
        Perimeter Scholars International\\
        and the requirements for the degree of\\
        Master of Science

        \vspace*{2.0cm}

        Waterloo, Ontario, Canada, June 2014 \\

        \vspace*{1.0cm}

        \copyright\ Jason Wien 2014 \\
        \end{center}
\end{titlepage}

\pagestyle{plain}
\setcounter{page}{2}

\cleardoublepage 


\begin{center}\textbf{Author's Declaration}\end{center}

  \noindent
As this essay was written concurrently with the paper \cite{Headrick2014}, much of the material appears in both works. In particular, sections 2, 3, and 4 of \cite{Headrick2014} significantly overlap with section \ref{holoholes}, chapters \ref{time} and \ref{new}, and appendix \ref{planarS} of this essay. This overlapping material represents original research conducted for the fulfillment of the PSI essay requirement under the supervision of Rob Myers, however here it is presented with a slightly different perspective. 

  \bigskip
  
  \noindent
I understand that my essay may be made electronically available to the public.

\cleardoublepage


\begin{center}\textbf{Abstract}\end{center}

Recently it has been proposed that the Bekenstein-Hawking formula for the entropy of spacetime horizons has a larger significance as the leading contribution to the entanglement entropy of general spacetime regions, in the underlying quantum theory \cite{Bianchi2012}. This `spacetime entanglement conjecture' has a holographic realization that  equates the entropy formula evaluated on an arbitrary space-like co-dimension two surface with the differential entropy of a particular family of co-dimension two regions on the boundary.  The differential entropy can be thought of as a directional derivative of entanglement entropy along a family of surfaces. 

This holographic relation was first studied in \cite{Balasubramanian2014} and extended in \cite{Myers2014}, and it has been proven to hold in Einstein gravity for bulk surfaces with planar symmetry (as well as for certain higher curvature theories) in \cite{Headrick2014}. In this essay, we review this proof and provide explicit examples of how to build the appropriate family of boundary intervals for a given bulk curve. Conversely, given a family of boundary intervals, we provide a method for constructing the corresponding bulk curve in terms of intersections of entanglement wedge boundaries. We work mainly in three dimensions, and comment on how the constructions extend to higher dimensions.

\cleardoublepage


\begin{center}\textbf{Acknowledgements}\end{center}

I would like to chiefly thank my advisor, Rob Myers, who provided much needed direction and advice in completing this research project. In addition, I would like to thank our collaborator Matt Headrick, who provided crucial insights for our work. 

Further, I would like to thank the Perimeter Institute for Theoretical Physics for their generous support during my time as a student in the Perimeter Scholars International program. In particular, I would like to thank the PSI fellows for their advice and mentoring during my studies. 

Finally, I would like to thank Rabbi Moishy and Rivky Goldman and family for providing guidance and inspiration, and for being like family during my studies in Waterloo. 

\cleardoublepage

\renewcommand\contentsname{Table of Contents}
\tableofcontents
\cleardoublepage
\phantomsection

\addcontentsline{toc}{chapter}{List of Figures}
\listoffigures
\cleardoublepage
\phantomsection		

\pagenumbering{arabic}


\chapter{Background Material}
\label{intro}
A promising window into the quantum nature of gravity is the fact that spacetime horizons carry an entropy given by the Bekenstein-Hawking formula
\ban{
S_{BH}=\frac {\mathcal A}{4G_N} \label{BH}
}
where $\mathcal A$ is the area of the horizon and $G_N$ is Newton's constant for a $d+1$ dimensional spacetime \cite{Bekenstein1972, Bekenstein1973, Hawking1974, Hawking1975}.\footnote{Additionally we set $\hbar = c = k_B=1$.} In addition, a similar formula has been derived for higher curvature theories of gravity \cite{Wald1993, Jacobson1994, Iyer1994}. Providing a microscopic description of this gravitational entropy in terms of underlying quantum degrees of freedom has long been considered a necessary checkpoint for any viable theory of quantum gravity. Some of the earliest proposals for this microscopic description \cite{Sorkin1983,Bombelli1986,Srednicki1993,Frolov1993} suggested that this entropy could be related to quantum correlations between the two regions separated by the horizon. In modern language, this quantum entropy is called entanglement entropy.

Entanglement entropy is a general feature of a quantum system associated with separating the fundamental degrees of freedom into two subsystems. Explicitly let $\rho_A$ be the reduced density matrix associated with subsystem A, obtained by summing over the degrees of freedom in the complement of A. The entanglement entropy is given by the von Neumann entropy of the reduced density matrix\footnote{For a quantum field theory, this quantity is UV divergent and is regulated with a short distance cut-off. }
\ban{
S_{EE} =- \text{Tr}[ \rho_A \log \rho_A] \,.
}

Any quantum theory of gravity should include an entanglement entropy associated with separating its quantum gravitational degrees of freedom. In the low energy limit these degrees of freedom are given by the geometry of spacetime itself, so we will refer to this entropy as characterizing `spacetime entanglement.' It is conjectured that in a complete theory, the notion of spacetime entanglement should be defined not only for horizons but for an arbitrary bipartition of spacetime. 

In this essay, we use holography to construct an interpretation of the Bekenstein-Hawking entropy formula evaluated on an arbitrary `hole' in a holographic spacetime as the differential entropy of a family of intervals on its boundary. We hope this holographic construction provides a hint for understanding spacetime entanglement for general regions. Unless otherwise noted, our discussion is restricted to three dimensional spacetimes.

\section{The Spacetime Entanglement Conjecture}

Recently, the idea that the Bekenstein-Hawking entropy formula applies more generally has been stated explicitly as the `spacetime entanglement conjecture' for a $d+1$ dimensional spacetime\cite{Bianchi2012}:
\begin{quote}
\emph{In a theory of quantum gravity, any states describing a smooth spacetime geometry manifest the following property: for any sufficiently large region, the entanglement entropy between the degrees of freedom describing the given region with those describing its complement is finite and to leading order, is given by}
\end{quote}
\ban{
S_\text{grav}= 2\pi \frac {\mathcal A}{\ell_P^{d-1}}
}
where $\mathcal A$ is the area of the `entangling surface' dividing the two regions\footnote{Note that in general the entangling surface will be co-dimension two, so we use `area' generally.} and $\ell_p$ is the Planck length, with $\ell_p^{d-1}= 8\pi G_N$. Note that this formula is simply the Bekenstein-Hawking entropy formula written in fundamental units. One of the implicit assumptions of this conjecture is that the low energy theory reduces to Einstein gravity to leading order. 

In particular, it was observed that the entanglement entropy of a generic region $A$ in a quantum field theory takes the form
\ban{
S_{EE}=c_0 \frac{R^{d-1}}{\delta^{d-1}}+c_2 \frac{R^{d-3}}{\delta^{d-3}}+\cdots
}
where $R$ is some scale which characterizes the geometry of the boundary of $A$, and $\delta$ is a short distance cutoff. For quantum gravity, one expects $\delta \sim \ell_P$, and so this expansion hints at the association of the leading term with the Bekenstein-Hawking formula. In \cite{Bianchi2012}, this conjecture was supported by various lines of evidence from perturbative quantum gravity as well as from simplified models of induced gravity and loop quantum gravity. Additionally, some interesting evidence for this conjecture comes from holography. This subject will be the main focus of this essay.

\section{Holographic Entanglement Entropy}

Holography provides a natural setting for studying spacetime entanglement.\footnote{See \cite{Mcgreevy2009,Horowitz2006} for reviews of the AdS/CFT correspondence and holography.} In holography, it is conjectured that the degrees of freedom of quantum gravity can be completely described in terms of a set of degrees of freedom defined on its boundary. This notion of holography was first proposed in \cite{Hooft1993, Susskind1995}, with an explicit realization given in \cite{Maldacena1998} as a correspondence between type IIB string theory and $\mathcal N=4$ super-Yang-Mills gauge theory. 

In a certain limit, this realization can be thought of as a duality between a five dimensional anti-de Sitter spacetime (the bulk) and a four dimensional conformal field theory on its boundary, or the AdS/CFT correspondence.  Indeed, a number of models of quantum gravity have been shown to have an AdS/CFT correspondence in an appropriate limit. Schematically, this correspondence relates the partition function of the conformal field theory to that of the gravity theory. We can compute the gravity partition function via the saddle-point approximation \cite{Gubser1998, Witten1998}, and therefore write 
\ban{
\mathcal Z_{\text{CFT}}\simeq \left. e^{-S_{\text{grav}}(\text{AdS})}\right.
}
where $S_{\text{grav}}$ is the supergravity action consisting of Einstein gravity coupled to a variety of matter fields in the bulk. Over the past fifteen years, this conjecture has been shown to hold for many settings and applications, building up a general dictionary for this gauge/gravity duality. 

This holographic dictionary in principle provides a method for computing $n$-point functions of various operators in the boundary theory as a straightforward gravitational calculation in the bulk, and therefore it is often used to study regimes of quantum systems inaccessible to standard approaches. In this essay however, we take the complementary perspective, that is we seek to translate a poorly understood concept in quantum gravity into the language of the boundary field theory.

\begin{figure}[h!]
\begin{center}
\includegraphics[width=0.5\textwidth]{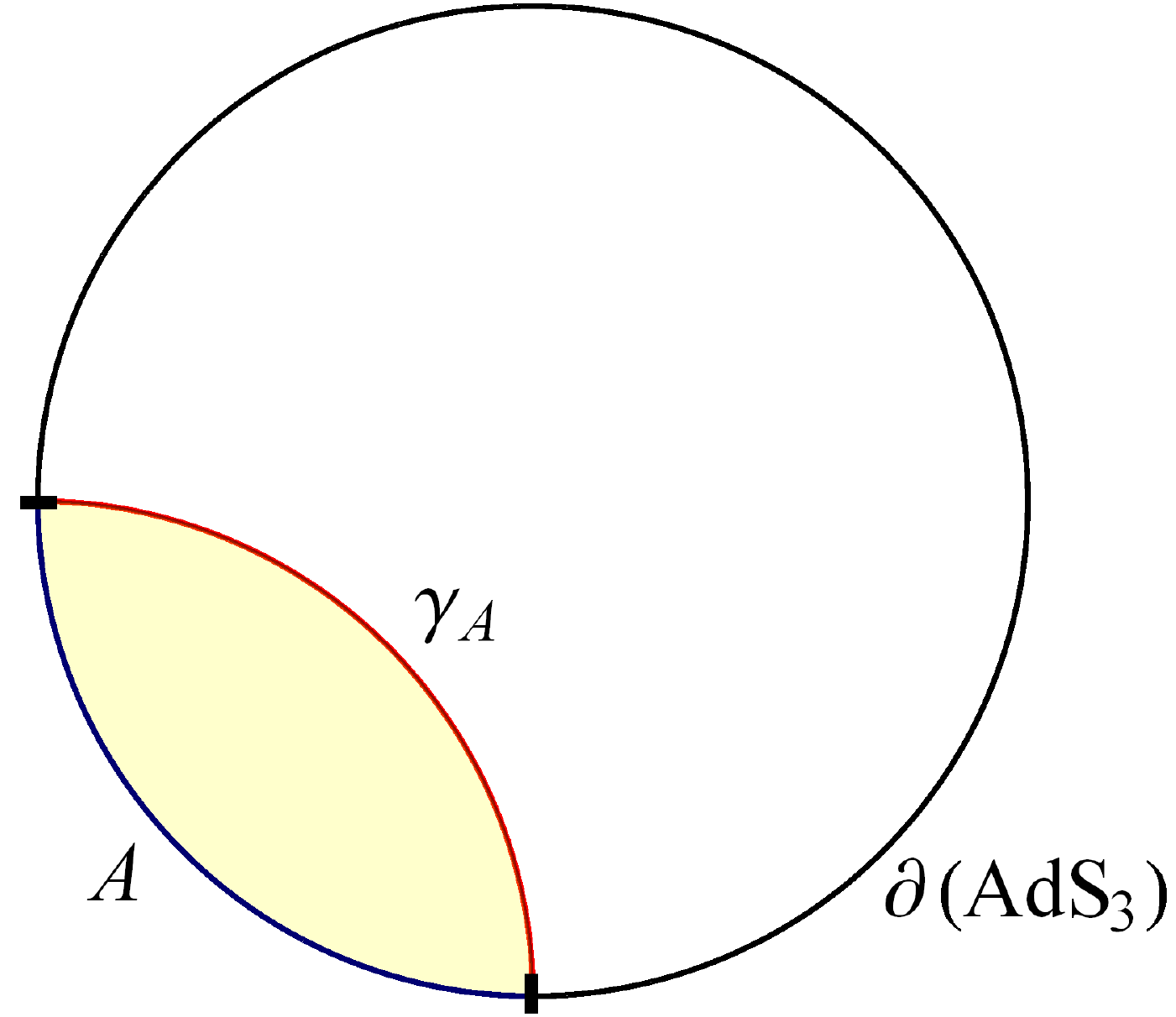}
\caption{The extremal curve $\gamma_A$ is shown above for an interval $A$ on the boundary of AdS$_3$, drawn on a constant time slice.}
\label{HEE}
\end{center}
\end{figure}

A powerful tool to come out of holography has been the Ryu-Takayanagi (RT) formula for holographic entanglement entropy.\footnote{See \cite{Ryu2006a,Nishioka2009,Takayanagi2012} for overviews of holographic entanglement entropy.} First proposed in \cite{Ryu2006}, this formula states that the entanglement entropy of a region $A$ on the boundary of a $d+1$ dimensional holographic spacetime is
\ban{S(A)= \frac{\mathcal A(\gamma_A)}{4G_N^{(d+1)}} \label{RT}}
where $\gamma_A$ is the $d-1$ dimensional extremal area surface extending into the bulk whose intersection with the boundary is the boundary of $A$. An example of a region and its corresponding extremal curve is shown for AdS$_3$ in fig. \ref{HEE}. 

This formula has been explicitly checked for many cases where the entanglement entropy in the boundary field theory is known, and a general argument for its validity was given by \cite{Lewkowycz2013}. Note however that the RT formula only applies on a constant time slice of a static spacetime. A covariant generalization was proposed in \cite{Hubeny2007}, but this proposal has not been as extensively studied. It remains an open problem to establish general arguments for its validity. 

Holographically, the bulk spacetime described by Einstein gravity emerges as a classical limit of an underlying quantum theory. Applying the spacetime entanglement conjecture, we therefore expect the Bekenstein-Hawking formula evaluated on any bipartition of spacetime to have a holographic interpretation in terms of the boundary theory. One of the simplest realizations of this idea is the observation that the holographic prescription given by the RT formula associates the entanglement entropy of a region $A$ with the Bekenstein-Hawking entropy of the surface $\gamma_A$. 

A main limitation of this approach however is that it only applies to extremal surfaces in the bulk which are homologous to some boundary region. More generally, we would like to be able to construct a boundary observable for the Bekenstein-Hawking entropy for any bipartition of spacetime, \ie an arbitrary co-dimension two `hole.'

\section{Holographic Holes and Differential Entropy}
\label{holoholes}

Recently \cite{Balasubramanian2014} showed that a for closed curve on a constant time-slice of AdS$_3$, the Bekenstein-Hawking entropy formula evaluated on this curve is equal to a quantity on the boundary called differential entropy. Given a family of $n$ intervals $\{I_k\}$ which cover a time slice of the boundary, the differential entropy is given by
 \ban{
E=\lim_{n\to \infty}\sum_{k=1}^n \[\, S(I_k)-S(I_k\cap I_{k+1})\,\]
\,. \label{residue}}
We review this `hole-ographic' construction in section \ref{time}. 

For more general cases, we find that instead of considering an infinite family of intervals on a two dimensional boundary, it is more convenient to consider the intervals directly in the continuum limit, as defined by two curves which we denote $\gamma_L(\la)$ and $\gamma_R(\la)$. These curves denote the endpoints of an interval labeled by $\la$. We pass between the discrete and continuum descriptions by associating for each interval $I_k$ the parameter $\la_k=\frac k n$ taking $n\to \infty$, so that each interval is parameterized by $\la \in [0,1]$. Additionally, we denote the entanglement entropy of an interval $k$ by $S(\gamma_L(\la_k), \gamma_R(\la_k))$.\footnote{For $d$ dimensional boundaries with $d>2$ we instead consider a $d-1$ dimensional strip that consists of an interval that is translated along the $d-2$ `planar' spatial dimensions. Explicitly we have $\gamma_L(\la,\sigma_i)=\{t_L(\la),x_L(\la), \sigma_1, \cdots, \sigma_{d-2}\}$. We explicitly characterize this `planar symmetry' in appendix \ref{planarS}.}

\begin{figure}[h!]
\centering
\subfloat[]{\includegraphics[width=0.55\textwidth]{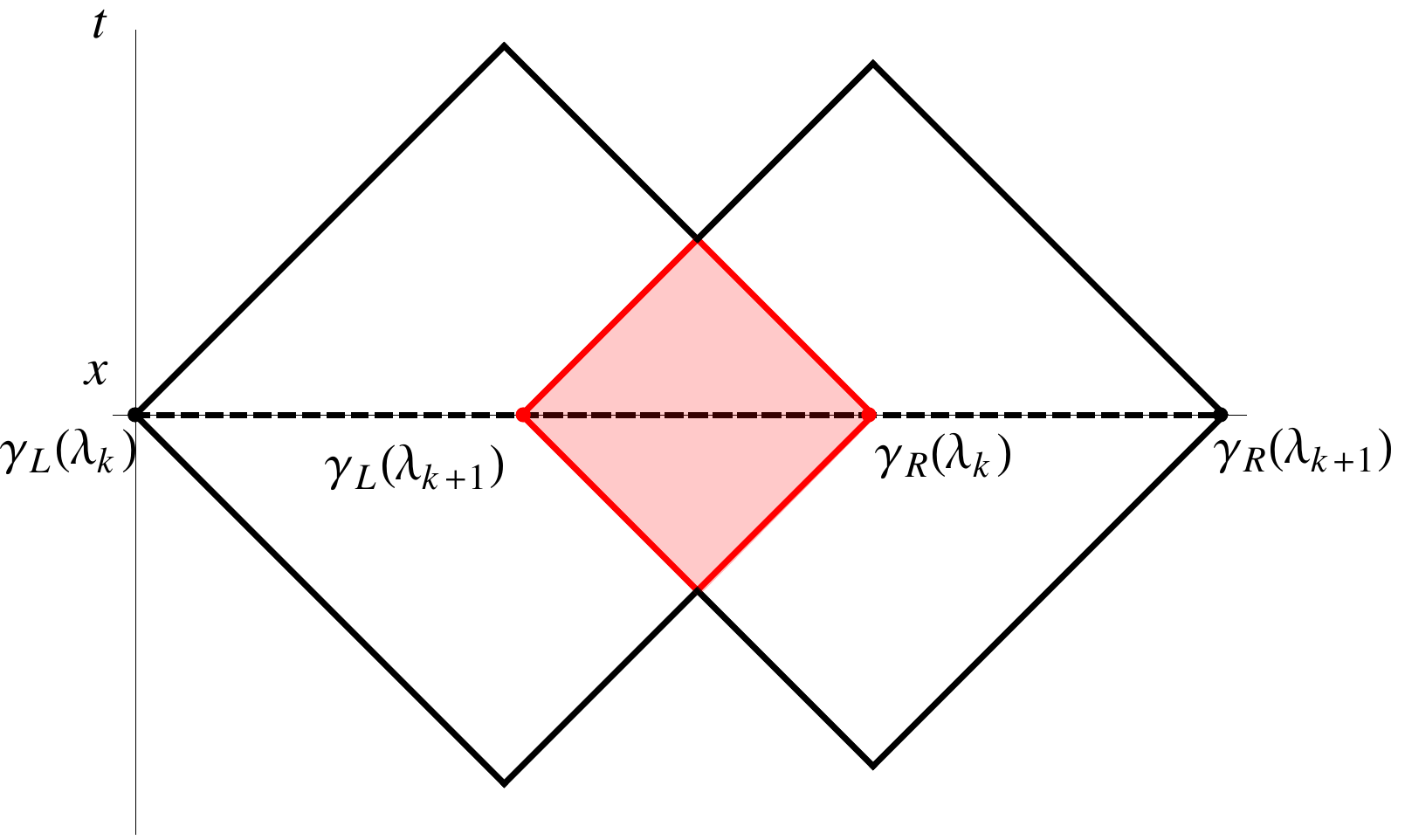}}\qquad
\subfloat[]{\includegraphics[width=0.55\textwidth]{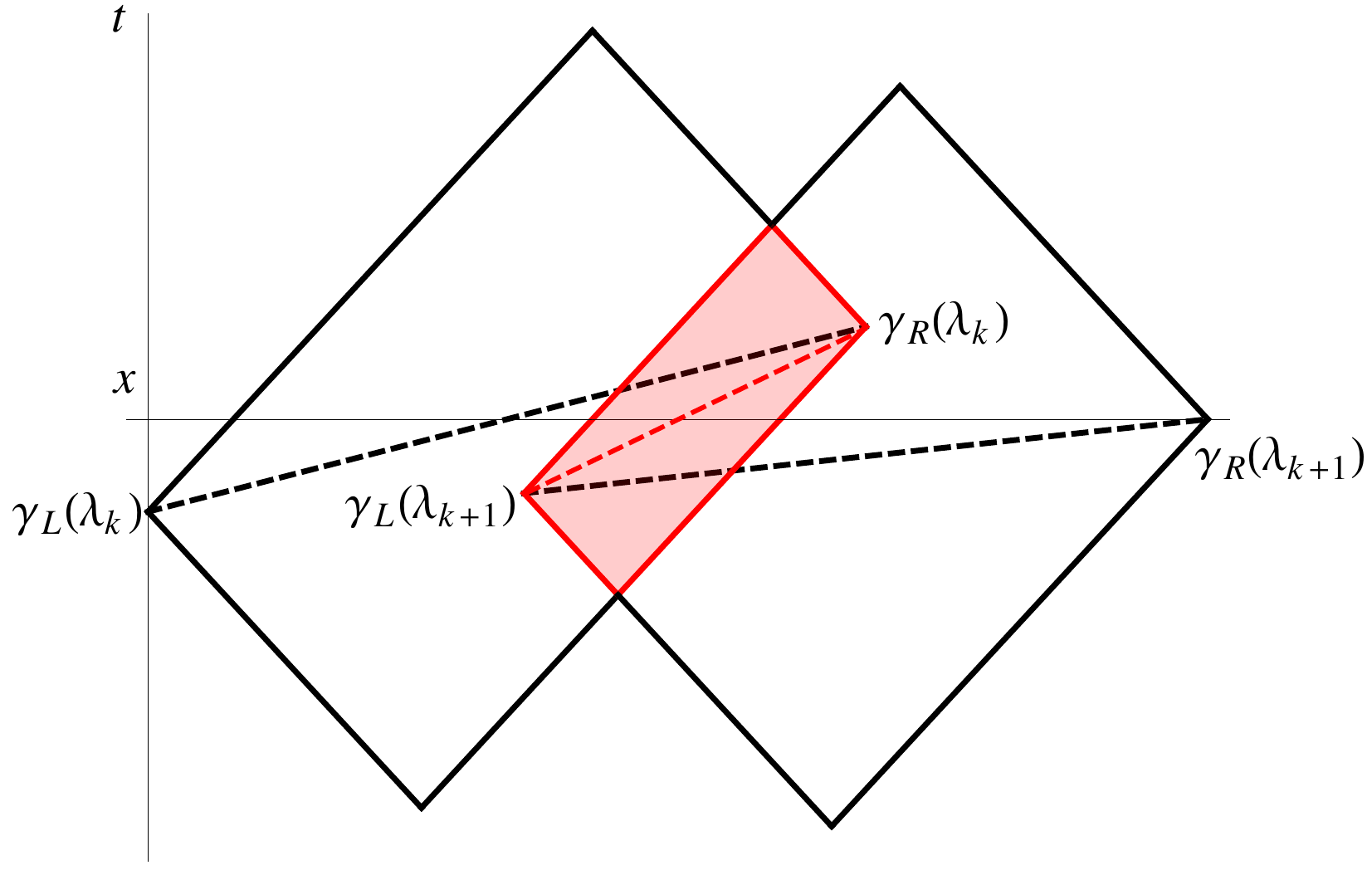}}
\caption{The causal diamond for the interval $I_k \cap I_{k+1}$ for two intervals on a constant time slice is shaded in red above in (a). When the intervals do not lie on the same time slice as in (b), the appropriate interval has endpoints $\gamma_L(\la_{k+1})$ and $\gamma_R(\la_k)$. }
\label{LRintersect}
\end{figure}

Note that if we do not constrain the neighboring intervals $I_k$ and $I_{k+1}$ to lie on the same time slice, there may be some question as to what we mean by $I_k\cap I_{k+1}$. As illustrated in figure \ref{LRintersect}(a), when the intervals lie on the same time slice, the intersection of the intervals $I_k\cap I_{k+1}$ has endpoints $\gamma_L(\la_{k+1})$ and $\gamma_R(\la_k)$. In the general case we take these endpoints to define the intersection, \ie we take $S(I_k \cap I_{k+1})$ to mean $S(\gamma_L(\la_{k+1}), \gamma_R(\la_k))$. This definition becomes intuitively clear when we consider the intersection of the causal diamonds of $I_k$ and $I_{k+1}$ as in figure \ref{LRintersect}(b). 

In the continuum limit we have
\ban{
\lim_{\substack{n\to \infty}}
\[\, S(I_k)-S(I_k\cap I_{k+1})\,\]&=
S(\gamma_L(\la), \gamma_R(\la))-S( \gamma_L(\la+d\la), \gamma_R(\la))
\nonumber\\
&=
-\frac{dS(\gamma_L(\la),\gamma_R(\la))}{d\gamma^a_L(\la)}\,\frac{d\gamma^a_L(\la)}{d\la}\,d\la\,.
\label{differ}
} 
Therefore \reef{residue} becomes
 \ban{
E&=-\int_0^1d\la\ \frac{dS(\gamma_L(\la),\gamma_R(\la))}{d\gamma^a_L(\la)}\,
\frac{d\gamma^a_L(\la)}{d\la}
\label{left}\\
&=\int_0^1d\la\ \frac{dS(\gamma_L(\la),\gamma_R(\la))}{d\gamma^a_R(\la)}\,
\frac{d\gamma^a_R(\la)}{d\la}\,.
\label{right}
 }
The second expression comes from considering instead the continuum limit of\\
$ S(I_k)-S(I_{k-1}\cap I_{k})$. Assuming periodic boundary conditions, \ie $S(\gamma_L(0),\gamma_R(0))=S(\gamma_L(1),\gamma_R(1))\,$, we can integrate by parts to show the equality of these two expressions. As evident from the expression, we can think of differential entropy as the directional derivative of the entropy functional along a family of intervals, hence the name. 

In the remainder of this essay, we discuss holographic constructions relating the differential entropy of a family of boundary intervals with the gravitational entropy of a bulk curve. In chapter \ref{time}, we review the construction of \cite{Balasubramanian2014} and extend it to holes which can also vary in time. Additionally, we review the proof of \cite{Headrick2014} which establishes this holographic relation in more generality. In chapter \ref{new}, we show how one can build the bulk curve by considering intersections of neighboring entanglement wedges of a family of boundary intervals in a continuum limit. We close with directions for future research in chapter \ref{discuss}. 

Multiple appendices are attached. In appendix \ref{geom} we fill in some of the details glossed over in the geometric arguments in chapter \ref{new}. Additionally, with a characterization of `planar symmetry' given in appendix \ref{planarS}, in appendix \ref{AdSn} we construct a higher dimensional generalization of chapter \ref{time}. Further, we comment on how this framework extends to higher curvature theories of gravity in appendix \ref{lovelock}. Finally in appendix \ref{scaledE}, we show that generic associations of a bulk curve with a family of boundary intervals yields a differential entropy which is scaled relative to the gravitational entropy in a natural way.

\chapter{Holographic Holes in AdS$_3$}
\label{time}
In this chapter, we review the discussion of \cite{Balasubramanian2014,Myers2014} and generalize to arbitrary space-like bulk surfaces which can vary in time. This construction motivates the holographic lemma reviewed in section \ref{general}. 
To simplify the discussion, we outline the construction explicitly for AdS$_3$, however, as we will see this procedure readily extends to higher dimensions, to other holographic backgrounds (\ie backgrounds that are not asymptotically AdS) and to
certain classes of higher curvature gravity theories by the general argument of section \ref{general}. The example of applying this `hole-ographic' construction to time varying holes in higher dimensions can be found in appendix \ref{AdSn}. 

Given a space-like curve in AdS$_3$, we construct a family of boundary intervals whose differential entropy is equal to the gravitational entropy of the original curve. We will work in Poincar\'e coordinates with metric
\ban{ds^2= \frac {L^2}{Z^2}\(dZ^2 -dT^2 +dX^2\)\label{3metric}}
where $L$ is the AdS radius. Let the initial curve in the bulk be specified by the parameterization $\bulkc_B(\la)=\lbrace Z(\la), X(\la), T(\la) \rbrace$ where $0\le\la\le1$. In addition, we impose periodic boundary conditions and rescale our parameterization so that $\bulkc_B(\la=0)=\bulkc_B(\la=1)$. As described in the introduction, we specify the corresponding
family of intervals on the asymptotic boundary at $z=0$ by the two endpoint curves: $\ga_L(\la)=\lbrace x_L(\la), t_L(\la) \rbrace$ and $\ga_R(\la)=\lbrace  x_R(\la), t_R(\la) \rbrace$.  Implicitly, here and throughout this essay, we are imposing that the $x$ direction is periodic with period $\Delta x=\ell$. One should think of the latter as some
infrared regulator scale, \ie it ensures that the proper length of the bulk curves considered here are finite. We assume that $\ell$ is always much
larger than the proper length of any of the intervals defined by $\ga_L$ and $\ga_R$.

The quantities we wish to compute are defined via volume functionals in Einstein gravity, and so this setup has a built-in notion of `reparameterization invariance,' which we apply both to the Bekenstein-Hawking formula evaluated on the bulk curve and on the extremal curves determining the entanglement entropy in the boundary theory. Under reparameterization of $\bulkc_B(\la)$ via $\lambda \to \tilde \lambda$, the entropy of the hole given by the Bekenstein-Hawking formula \reef{BH} is unchanged as the volume functional keeps the same form, \ie
\ban{S_{BH}=\frac 1 {4 G_N} \int_0^1  \sqrt{g_{\mu\nu}\pd{x^\mu}\lambda\pd{x^\nu}\lambda} d\lambda \ =\  \frac 1 {4 G_N} \int_0^1  \sqrt{g_{\mu\nu}\pd{x^\mu}{\tilde \lambda}\pd{x^\nu}{\tilde \lambda}} d\tilde \lambda\,.
\label{reparam1}}

Similarly we have reparameterization invariance for an extremal curve in the bulk, which determines the 
holographic entanglement entropy for an interval at fixed $\la$. Let $s$ be the `time' parameter
on these extremal curves, \ie  $\Gamma(s;\la)=\lbrace z(s;\la), x(s;\la), t(s;\la) 
\rbrace$ with the boundary conditions $\Gamma(s=0;\la)=\lbrace 0, 
\gamma^a_L(\la)\rbrace$ and $\Gamma(s=1;\la)=\lbrace 0, \gamma^a_R(\la)\rbrace$.
Then, since the entropy functional is analogous to that above, reparameterizations $s\to\tilde s$ do not
 change the entropy of the interval at any given $\lambda$.

\section{Holes at Constant $Z$}
\label{VarTConstZ}
Next we show explicitly how to construct an appropriate family of intervals $\gamma_L(\la)$, $\gamma_R(\la)$ from the initial curve $\bulkc_B(\la)$ in the bulk, beginning with a re-derivation of the results of \cite{Balasubramanian2014}. For each $\lambda$, we follow the extremal curve tangent to $\bulkc_B(\lambda)$ to the boundary, and the intersection of each extremal curve with the boundary defines the endpoints $ \gamma_L(\lambda)$ and $ \gamma_R(\lambda)$. Stated in this way, this  prescription straightforwardly extends to more general cases.\footnote{\eg by considering tangent surfaces in higher dimensions.}

\begin{figure}[h!]
\begin{center}
\includegraphics[width=0.6\textwidth]{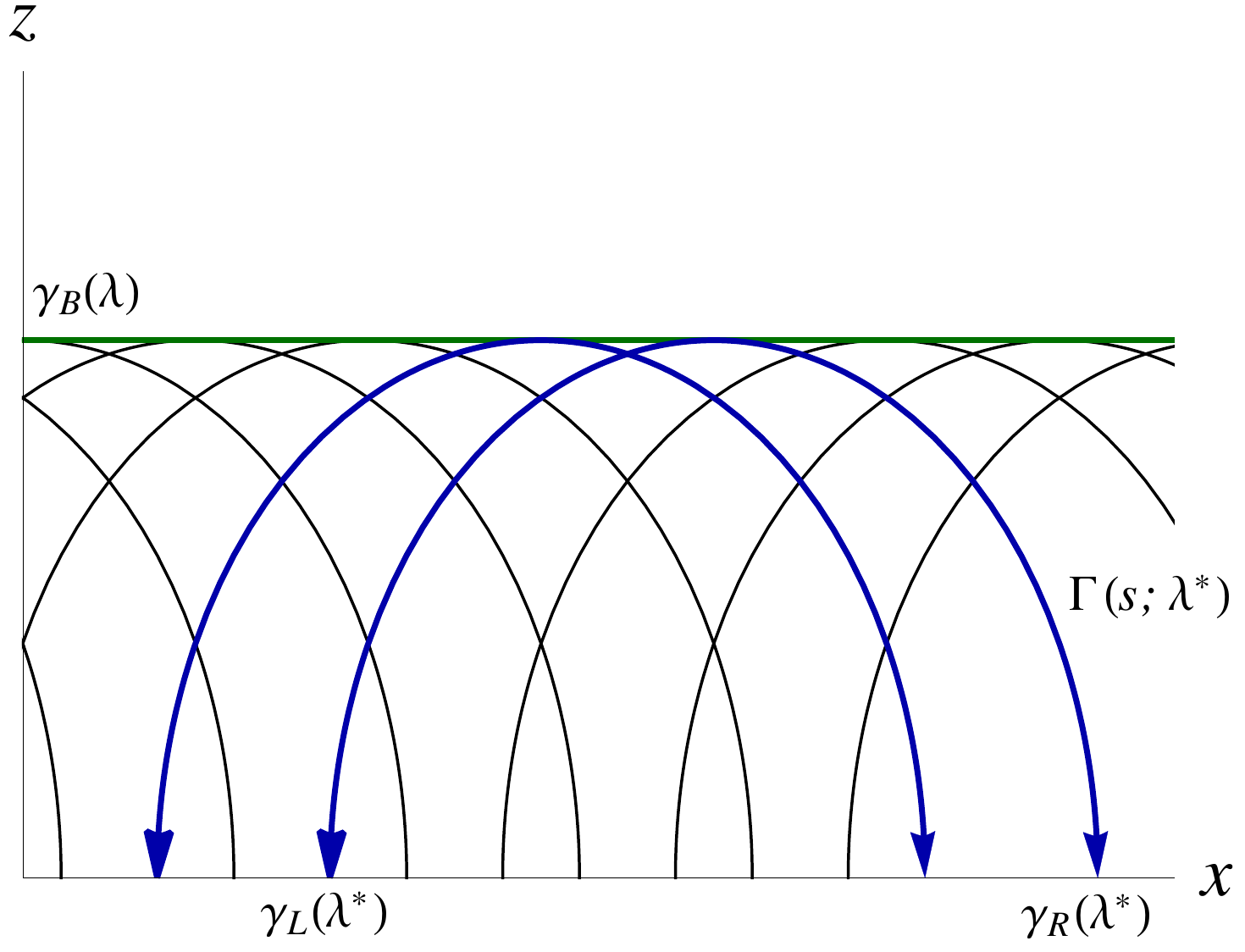}
\caption{The bulk curve $\bulkc_B(\la)$ is shown above in green, along with the tangent geodesics at each point. One such geodesic $\Gamma(s;\lambda^*)$ is highlighted in blue, along with a neighboring geodesic at $\lambda^*-d\lambda$. The points $\gamma_L(\lambda^*)$ and $\gamma_R(\lambda^*)$ are explicitly drawn on the boundary at $z=0$. }\label{intervals}
\end{center}
\end{figure}

For simplicity, let us first consider a bulk curve $\bulkc_B(\la)$ at constant $z=Z_0$ and $t=T_0$ \ie $\bulkc_B(\la) = \lbrace Z_0, \ell\la, T_0 \rbrace$ --- recall that $\la\in[0,1]$ and $\ell$ is the period in the $x$ direction.
In this case the tangent curve is given by a semicircle parameterized by 
\ban{ \Gamma(s; \la) = \lbrace  Z_0 \sin s,\ell\lambda+ Z_0 \cos s,T_0\rbrace  \label{geod}}
where $s\in [0, \pi]$. Therefore we have 
\ban{\gamma_L (\lambda) = \coords{ \ell\lambda - Z_0,T_0} \hspace{0.5cm}\text{and}\hspace{0.5cm}  \gamma_R(\lambda)=\coords{\ell\lambda+Z_0,T_0} \label{bound0}\,. }
The general setup is illustrated in figure \ref{intervals}.

The entanglement entropy of a single interval is given holographically by \cite{Ryu2006}
\ban{S(\gamma_L(\lambda), \gamma_R(\lambda))=\frac L {2G_N} \log \[\frac {x_R(\lambda)-x_L(\lambda)}\delta
 \]=\frac L {2G_N} \log \[\frac {2Z_0}\delta \]\label{entropy0}}
where $\delta$ is the short-distance cut-off for the boundary CFT.
We can compute the differential entropy \reef{differ} to get
\ban{E= \frac L{4G_N}\, \int _0^1 \frac{\ell}{Z_0}\,d\lambda \,. \label{diff0}}
Comparing this to gravitational entropy \reef{BH} applied to $\bulkc_B$ we have
\ban{S_{BH}=\frac L{4G_N}\, \int_0^1 \frac{\ell}{Z_0}\,d\lambda \label{area0}}
and hence $E=S_{BH}$. 

\vspace{0.2cm}

We now let the bulk curve $\bulkc_B(\la)$ vary in time and be parameterized by 
\ban{
\bulkc_B (\lambda) = \lbrace Z_0, \ell\lambda, T(\lambda) \rbrace\,.
}
For each point on $\bulkc_B(\la)$, we can construct the tangent extremal curve by following a geodesic in the direction of the tangent vector to the boundary. At a given $\lambda$, the tangent vector is proportional to 
\ban{
u(\lambda)= \lbrace 0,\ell ,T'(\lambda)\rbrace \,.
}
To find the geodesic along this tangent vector, we take advantage of the Lorentz symmetry in the $x^\mu$ coordinates of AdS space. First we boost by angle $\beta(\la) = \log {\sqrt{\frac{\ell+ T'(\la)}{\ell- T'(\la)}}}$ so that the tangent vector has vanishing time-like component. In this boosted frame, the correct geodesic is given by $\Gamma^*(s;\la)=\lbrace Z_0 \sin s, \ell\lambda+ Z_0 \cos s, T(\la)\rbrace$. 
We apply the inverse boost to construct the geodesic tangent to the bulk curve in the original coordinate system:
\ban{
\G(s; \la)=\coords{ Z_0 \sin s,\ell\la+\frac{\ell\ Z_0 \cos s}{\sqrt{\ell^2-T'(\la)^2}},T(\la)+\frac{ T'(\la)Z_0 \cos s}{\sqrt{\ell^2-T'(\la)^2}}} \label{geod1} \,.
}
This extremal curve intersects the AdS$_3$ boundary at $s=0$ and $s=\pi$, and so the family of intervals is given by 
\ban{\gamma_{R,L} (\lambda) = \coords{ \ell\lambda\pm \frac{\ell\, Z_0}{\sqrt{\ell^2-T'(\la)^2}}, T(\la)\pm \frac{ T'(\la)Z_0}{\sqrt{\ell^2-T'(\la)^2}}}\label{family1}}
where the + and -- signs are chosen for $\gamma_R$ and $\gamma_L$, respectively.

To compute the entanglement entropy of each interval, we compute it in the boosted frame, where the result is known \reef{entropy0}, and carry it over to the original coordinates by Lorentz symmetry. Hence
\ban{S(\gamma_L(\lambda), \gamma_R(\lambda))=\frac L{2G_N}\, \log \[\frac {|\gamma_R-\gamma_L|}{\delta} \]=\frac L{4G_N}\, \log \[\frac {(x_R-x_L)^2-(t_R-t_L)^2}{\delta^2} \] \label{Sboost}}
Substituting the expressions for the endpoint curves \reef{family1} into our formulae, we have
\ban{E = \frac {L}{4G_N} \int _0^1 \frac 1{Z_0} \sqrt{\ell^2-T'(\la)^2}\ d\la\,. \label{ent1}}
Note that there is no total derivative contribution here since $Z'(\la)=0$ --- compare with eq.~\reef{entropy2}.
The gravitational entropy of $\bulkc_B(\la)$ given by eq.~\reef{reparam1} is 
\ban{S_{BH}=\frac {L}{4G_N} \int _0^1 \frac 1{Z_0} \sqrt{\ell^2-T'(\la)^2}\ d\la \label{vol1}\,.}
Comparing eqs.~\reef{ent1} and \reef{vol1}, we see that in this case $E=S_{BH}$. Note that $\bulkc_B(\la)$ is assumed to be space-like everywhere, so $|T'(\la)^2|<\ell$. 

\section{An Arbitrary Hole}\label{arbhole}

\begin{figure}[h!]
\begin{center}
\includegraphics[width=0.6\textwidth]{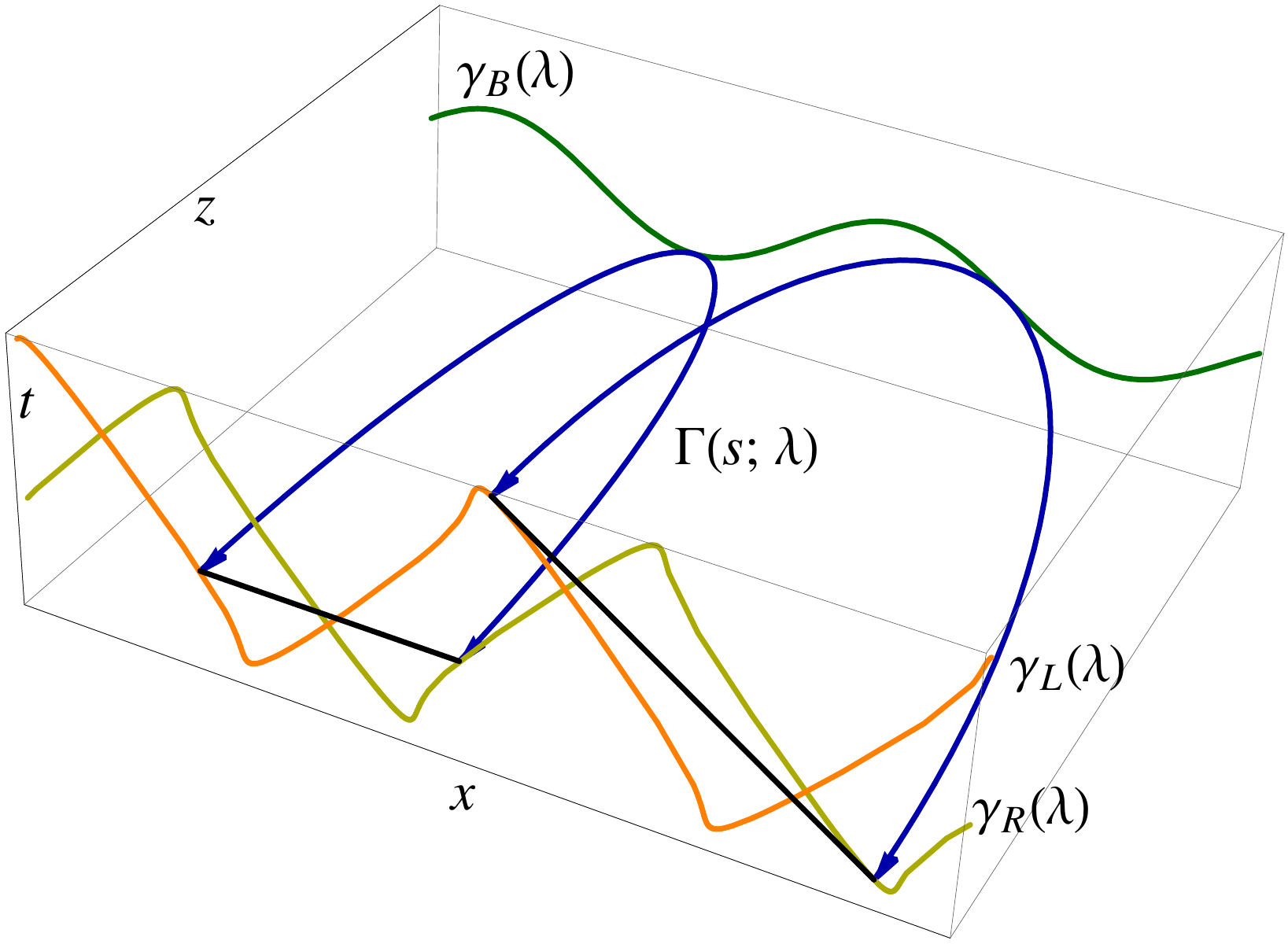}
\caption{For each point on the bulk curve $\bulkc_B(\la)$, the intersection of the extremal curve $\Gamma(s;\lambda)$ with the boundary defines an interval between $\gamma_L(\la)$ and $\gamma_R(\la)$. We take the family of intervals as described by the curves $\gamma_L(\la),\gamma_R(\la)$ shown in yellow and orange respectively. The differential entropy of this family of intervals equals the gravitational entropy of the bulk curve.}
\label{3Dintervals}
\end{center}
\end{figure}

We now consider an arbitrary bulk curve $\bulkc_B(\la)=\coords{Z(\lambda),X(\lambda), T(\lambda)}$ with the condition that its tangent vector is space-like everywhere. To find the tangent extremal curve at a point, we again begin by boosting the tangent vector by $\beta(\la) = \log \sqrt{\frac {X'(\la)+T'(\la)}{X'(\la)-T'(\la)}}$ so it is completely space-like. In the boosted coordinates, the tangent vector is proportional to 
\ban{u^*(\lambda)=\coords{Z'(\la), \sqrt{X'(\la)^2-T'(\la)^2}, 0} \label{tan2}\,.}
As constant time geodesics in AdS$_3$ are given by semicircles, we can use Euclidean geometry in the ($z,x$)-plane to characterize the extremal curve. The tangent vector $u^*(\la)$ lies on a semi-circle, so following its normal vector $n^*(\la)$ to the boundary gives its center. We choose the length $n^*(\la)$ such that $\bulkc_B(\lambda)+n^*(\la)$ lies on the boundary, so the coordinate radius of the semi-circle containing the geodesic is equal to $|n^*(\la)|$. We have
\ban{
n^*(\lambda)&= \frac{Z(\la)}{\sqrt{X'(\la)^2- T'(\la)^2}} \coords{  -\sqrt{X'(\la)^2- T'(\la)^2},  Z'(\la) ,0}\,.
}
So $c^*(\lambda) = \coords{0, X(\lambda) +\frac {Z(\la)Z'(\la)}{\sqrt{X'(\la)^2- T'(\la)^2}},T(\la)}$ is the center of the semi-circle in the boosted coordinates and $r^*(\lambda)\equiv Z(\la)\sqrt{1+ \frac { Z'(\la)^2}{X'(\la)^2-T'(\la)^2}}$ is the radius. Therefore we can parameterize this semicircle and boost back to the original coordinate system to get the tangent extremal curve as 
\ban{
\notag  \G(s; \la)= &\left\{r^*(\la)  \sin s,X(\la)+ \frac{Z(\la) Z'(\la)X'(\la) }{{X'(\la)^2- T'(\la)^2}}+\frac{X'(\la)\, r^*(\la)}{\sqrt{X'(\la)^2- T'(\la)^2}}\cos s,\right.\\
&\hspace{0.6cm}\left.T(\la)+ \frac{Z(\la) Z'(\la)T'(\la) }{{X'(\la)^2- T'(\la)^2}}+\frac{T'(\la)\, r^*(\la)}{\sqrt{X'(\la)^2- T'(\la)^2}}\cos s\right\}\,.
\label{housefire}}
The bulk curve and some tangent extremal curves are shown in figure \ref{3Dintervals}. 

Given this parameterization, it is straightforward to compute the differential entropy via \reef{differ} as
\ban{
\notag E= \frac L{4G_N}\int_0^1 d\la & \left( \frac 1 {Z(\la)}{ \sqrt{X'(\la)^2-T'(\la)^2+Z'(\la)^2}}\right.\notag\\
&\left.+\frac{Z''(\la)}{\sqrt{X'(\la)^2-T'(\la)^2+Z'(\la)^2}} +  \frac{Z'(\la)}{X'(\la)^2-T'(\la)^2}\frac{T'(\la)T''(\la)-X'(\la)X''(\la)}{\sqrt{X'(\la)^2-T'(\la)^2+Z'(\la)^2}}\right) \notag\\
\notag =\frac L {4G_N}\int_0^1d\la&  \frac 1{ Z(\la)}{ \sqrt{X'(\la)^2-T'(\la)^2+Z'(\la)^2}}  +\left.\frac L{4G_N} \sinh^{-1}\left({\frac{ Z'(\la)}{\sqrt{X'(\la)^2- T'(\la)^2}}}\right)\right|_{0}^1\\
=\frac L {4G_N}\int_0^1d\la & \frac 1 {Z(\la)}{ \sqrt{X'(\la)^2-T'(\la)^2+Z'(\la)^2}} \label{entropy2} 
}
where the boundary term vanishes by the periodic boundary conditions for $\bulkc_B(\la)$. Computing the gravitational entropy for $\bulkc_B(\la)$, we have
\ban{
S_{BH}=\frac L{4G_N}\int_0^1 d\lambda  \frac 1 {Z(\la)} \sqrt{X'(\la)^2-T'(\la)^2+Z'(\la)^2} \label{vol2}
}
Therefore we see that for any space-like curve in AdS$_3$, $E=S_{BH}$. Note that in the case where $T'(\la)=0$, eqs. \reef{entropy2} and \reef{vol2} reduce to formulas found in \cite{Myers2014} for constant time bulk curves. In addition, this result extends straightforwardly to time-varying surfaces with `planar symmetry' in higher dimensions, and the details can be found in appendices \ref{planarS} and \ref{AdSn}.

\section{A Holographic Lemma}
\label{general}

In this section, we review the proof found in \cite{Headrick2014} of a holographic lemma which sheds light on the correspondence explored in this chapter. The discussion is actually quite general and allows for holes with planar symmetry in higher dimensions, general holographic backgrounds, and certain higher curvature bulk theories.\footnote{See appendices \ref{planarS} and \ref{AdSn} for a discussion of the higher dimensional generalization, and appendix \ref{lovelock} for a discussion of higher curvature theories.} 

The main innovations leading to this proof come from lessons found  in the constructions of the previous sections. First, we can think of the family of boundary intervals as characterized by the curves of endpoints $\gamma_L(\la)$ and $\gamma_R(\la)$, and second we note that the appropriate family of boundary intervals have extremal curves which are tangent to the bulk curve at each point. Additionally, note that solving for an extremal curve for a given interval reduces to extremizing the `action' given by the entropy functional as in eq.~\reef{reparam1}. Therefore we can consider the extremal curve for a given interval $\la$ as the classical trajectory for boundary conditions specified by $\gamma_L(\la)$ and $\gamma_R(\la)$, at the `time' parameter $s_i$ and $s_f$ respectively. As such, we will use the machinery of classical mechanics to prove a holographic lemma relating the differential entropy a family of boundary intervals to the gravitational entropy of a bulk curve. 

Consider an action
\ban{
S=\int_{s_i}^{s_f} ds \, \cL(q^a, \partial_s q^a)
}
where $\cL$ depends only on the coordinate functions and their first derivatives, and is manifestly reparameterization invariant under $s \to \tilde s$. 
Further consider a family of boundary conditions given by $\{s_i(\la), q_i^a(\la)\}$ and $\{s_f(\la), q_f^a(\la)\}$ that form a closed loop \ie $s_{i,f}(0)=s_{i,f}(1)$ and $q^a_{i,f}(0)=q^a_{i,f}(1)$. With this condition, we consider perturbing the endpoints of the classical trajectory. The change in action is
\ban{
\delta S_{on}= p_f^a \delta q_f^a- H_f \delta s_f - p_i^a \delta q_i^a + H_i \delta s_i + {\int ds [\text{eom}\cdot \delta q]} \label{vary}
}
where $p^a_{i,f}= \partial \cL/ \partial \dot q ^a|_{s=s_{i,f}}$. The equations of motion (eom) vanish because we are considering an on-shell trajectory. Additionally for a reparameterization invariant theory, the Hamiltonian vanishes. Therefore we have 
\ban{
\delta S_{on} = p_f^a \delta q_f^a - p_i^a \delta q_i^a \label{vary1}
}
and we can write
\ban{
\pd{S_{on}}{q_f^a} = p_f^a \hspace{1cm}\text{and}\hspace{1cm} \pd{S_{on}}{q_i^a} =- p_i^a \label{lemma2}\,.
}
Finally, integrating \reef{vary1} over the family specified by $\la$ produces a vanishing result as we are integrating a total derivative over closed boundary conditions. Therefore we have the following lemma 
\ban{
\int_0^1 d\la \, p_f^a \, \pd{q^a_f}\la=\int_0^1 d\la \, p_i^a \, \pd{q^a_i}\la \label{lemma}
}
where $p_{i,f}^a = \partial \cL/ \partial \dot q^a |_{s=s_{i,f}}$. 

In the context of holographic entanglement entropy in Einstein gravity, the action is reparameterization invariant as in eq.~\reef{reparam1}. Additionally only first derivatives of the coordinate functions appear, so we can apply the results of this lemma to differential entropy.

\begin{figure}[h!]
\begin{center}
\includegraphics[width=0.6\textwidth]{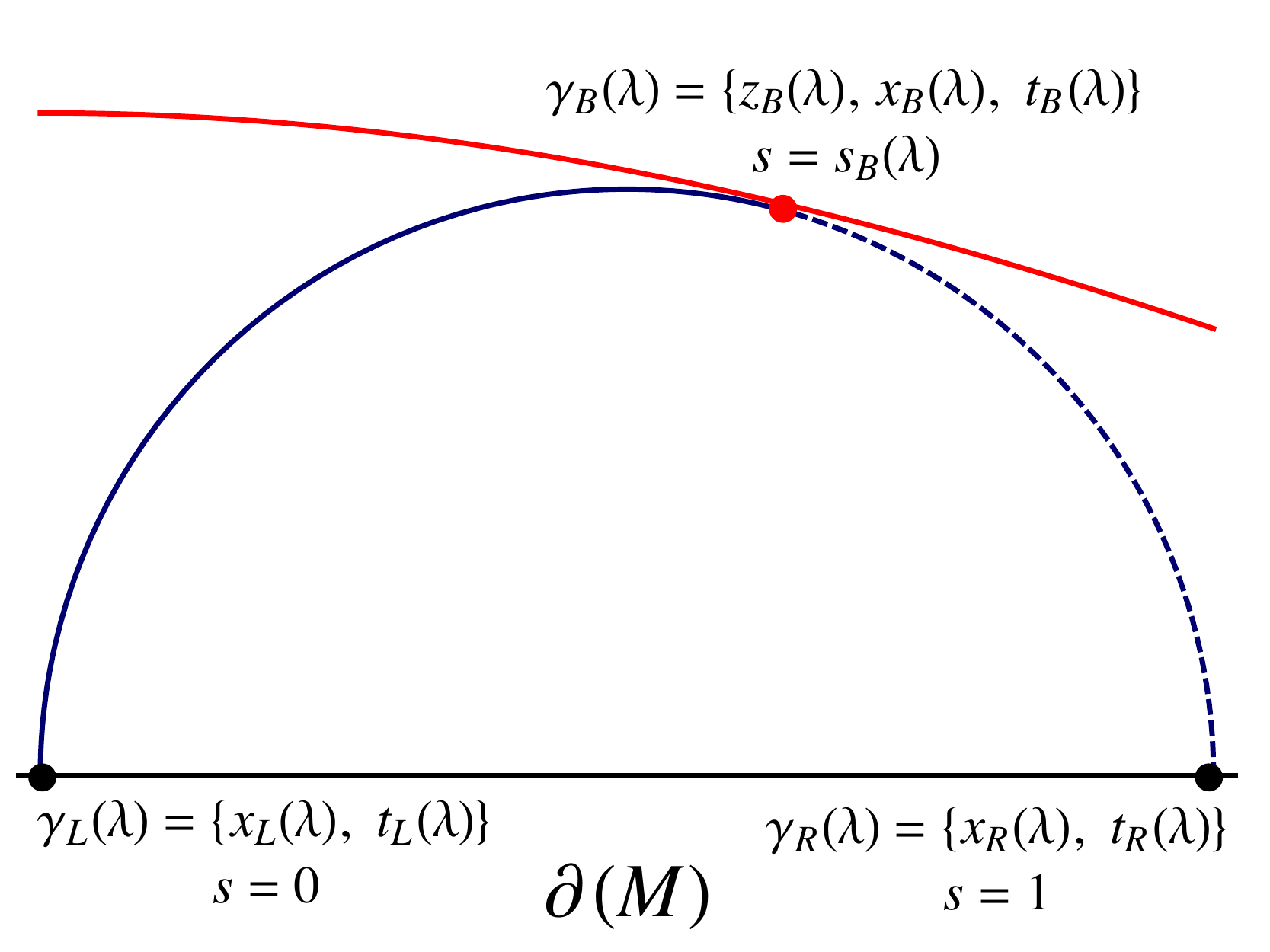}
\caption{We consider evaluating the action only on a portion of the classical trajectory from the boundary to the bulk surface of interest. In the case of tangent vector alignment we can apply the lemma \reef{almost} to equate the differential entropy of the family of boundary intervals with the Bekenstein-Hawking of the bulk curve.}\label{lemmaDiff}
\end{center}
\end{figure}

For a three dimensional holographic spacetime, the family of boundary conditions become the family of endpoints for boundary intervals, with the extremal curves representing the classical trajectories. Additionally, we consider a bulk curve given by $\gamma_B(\la)=\{z_B(\la), t_B(\la),x_B(\la)\}$.\footnote{For higher dimensions, we consider a bulk surface with planar symmetry along $d-2$ spatial directions denoted $\sigma_i$. Explicitly let $\gamma_B(\la,\sigma_i)=\{z_B(\la), t_B(\la),x_B(\la), \sigma_1, \cdots, \sigma_{d-2}\}$.} Let $s_B(\la)$ denote the parameter at which the extremal curve at $\la$ intersects the bulk surface, as illustrated in figure \ref{lemmaDiff}. 
We can apply \reef{lemma} to this setup to get
\ban{
-\int_0^1 d\la \deriv{S(\gamma_L(\la), \gamma_R(\la))}{\gamma^\mu_L(\la)} \deriv{\gamma^\mu_L(\la)}\la&=-\int_0^1 d\la \deriv{S(\gamma_L(\la), \gamma_B(\la))}{\gamma^\mu_L(\la)} \deriv{\gamma^\mu_L(\la)}\la\notag\\
&=\int_0^1 d\la \deriv{S(\gamma_L(\la), \gamma_B(\la))}{\gamma^\mu_B(\la)} \deriv{\gamma^\mu_B(\la)}\la\notag\\
&=\int_0^1 d\la \pd{\cL}{\dot \gamma_B^\mu(\la)} \deriv{\gamma_B^\mu(\la)}\la \label{almost}
}
where $\dot \gamma_B^\mu=\partial_s \gamma_B^\mu$ is the tangent vector to the trajectory at $s_B(\la)$. The first equality comes from the fact that the  change in action by shifting $\gamma_R(\la)\to \gamma_B(\la)$ does not depend on $\gamma_L(\la)$.

Additionally for a reparameterization invariant theory, the Hamiltonian vanishes, and we have the identity $\pd \cL { \gamma_B'{}^\mu} \gamma'_B{}^\mu = \cL(\gamma_B,\gamma'_B)$. Therefore if we impose `tangent vector alignment,' \ie $ \partial_\la{\gamma_B^\mu(\la)}=\alpha(\la)\dot \gamma_B^\mu (\la) $, we have
\ban{
-\int_0^1 d\la \deriv{S(\gamma_L(\la), \gamma_R(\la))}{\gamma^\mu_L(\la)} \deriv{\gamma^\mu_L(\la)}\la=\int_0^1 d\la \, \cL(\gamma_B(\la), \partial_\la \gamma_B(\la)) \label{final}
}
The left hand side is the differential entropy of the boundary intervals and the right hand side is the Bekenstein-Hawking entropy of the bulk curve, so we have the result $E=S_{BH}$. 

Note that tangent vector alignment plays a central role in this argument. We can interpret this condition as a method for choosing appropriate boundary intervals given a bulk curve, in that the extremal curve for the interval $\la$ must be tangent to the bulk curve at $\la$. Alternatively, one may understand this condition as a prescription for how to build a bulk curve given a family of boundary intervals. We explore this complementary perspective in the next chapter.

\chapter{Boundary-to-Bulk Construction}
\label{new}

In chapter \ref{time}, we began with a bulk curve and showed how to construct
a family of boundary intervals whose differential entropy is equal to
the gravitational entropy of the original curve. However, in light of the discussion of section \ref{general}, we expect that this process can be reverse engineered. That is,  given a family of (space-like) boundary intervals, we seek to construct a bulk curve whose gravitational entropy is equal to the differential entropy of the original intervals. Trivially,
there are many bulk surfaces which yield the correct value of the gravitational entropy, and so implicitly we require that the bulk curve is constructed in some sort of natural way. 

In the construction in the previous chapter, each boundary interval was associated with a point on the bulk curve at which its extremal curve was tangent. Reversing this observation, we expect that the bulk curve can be thought of as a set of points taken from each extremal curve subject to a `tangent vector alignment' condition. Of course, it is non-trivial that this set will produce a smooth curve! 

Following the notation of the previous sections, we begin with a family of boundary intervals defined by
the endpoint curves, $\gamma_L(\la)$ and $\gamma_R(\la)$, and we parameterize the corresponding extremal curves by $\Gamma(s;\la)$ with the boundary conditions $\Gamma(s=-1;\la)=\gamma_L(\la)$ and $\Gamma(s=1;\la)=\gamma_R(\la)$.\footnote{For convenience we have changed the `time' on the extremal curve so that $s\in [-1,1]$.} To construct the bulk curve $\bulkc_B(\la)$ with the appropriate gravitational entropy, we take a point $s_B(\la)$ from each extremal curve for each value of 
$\la\in [0,1]$. Hence we can parameterize this curve as $\bulkc_B(\la)=\Gamma(s_B(\la);\la)$ for some function $s_B(\la)$. The problem at hand therefore reduces to finding a function $s_B(\la)$ so that the gravitational entropy of $\gamma_B(\la)$ is equal to the differential entropy of the original boundary intervals. 

One may suspect that for generic families of boundary intervals there is no solution for the tangent vector alignment condition, and indeed we find this to be the case. Instead, we note that a slight generalization of the hole-ographic construction will allow us to construct a natural bulk curve for families of boundary intervals obeying natural geometric constraints. We will discuss this generalization quite thoroughly in section \ref{generic}, first building intuition in the simpler case of a family of boundary intervals on a constant time slice. At the end of this chapter, we explicitly find the solution for the particular case of AdS$_3$ in section \ref{AdS}, using this concrete example to characterize and explore some general features of the construction. 

To simplify the discussion, we limit our analysis to general holographic spacetimes in three dimensions, however the construction extends straightforwardly to higher dimensional backgrounds with planar symmetry.  Generally, we restrict our attention
to the situation where the bulk is described by Einstein gravity, for which the appropriate entropy functional is simply the 
Bekenstein-Hawking entropy, as in eq.~\reef{BH}.

To understand how the previous hole-ographic construction can be generalized, we note that a key step in the
proof of section \ref{general} came from requiring
 \be
\frac{\partial \cL}{\partial \dot \bulkc_B^\mu}\,
\frac{d\bulkc^\mu_B(\la)}{d\la}=\frac{\partial \cL}{\partial (\partial_\lambda \bulkc_B^\mu)}\,
\frac{d\bulkc^\mu_B(\la)}{d\la}\,.
 \label{step2}
 \ee
For Einstein gravity, $\left.\cL(x^\mu,\partial_\la x^\mu)\right|_{\bulkc_B}=|\partial_\la \bulkc_B|/(4G_N)=\sqrt{g_{\mu\nu}\,
\partial_\la \bulkc_B^\mu \partial_\la \bulkc_B^\nu}/(4G_N)$ and hence $\partial \cL/\partial (\partial_\lambda \bulkc_B^\mu)
=g_{\mu\nu}\partial_\la \bulkc_B^\nu/(4G_N\, |\partial_\la \bulkc_B|)$. Therefore eq.~\reef{step2} yields
 \be
\frac{\dot \bulkc_B^\mu}{|\dot \bulkc_B|}\,g_{\mu\nu}\,
\bulkc_B'{}^{\!\!\nu}=\frac{\bulkc_B'{}^{\!\!\mu}}{|\bulkc_B'|}\,g_{\mu\nu}\,
\bulkc_B'{}^{\!\!\nu}\,.
 \label{step3}
 \ee
where, for simplicity, we have introduced the notation  $f'(\la)=\partial_\la f(\la)$.
In section \ref{general}, we solved this equation by imposing tangent vector alignment, \ie $\bulkc_B'{}^{\!\!\mu}/|\bulkc_B'|=\dot \bulkc_B^\mu/|\dot \bulkc_B|$. However, we note that the general solution in fact takes the form
  \be
\frac{\bulkc_B'{}^{\!\!\mu}}{|\bulkc_B'|}=\frac{\dot \bulkc_B^\mu}{|\dot \bulkc_B|}+k^\mu\qquad
{\rm with}\ \ k\cdot \bulkc_B'=0 \,.
 \label{sol3}
 \ee
Further, it is straightforward to show that the extra vector satisfies
  \be
k\cdot k=0\quad
{\rm and}\quad k\cdot\dot \bulkc_B=0 \,.
 \label{sol4}
 \ee
Because of the first condition in eq.~\reef{sol4}, we refer to this solution as `null vector alignment,' and this more general solution will form the basis of our generalized hole-ographic construction. 
Of course, the key point is that with the general solution \reef{sol3}, one still finds that eq.~\reef{final} still holds. That is, the gravitational entropy evaluated on the bulk curve is equal to the differential entropy of the family of boundary intervals.

It will be more convenient to phrase the discussion in terms of the extremal curves $\Gamma(s;\la)$ rather than the bulk curve $\bulkc_B(\la)$. Hence we would like to change variables from  $\dot \bulkc_B(\la)$ and $\bulkc_B'(\la)$ to $\dot \Gamma(s;\la)|_{s=s_B(\la)}$ and $\Gamma'(s;\la)|_{s=s_B(\la)}$. In fact, this is
straightforward to realize.  First we note
\be
 \bulkc'_B{}^{\!\!\mu}(\la)=\left. \Gamma'{}^{\mu}(s;\la)\right|_{s_B(\la)}+ \dot \Gamma^\mu(s;\la)|_{s_B(\la)}\ s_B'(\la)
\quad {\rm and}\quad 
\dot \bulkc_B^\mu(\la)=\dot \Gamma^\mu(s;\la)|_{s_B(\la)}\,.
\label{note99}
\ee
It is then straightforward to show that eq.~\reef{step3} is equivalent to the following 
\be
\dot \Gamma\cdot\Gamma'=|\dot\Gamma|\,|\Gamma'|\ . \label{intersect}
\ee
This new condition has essentially the same form as the constraint \reef{step3}
written in different variables, and so the general solution also has the same form as \reef{sol3}
\ban{
\frac{ \Gamma'{}^\mu}{| \Gamma'|}=\frac{\dot \Gamma^\mu}{|\dot \Gamma|}+k^\mu\qquad
{\rm with}\ \ k\cdot\Gamma'=0\,,\ \ 
k\cdot k=0\, , \ \ 
{\rm and}\ \ k\cdot\dot \Gamma=0 \,.
\label{parallel}
}
That is, in terms of the new variables, the null vector alignment condition can be written in precisely the same way as before.

Finally to further facilitate the discussion, we introduce an orthonormal basis at each point on the extremal curve consisting of the tangent vector $\hat u(s;\la)=\dot \Gamma(s;\la)/|\dot \Gamma(s;\la)|$ and two orthogonal unit vectors $\hat n_1(s;\la)$ and $\hat n_2(s;\la)$.\footnote{We choose $\hat n_1(s;\la)$ to be space-like and to lie in the plane of the 
extremal curve, with $\hat n_1\cdot\hat n_1=1$ and $\hat n_1\cdot\hat u=0$. Further $\hat n_2(s;\la)$ is time-like
and orthogonal to the plane of the extremal curve,  with $\hat n_2\cdot\hat n_2=-1$ and $\hat n_2\cdot\hat u=0=
\hat n_2\cdot\hat n_1$.} 
Additionally, a central role is played by the `separation vector' $\Gamma'{}^\mu$, which indicates how the extremal curve at $\la$ is displaced in moving to the neighboring curve at $\la+d\la$. Below, it will be convenient to project $\Gamma'{}^\mu$ into the
subspace normal to $\hat u(s;\la)$ and so we define
 \be
v^\mu_\perp(s;\la)\equiv\Gamma'{}^\mu(s;\la) - \Gamma'{}^\sigma(s;\la)\, \hat{u}_\sigma(s;\la)\ \hat{u}^\mu(s;\la)\,.
\label{proj0}
 \ee
With this notation, the condition which selects out the solution \reef{parallel} can be written as $|v_\perp(s_B(\la);\la)|=0$. This framework will be useful for understanding the general case. 

\section{On a Constant Time Slice}\label{flat intersection}

\begin{figure}[h!]
\begin{center}
\includegraphics[width=0.5\textwidth]{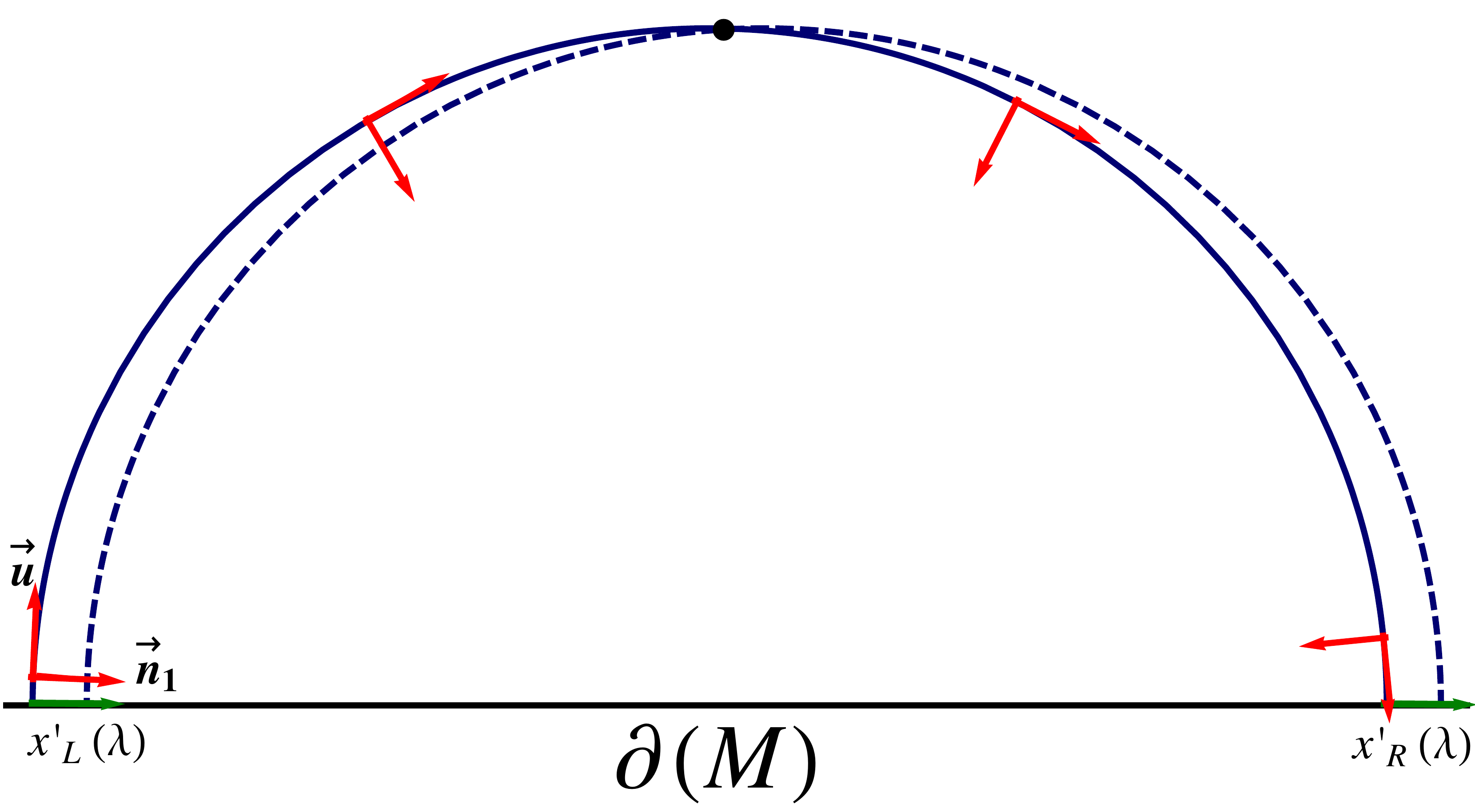}
\caption{Considering extremal curves in a constant time slice. At the intersection point, the component of $\Gamma'(s;\la)$ along $\hat n_1$ vanishes and $\Gamma'(s;\la)\propto \dot \Gamma(s;\la)$.}
\label{cross1}
\end{center}
\end{figure}

Let us begin by considering a family of  boundary intervals which all lie in a constant time slice. With this assumption, the $\hat n_2$ component of $\Gamma'(s;\la)$ will vanish and so $v_\perp(s;\la)$ will always be space-like. As such, the condition $|v_\perp(s_B(\la);\la)|=0$ can only be satisfied when this vector vanishes, \ie $v^\mu_\perp=0$.
As illustrated intuitively in figure \ref{cross1}, this vanishing occurs at the intersection point between the extremal curves at $\la$ and $\la+d\la$. Explicitly, at the intersection point,  the separation vector changes from pointing `inside' to pointing `outside' the extremal curve and therefore must vanish by continuity. Thus we can think of the bulk curve $\bulkc_B$ as consisting of the continuum limit of these intersection points. We sharpen this intuition in appendix \ref{geom}. 

\begin{figure}[h!]
\begin{center}
\includegraphics[width=0.5\textwidth]{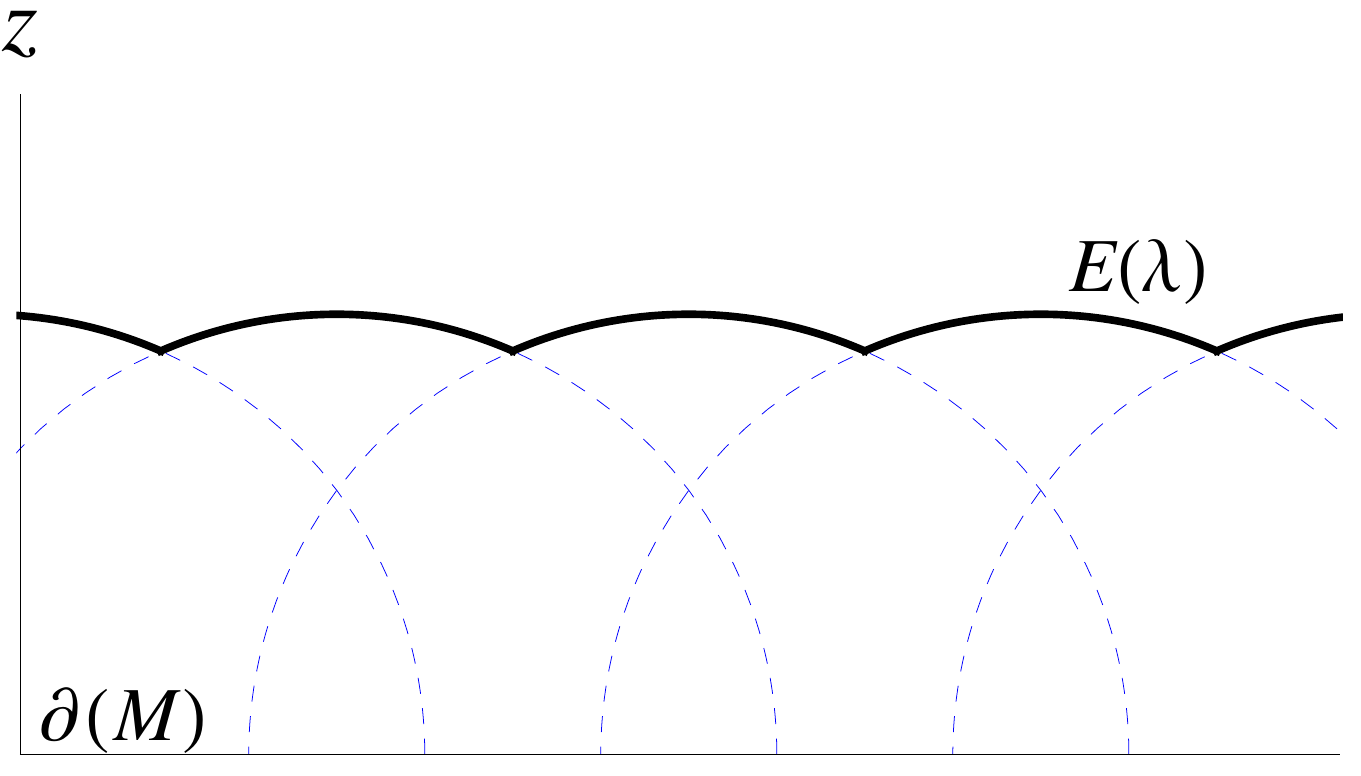}
\caption{We picture the outer envelope $E(\la)$ as built from the intersection of extremal curves of boundary intervals. Each extremal curve is drawn in dashed blue. In the continuum limit this curve consists essentially of only the intersection points, and by construction its gravitational entropy equals the differential entropy of the boundary intervals.
}
\label{envelope}
\end{center}
\end{figure}

Additionally, the bulk curve has a geometric interpretation in terms of the `outer envelope'  of the extremal curves as described in \cite{Myers2014}. We begin with a discrete set of $n$ boundary intervals parameterized by $\la_k=k/n$ --- see the discussion in the introduction. An approximation of the bulk curve is then constructed in a piece-wise fashion by taking the portion of each $\Gamma(s;\la_k)$ extending from the intersection with the extremal curve at $\la_{k-1}$ to the intersection with the extremal curve at $\la_{k+1}$, as illustrated in figure \ref{envelope}. Loosely, one can think of this approximate curve as the boundary of the union of spacetime regions enclosed by the extremal curves.\footnote{For families of intervals in AdS spacetime with $(x_L'+x_R') >0$, this picture is precise. However, as pointed out in \cite{Myers2014}, this picture breaks down in more generic situations.} 

In the continuum limit as $n\to \infty$, the gravitational entropy for the outer envelope equals the differential entropy of the family of boundary intervals, as each intersection point satisfies tangent vector alignment by the previous analysis. That is, in this limit, we may note that the individual portions of each extremal curve contributing to the outer envelope shrinks to zero size. Hence we can think that the final outer envelope is comprised essentially of the intersection points of $\Gamma(s;\la)$ and $\Gamma(s;\la+d\la)$ for each $\la$, taken in the limit as $d\la \to 0$. 

Note that in this framework, we are implicitly assuming that the neighboring extremal curves, \ie the extremal curves at $\la$ and $\la+d\la$,  intersect precisely once. Equivalently, we are assuming that the component of $\Gamma'{}^\mu(s;\la)$ along $\hat n_1(s;\la)$ vanishes exactly once. To construct a continuous bulk curve, there must be a solution for every $\la$, so this imposes a global constraint on the families of boundary intervals which have corresponding bulk curves. 

\begin{figure}[h!]
\centering
\subfloat[]{\includegraphics[width=0.45\textwidth]{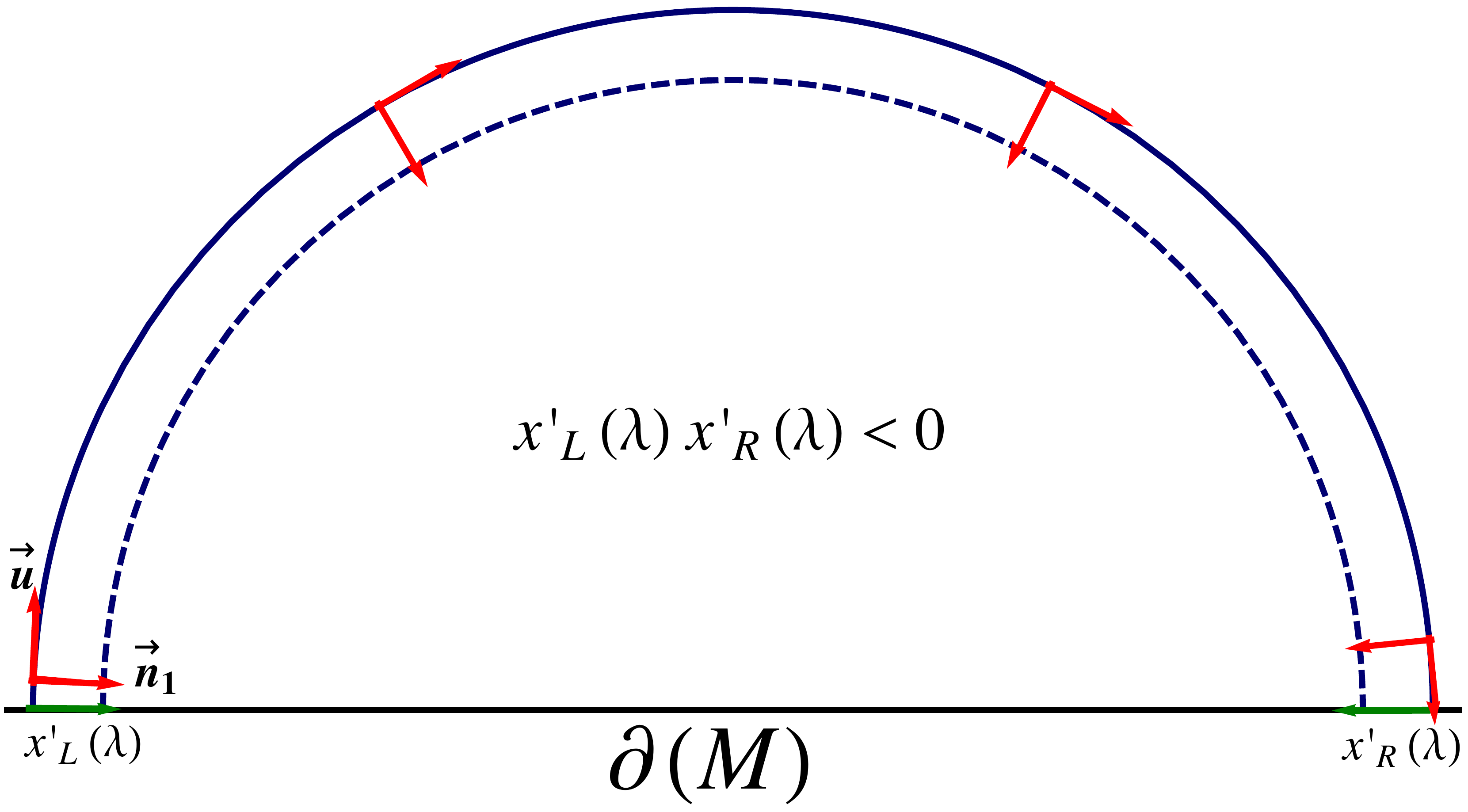}}\qquad
\subfloat[]{\includegraphics[width=0.45\textwidth]{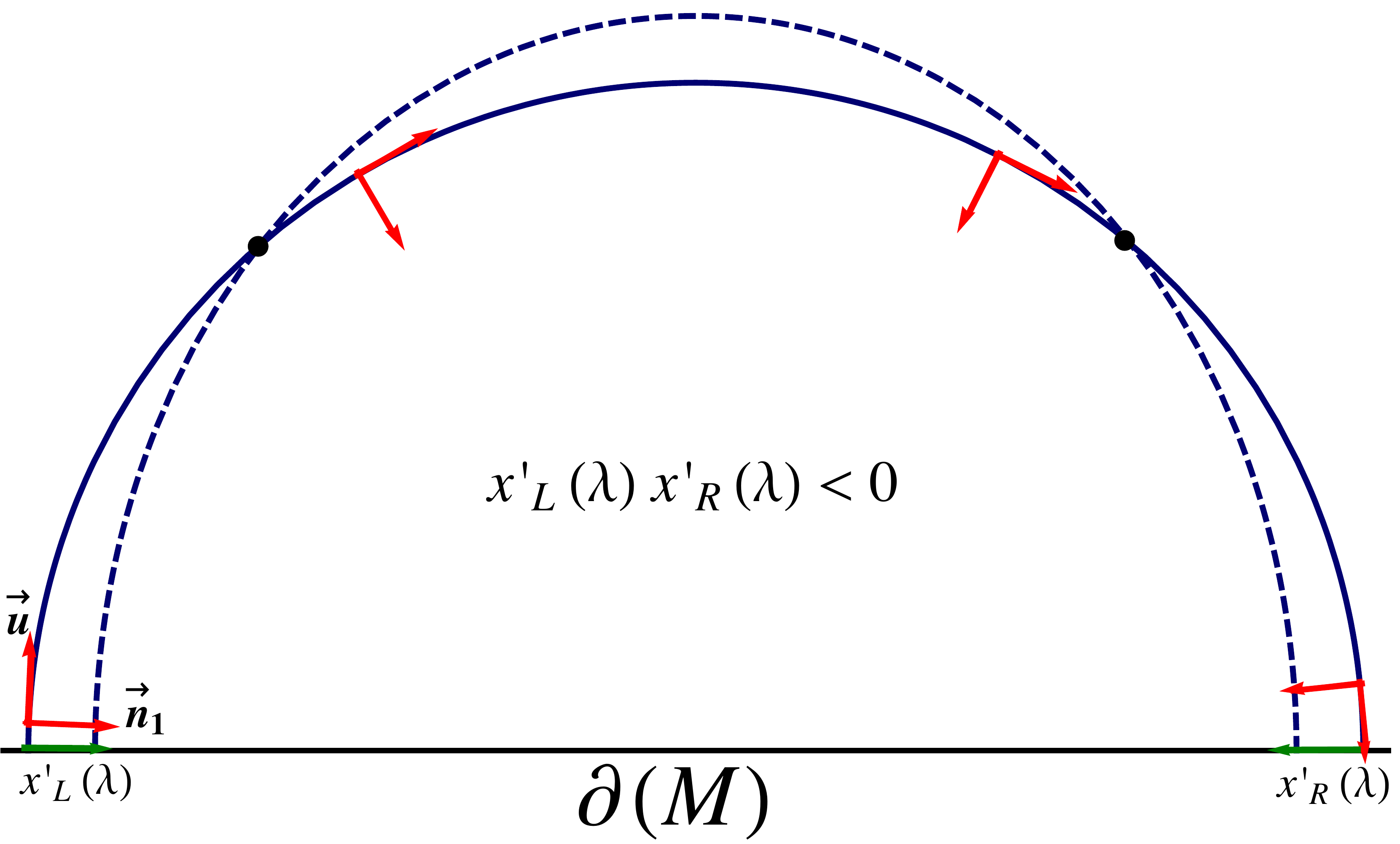}}
\caption{Considering extremal curves in a constant time slice. In the case $ x'_L(\la)\, x'_R(\la)<0$,
the curves given by $\Gamma(\lambda)$ and $\Gamma(\lambda+d\lambda)$ will not cross, as shown in (a). A logical
possibility is that there may be an even number of crossings, as shown in (b), however it turns out that this is not possible \cite{Headrick2013}. }
\label{cross2}
\end{figure}

As illustrated in figure \ref{cross2}a,  we can write this global constraint as \cite{Myers2014}
\be
x_R'(\la)\,x_L'(\la)>0\,.
\label{constrain88}
\ee
If $x_R'(\la)\,x_L'(\la)\le0$, one boundary region is entirely contained within the other and the same is true of the corresponding extremal curves in the bulk. Therefore if the constraint \reef{constrain88} is not satisfied then neighboring extremal curves will not intersect. A logical possibility that when $x_R'(\la)\,x_L'(\la)\le0$, the extremal curves may cross an even number of times,
as illustrated in figure \ref{cross2}b.
However, this possibility is ruled out by analysis in \cite{Headrick2013}.  

In particular, if we have one boundary region $I_{k+1}$ that is enclosed within a second
$I_k$, then the bulk region enclosed by the extremal surface $\Gamma(s;\la_{k+1})$, corresponding to $I_{k+1}$, is 
entirely contained within $\Gamma(s;\la_{k})$. Hence it is not possible for $\Gamma(s;\la_{k+1})$ to cross the curve $\Gamma(s;\la_{k})$ and there will be no intersections.

\section{Generic Families of Intervals}  \label{generic}

For a time varying family of boundary intervals with a corresponding bulk curve as constructed as in chapter \ref{time}, generically the extremal curves at $\la$ and $\la+d\la$ do not intersect. Therefore, we must generalize the previous construction. However, note we can make use of the null vector alignment condition \reef{parallel} with a non-vanishing null vector $k$. In this case, we note that loosely $s_B(\la)$ no longer characterizes the intersection point between neighboring extremal curves, but rather a point at which they are separated by null vector orthogonal to the extremal curves. Therefore, we seek to provide a geometric interpretation in which the bulk curve is constructed from intersections of each extremal curve with the `light sheet' of its neighboring curve. It turns out that the desired light sheets form the boundary of the so called `entanglement wedge' for each interval \cite{mattEW}. 

\begin{figure}[h!]
\begin{center}
\includegraphics[width=0.37\textwidth]{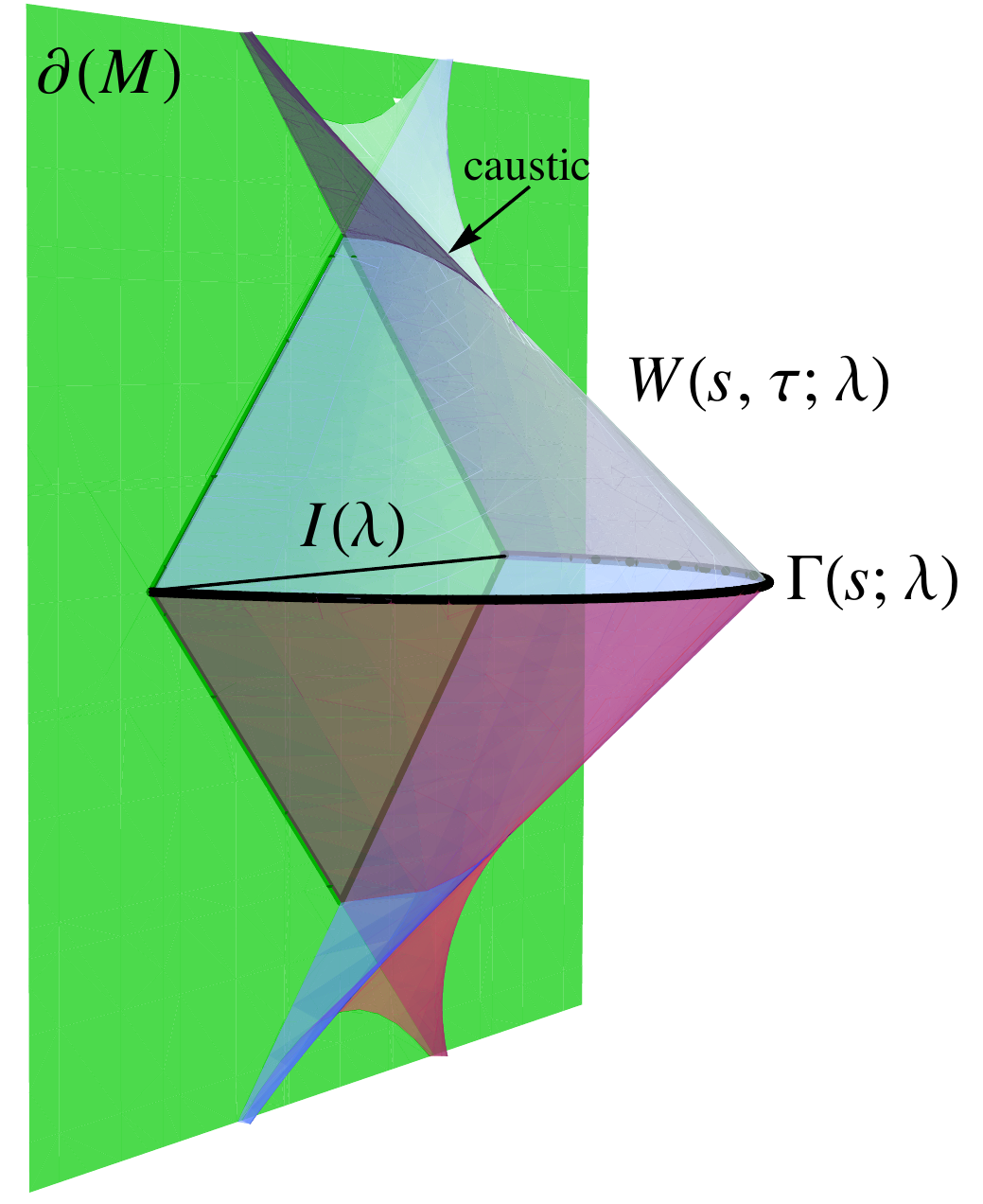}
\caption{The boundary of the entanglement wedge $W(s,\tau;\la)$ is shown above for the extremal curve $\Gamma(s;\la)$ corresponding to the interval $I(\la)$. The surface ends when the light rays emerging from $\Gamma(s;\la)$
either reach the asymptotic boundary or form caustics. }
\label{EW}
\end{center}
\end{figure}

First, we must properly define the notion of entanglement wedges, as described in \cite{mattEW}. Given a boundary region and a corresponding extremal surface in the bulk, the entanglement wedge
is defined as the domain of dependence or causal development of any spatial slice extending between these two.  In our
situation, we are interested in the boundary of the entanglement wedge $W(s,\tau;\la)$, which is formed by the (converging) light sheets projected orthogonally from 
the extremal curve $\Gamma(s;\la)$ toward the boundary. The light rays comprising these light sheets may reach the asymptotic boundary, however,  generically they will end with the formation of caustics, as illustrated in figure \ref{EW}. 
One remarkable feature of the entanglement wedges is that the intersection of $W(s,\tau;\la)$ with the asymptotic boundary is precisely the boundary of the causal development of the boundary interval, as shown in \cite{mattEW}.

\begin{figure}[h!]
\begin{center}
\includegraphics[width=0.6\textwidth]{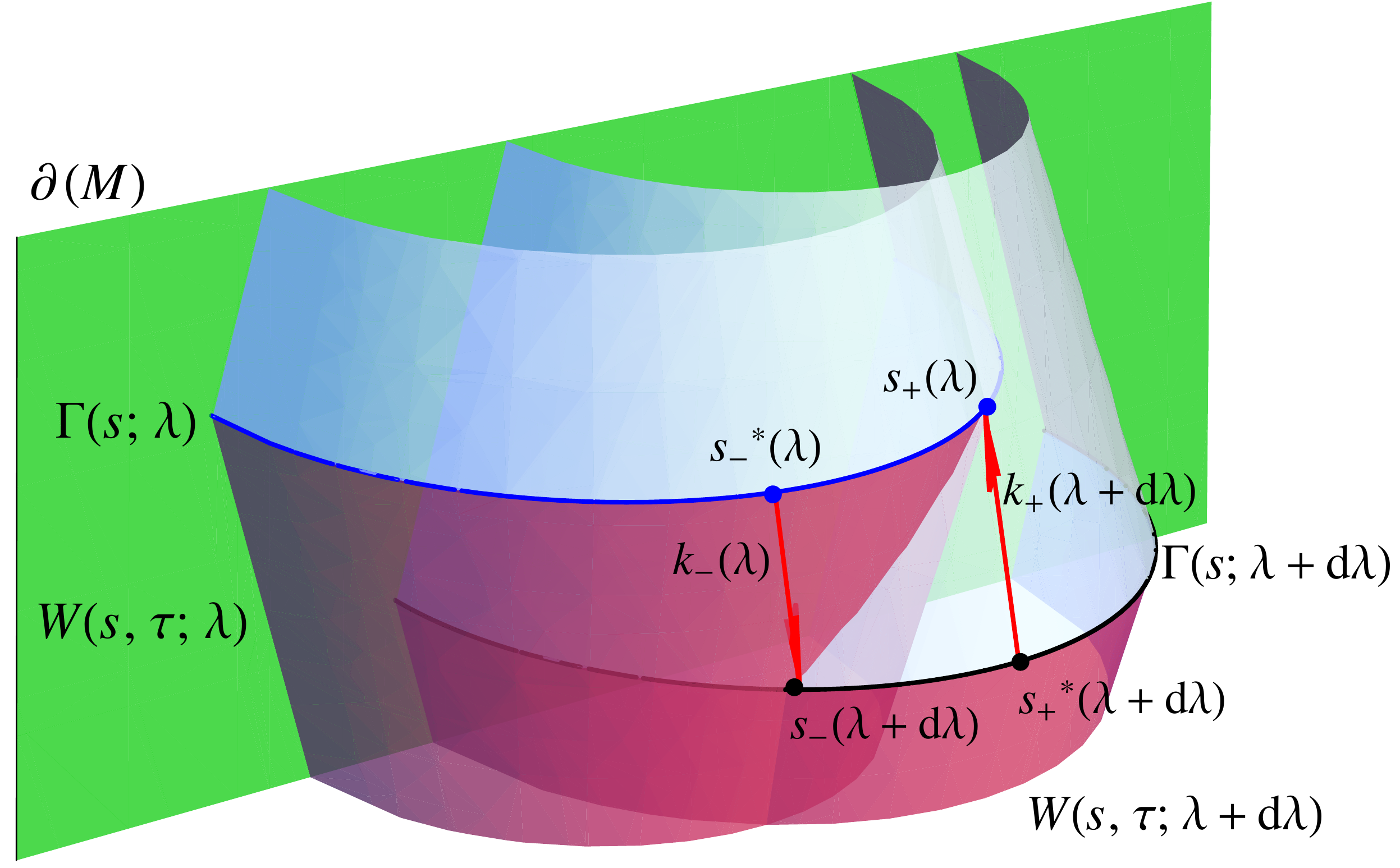}
\caption{The intersection of the surfaces $W(s,\tau;\la)$ and $W(s,\tau;\la+d\la)$ is shown above. The point $\Gamma(s^*_+(\la+d\la);\la+d\la)$ on the neighboring extremal curve is identified as being separated from the intersection point $\Gamma(s_+(\la);\la)$ by the null vector $k^\mu_+(\la+d\la)$. Similarly, the point $\Gamma(s^*_-(\la);\la)$ on the neighboring extremal curve is identified as being separated from the intersection point $\Gamma(s_-(\la+d\la);\la+d\la)$ by the null vector $k^\mu_-(\la)$. One can see intuitively that in the limit $d\la \to 0$, $s_-^*(\la)$ does not generically approach $s_+(\la)$.}
\label{EWI}
\end{center}
\end{figure}

Using entanglement wedges, we can extend the notion of intersection discussed in the previous section to build up a bulk curve. Denoting the intersection of $\Gamma(s;\la)$ with $W(s,\tau;\la \pm d \la)$ by $s_\pm(\la)$, we first argue that at $s_+(\la)$ the null vector alignment condition \reef{parallel} is satisfied and therefore we can build the bulk curve via $\bulkc_B(\la)=\Gamma(s_+(\la);\la)$. The notation and setup is illustrated in figure \ref{EWI}. 
We label the intersection point $s_+(\la)$ on $\Gamma(s;\la)$, write its separation from a point on $s^*_+(\la+d\la)$ on $\Gamma(s;\la+d\la)$ by the null vector $k_+(\la+d\la)$. Here,  $k_+(\la+d\la)$ is orthogonal to $\Gamma(s;\la+d\la)$ at $s^*_+(\la+d\la)$, or explicitly  $k_+(\la+d\la)\cdot \dot\Gamma(s^*_+(\la+d\la);\la+d\la)=0$. Intuitively we know when two extremal curves intersect we have $\Gamma'\propto \dot \Gamma$, and so if the intersection is off by a null vector we can loosely expect $\Gamma' \propto \dot \Gamma + k$.\footnote{As we are considering infinitesimally separated curves, we can safely assume that we are well outside of the range of caustic formation.} When the extremal curve intersects the neighboring entanglement wedge, this null vector is additionally orthogonal to $\dot \Gamma$ by construction, and so indeed all parts of the null vector alignment condition \reef{parallel} is satisfied. In this way we take the bulk curve to consist of the continuum limit of these `intersection' points. We confirm this rough intuition in appendix \ref{geom}.

\begin{figure}[h!]
\begin{center}
\includegraphics[width=0.45\textwidth]{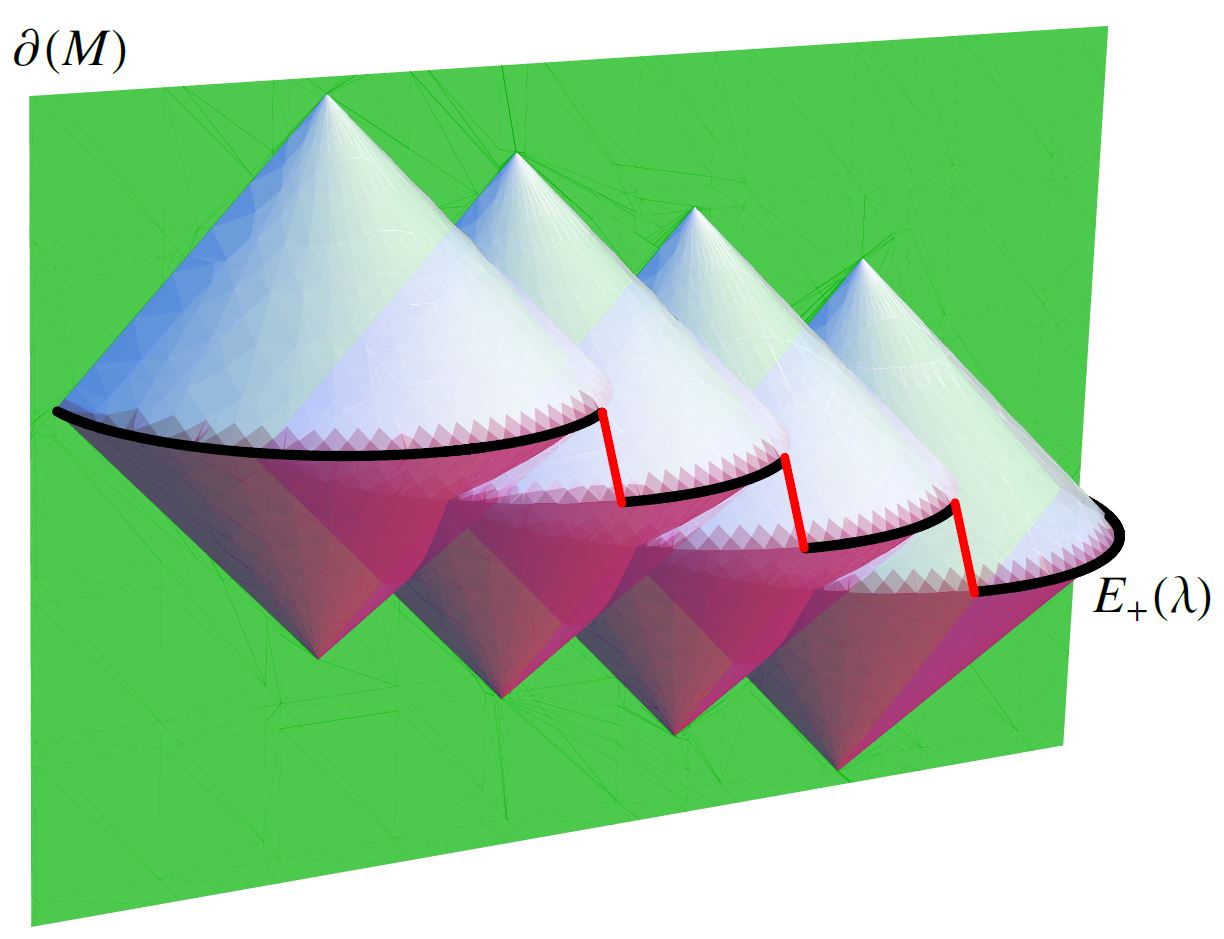}
\caption{We picture the outer envelope $E_+(\la)$ as being built from the pieces of the extremal curve between $s_+^*(\la)$ and $s_+(\la)$, connected by null segments on each entanglement wedge boundary. In the continuum limit this curve consists only of the intersection points $s_+(\la)$, and its gravitational entropy is equal to the differential entropy of the boundary intervals. A similar curve $E_-(\la)$ can be constructed, which generically differs from $E_+(\la)$.}
\label{3Denvelope}
\end{center}
\end{figure}

Additionally, we are led to an generalized notion of the outer envelope in this case. The piece-wise construction of the bulk curve consists of segments of the extremal curves extending between intersections with the boundaries of the
corresponding entanglement wedges. However, these segments do not form a contiguous curve but rather they are connected by $\mathcal O(d\la)$ null segments lying in the boundaries $W(s,\tau;\la)$.  Hence in the continuum limit, these null pieces vanish. A sketch of this construction 
is given in figure \ref{3Denvelope}.

Figure \ref{EWI} also shows the intersection of the extremal curve $\Gamma(s;\la+d\la)$ with the boundary $W(s,\tau;\la)$ of the entanglement wedge for $\Gamma(s;\la)$. Similarly, we label this intersection point $s_-(\la+d\la)$ on $\Gamma(s;\la+d\la)$
and it is connected to a point on $s^*_-(\la)$ on $\Gamma(s;\la)$ by the null vector $k_-(\la)$. Here,  $k_-(\la)$ is orthogonal to $\Gamma(s;\la)$ at $s^*_-(\la)$, \ie  $k_-(\la)\cdot \dot\Gamma(s^*_-(\la);\la)=0$. Again, we expect that at this intersection point, the null vector alignment condition \reef{parallel} is satisfied in the continuum limit, \ie $\Gamma'(s;\la)|_{s^*_-(\la)} \propto \dot \Gamma (s;\la)|_{s_-^*(\la)}+ k_-(\la)$. We also verify this result in appendix \ref{geom}.

One interesting feature of the present construction is that generally when both intersections exist, they do not coincide in
the continuum limit. That is, the difference $s_+(\la)-s_-^*(\la)$ is an order one quantity.\footnote{We show this explicitly in appendix \ref{geom}.} This feature may already be 
evident in figure \ref{EWI} but it will also become explicit in the examples in the following section. Therefore applying
the generalized notion of the outer envelope, we are led to a second distinct curve in the bulk. Hence for a broad
class of families of boundary intervals, the null vector alignment condition \reef{parallel} actually leads to the construction
of two bulk curves for which the gravitational entropy equals the differential entropy of the boundary intervals. Of course, as we
will discuss in a moment, both intersections may not exist or they may not both exist globally. That is, the boundary intervals must satisfy global constraints analogous to eq.~\reef{constrain88} in order to properly define a bulk surface.

\begin{figure}[h!]
\begin{center}
\includegraphics[width=0.55\textwidth]{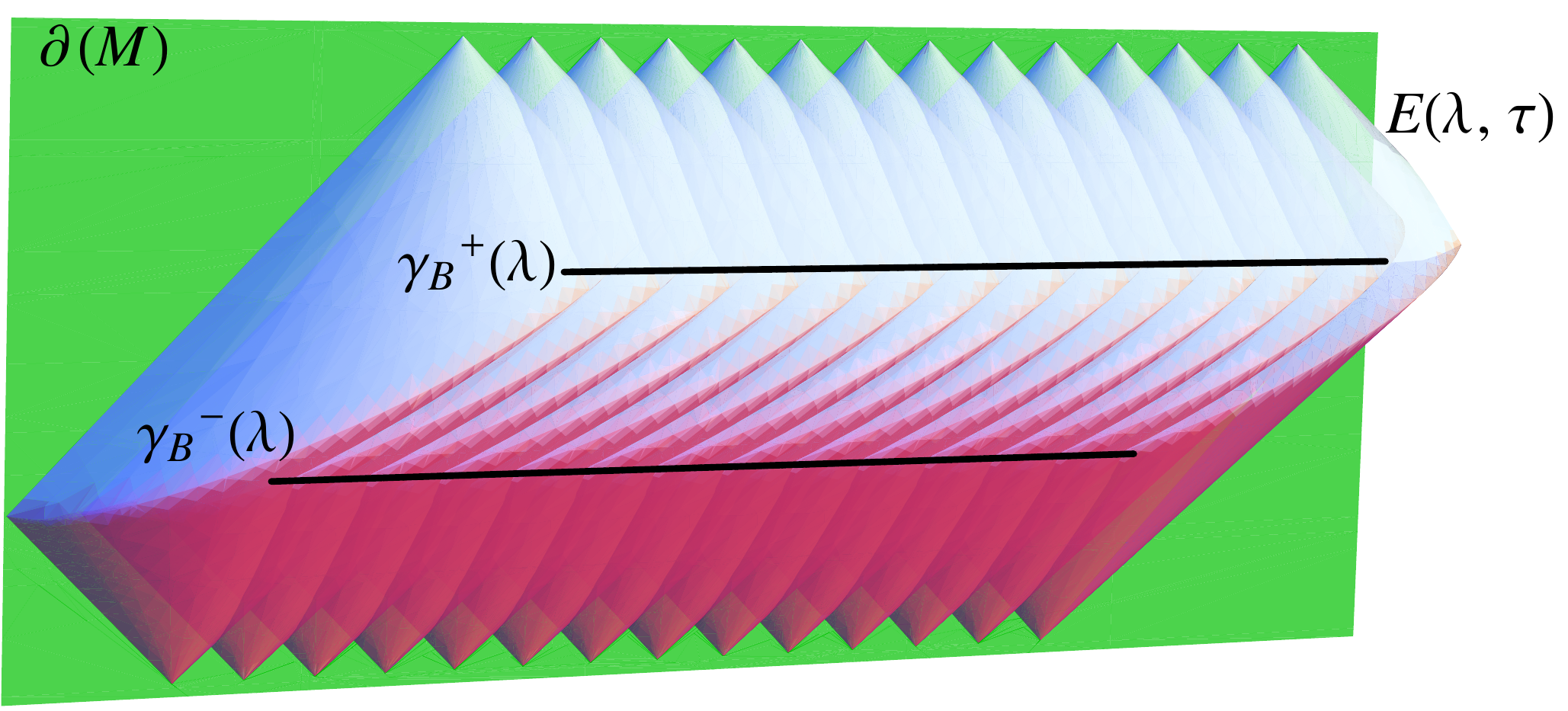}
\caption{(Colour online) The enveloping surface $E(\la, \tau)$ being built for a family of boundary intervals with a fixed width but slightly tilted in the ($t$,$x$)-plane in AdS$_3$. The two bulk curves, $\gamma_B^+(\la)$ and $\gamma_B^-(\la)$, correspond to the lines across which the normal vector makes the transition between space-like and null.}
\label{EWS}
\end{center}
\end{figure}

Further insight comes from extending the outer envelope to the `enveloping surface' $E(\la,\tau)$ which can loosely be thought of as  the boundary
of the union of all of the entanglement wedges.\footnote{Similar to the discussion of the outer envelope in \cite{Myers2014}, this picture is only precise for $\hat n_1(s_B(\la))\cdot a(\la)<0$, where $a^\mu(\la)$ is the proper acceleration along the bulk curve. 
This is a covariant generalization of the condition found for the constant time case \cite{Myers2014}. In higher dimensions, \ie bulk
dimensions greater than three, this condition becomes $\hat n_1(s_B(\la))\cdot K(\la)<0$, where $K^\mu(\la)$ is the trace of the extrinsic curvature on the bulk curve.} More precisely, this enveloping surface should be thought of as being composed of all of the segments of $W(s,\tau;\la)$ between the lines of intersection with $W (s,\tau;\la\pm d\la)$, as illustrated in figure \ref{EWS}. The bulk curves constructed with null vector alignment are then the lines on the enveloping surface across which normal vector make a transition between being space-like and null.\footnote{As the union of the entanglement wedges, the enveloping surface typically consists of five parts: First, `top' and `bottom' of the
entanglement wedges typically contains caustics --- see figure \ref{EW}. Hence the union of these cusps will produce regions at the top and bottom of the enveloping surface with a time-like normal. Second, the light sheets themselves make up sections of the enveloping surface with null normal vector. The regions with the future-pointing and past-pointing null normals correspond to the `upper' and `lower' parts of the enveloping surface respectively. Finally, the region between these null sections is comprised to the portions of the extremal surfaces running from $s_-(\la)$ to $s_+(\la)$. The union of all these geodesics will produce a surface with a space-like normal vector. The bulk curves then form the boundary between this space-like regions and the two null regions. With tangent vector alignment, $s_+(\la)=s_-(\la)$ and thus the space-like region shrinks to zero size. The bulk curve
is then the boundary between the upper and lower null regions. \labell{parcel0}} With tangent vector alignment, the `space-like' region shrinks to zero size and the normal vector is not well defined on the resulting bulk curve, \ie the normal makes a  transition between being future-pointing null and past-pointing null. 

\begin{figure}[h!]
\centering
\subfloat[]{\includegraphics[width=0.35\textwidth]{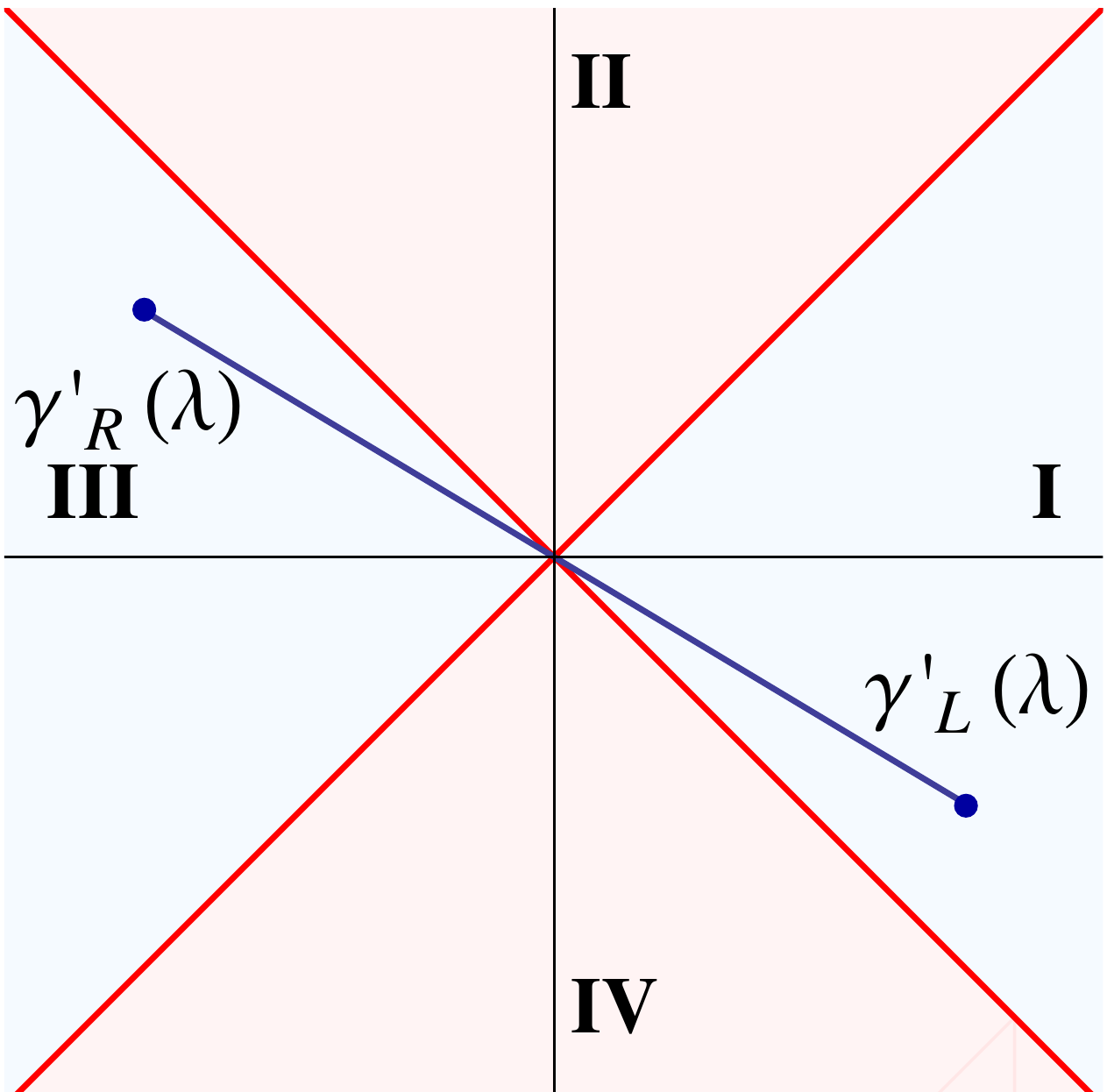}}\qquad
\subfloat[]{\includegraphics[width=0.35\textwidth]{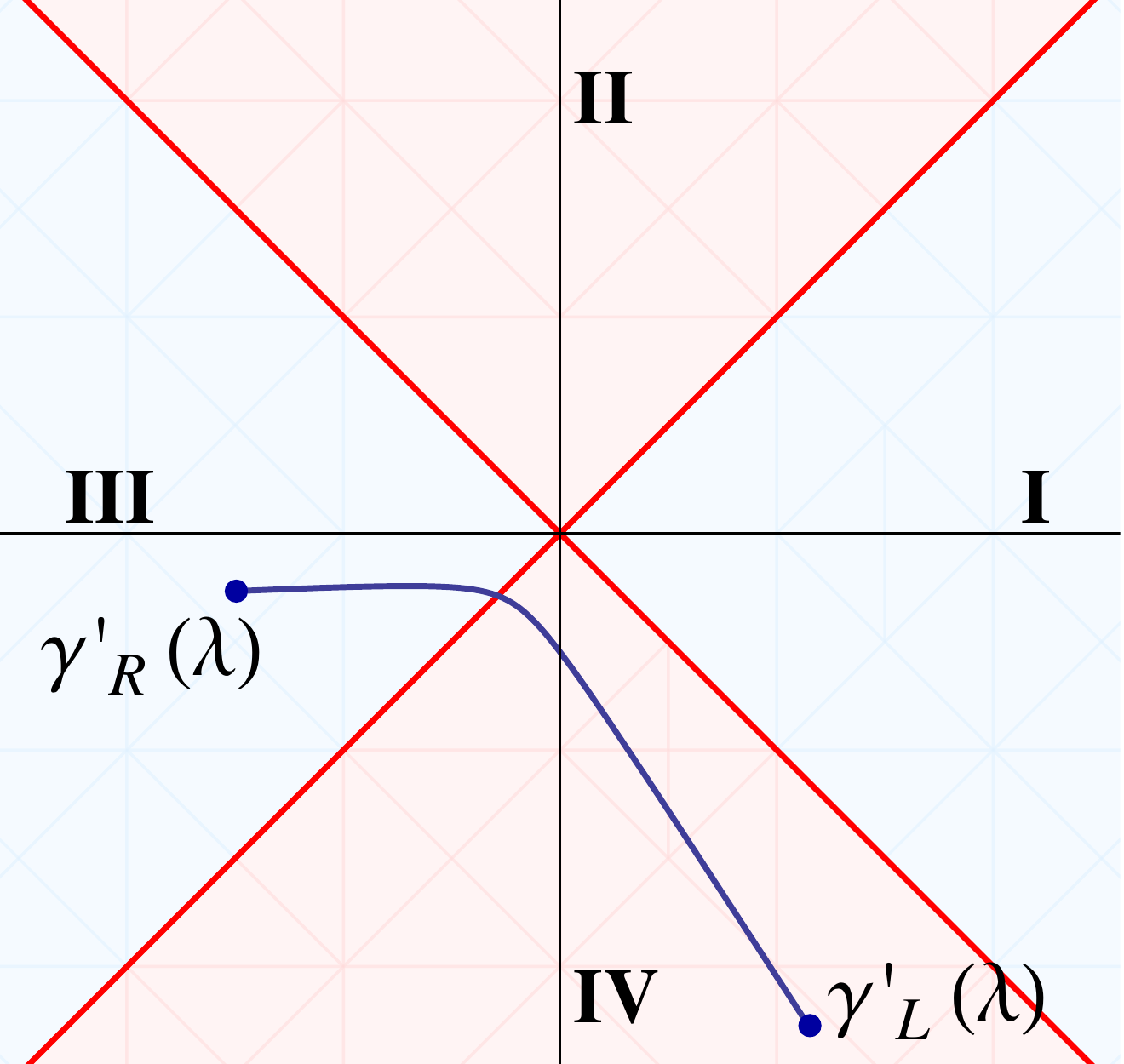}}\\
\subfloat[]{\includegraphics[width=0.35\textwidth]{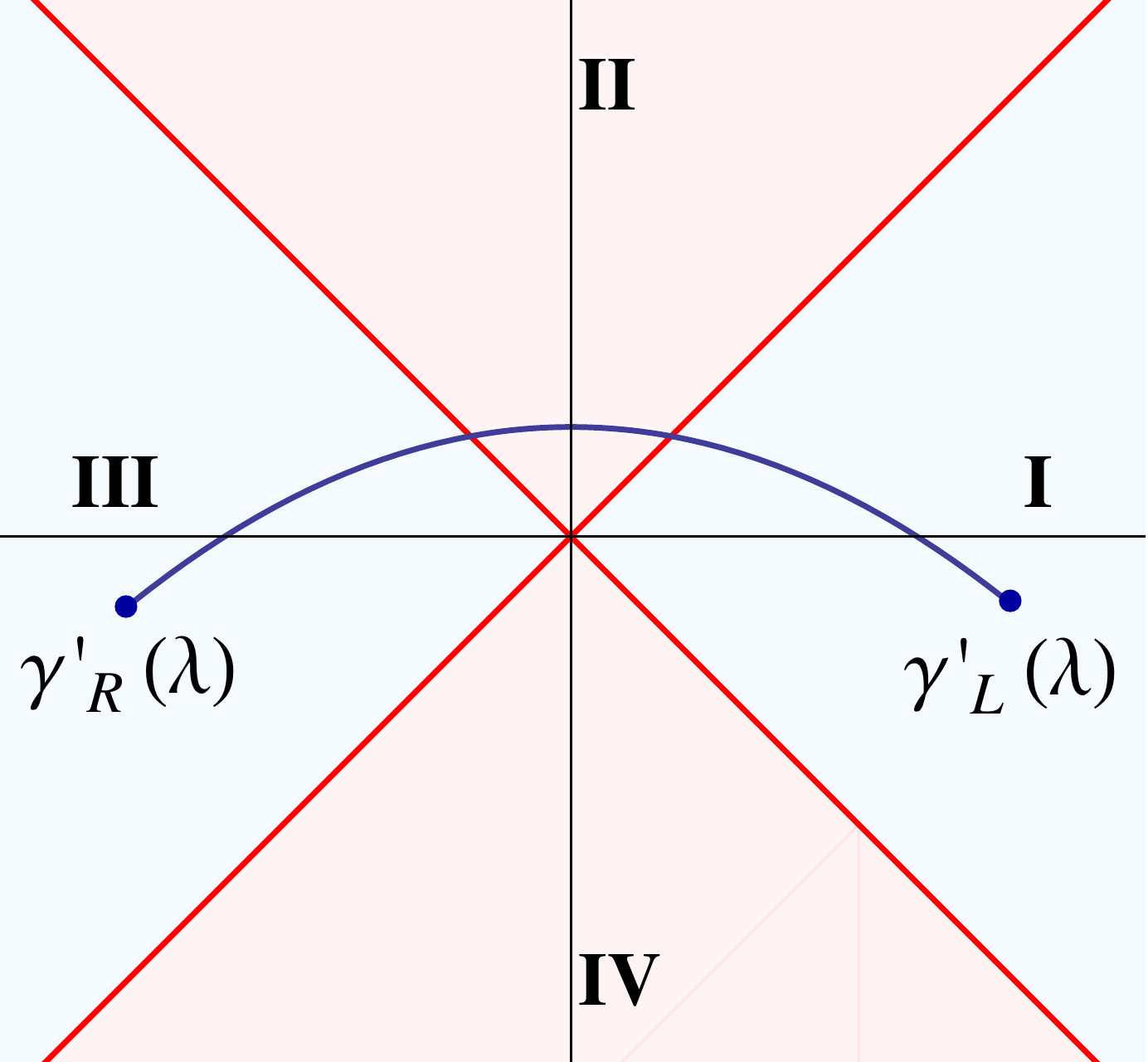}}\qquad
\subfloat[]{\includegraphics[width=0.35\textwidth]{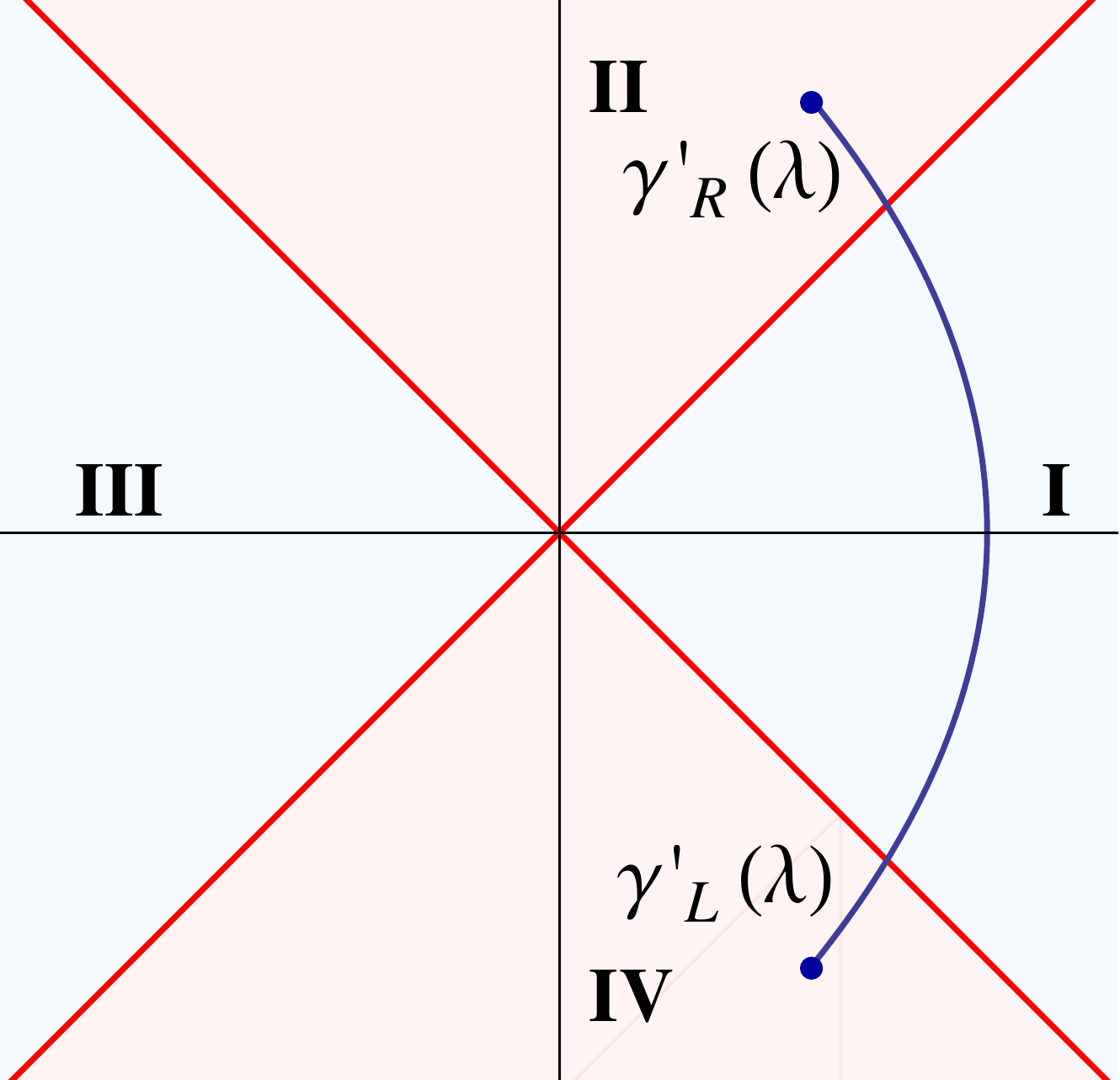}}\\
\caption{We draw the various trajectories of the projection of $\Gamma'{}^\mu(s;\lambda)$ into in the space normal to the tangent vector of the extremal curve $\Gamma(s;\la)$. Panel (a) illustrates tangent vector alignment as the trajectory goes through the origin. Panel (b) shows an example of a trajectory with only one solution. Panels (c) and (d) show examples of trajectories with two solutions which begin and end in space-like and time-like quadrants respectively.}\label{transit2}
\end{figure}

To better understand the possible intersections and the global constraints mentioned above, it is convenient to think of the `trajectory' along an extremal curve $\Gamma(s;\la)$ of the vector $v_\perp(s;\la)$ in the plane normal to the tangent vector $\hat u(\la)$ for a fixed $\la$. In figure \ref{transit2}, we illustrate a variety of these trajectories in this `transverse' plane.  In general, the trajectory starts at $v_\perp(\la) = \gamma_L'(\la)$ and ends at $v_\perp(\la) = \gamma'_R(\la)$, and in between wanders around in the transverse space in some way. Of course, we are particularly interested in the points, $s_B(\la)$, where the trajectory crosses the light cone since as noted
above, the condition $|v_\perp|=0$ corresponds to null vector alignment. In figure \ref{transit2}, we draw various trajectories through the transverse plane.

\begin{figure}[h!]
\centering
\subfloat[]{\includegraphics[width=0.4\textwidth]{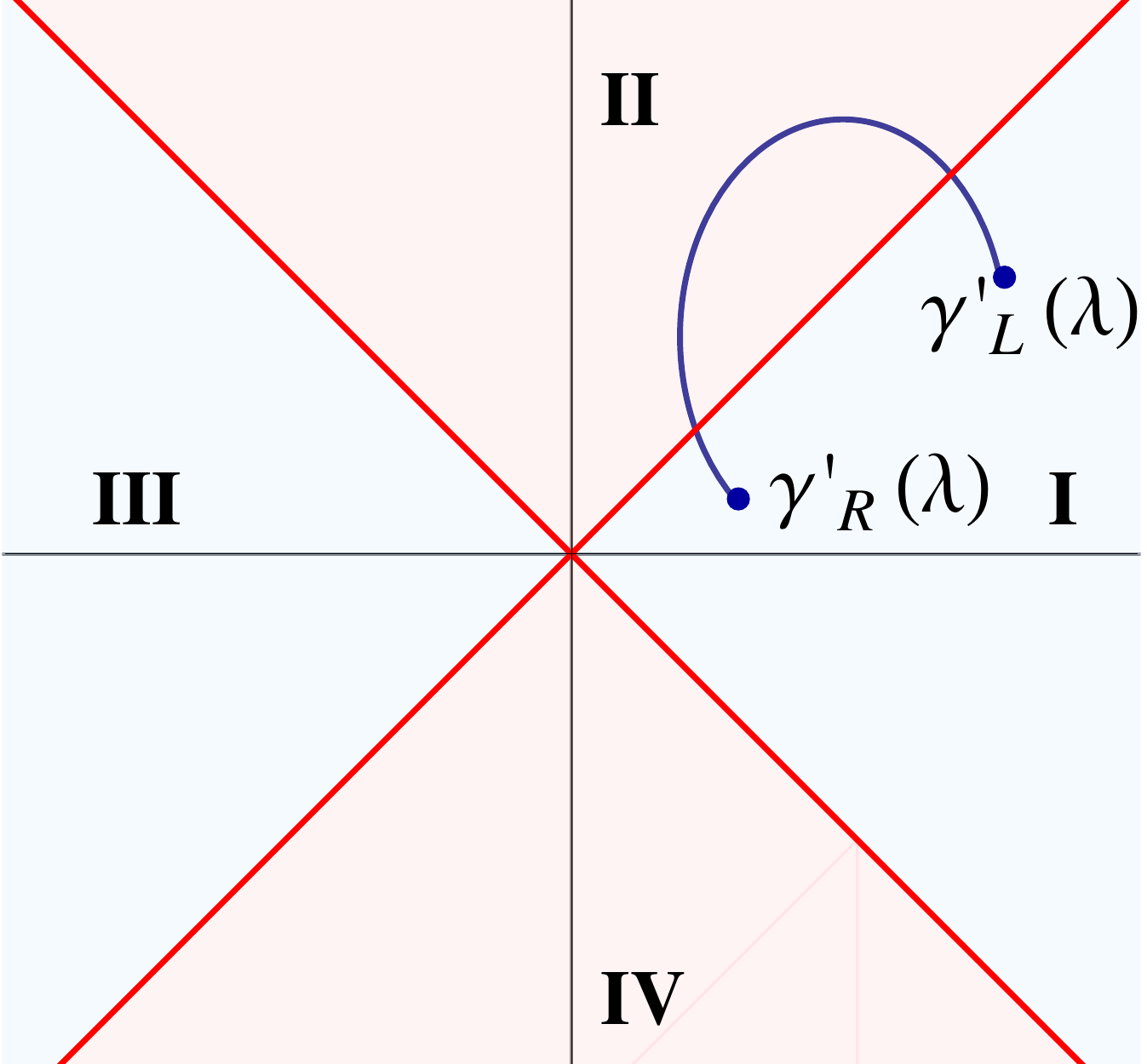}}\qquad
\subfloat[]{\includegraphics[width=0.4\textwidth]{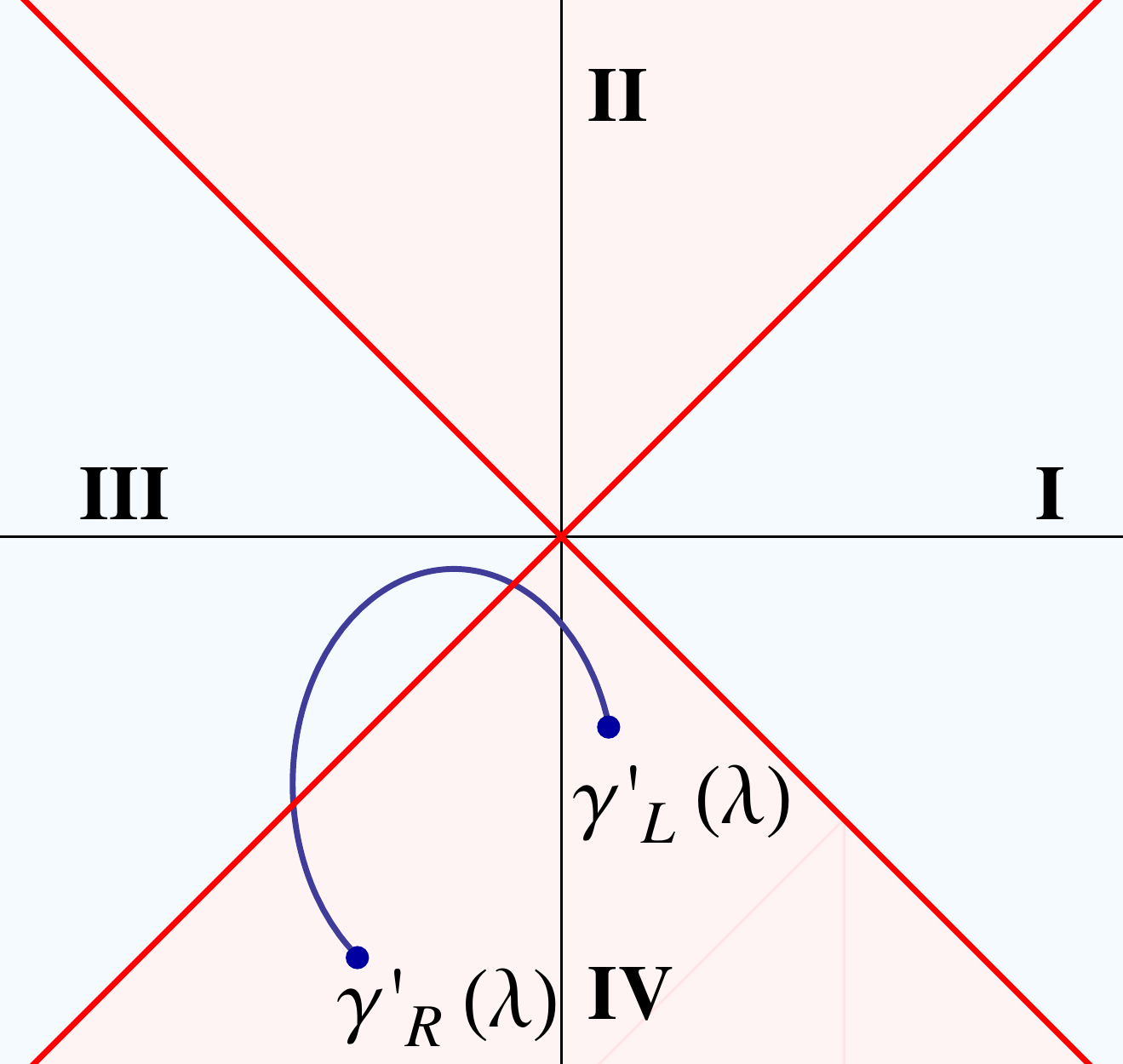}}
\caption{The trajectory drawn in (a) is ruled out by the covariant formulation \cite{Wall2012} of the previously mentioned argument from \cite{Headrick2013}. However, the trajectory drawn in (b) is only ruled out for AdS$_3$.}\label{transit2x}
\end{figure}

For there to be a solution to $|v_\perp|=0$, the trajectory must cross the light cone at least once.  We would like to translate this simple observation as the condition that the trajectory must begin and in different `quadrants,' as defined by the light cone
in the transverse space. For AdS$_3$ this statement is indeed necessary for there to be a solution, but in more general cases there can be more exotic trajectories. However, it is possible to rule out the covariant formulation of the constant time case, as illustrated in figure \ref{transit2x}. 

First we consider a trajectory like the one drawn in figure \ref{transit2x}a, where the endpoints $\gamma_L'(\la)$ and $\gamma_R'(\la)$ are both space-like and in
the same quadrant. These trajectories can be ruled out using the results of \cite{Wall2012}, which provides a covariant formulation of the argument used in the previous section. In particular, in the situation
illustrated, the interval $I(\la+d\la)$ is entirely contained within $I(\la)$ on some time-slice in the boundary. Hence, the corresponding extremal curves, $\Gamma(\la+d\la)$ and $\Gamma(\la)$ are everywhere space-like separated in the bulk.
Therefore $v^\mu_\perp$ must remain within the first quadrant along the entire trajectory and it cannot cross the
light cone, ruling out trajectories of the form illustrated in figure \ref{transit2x}a.   

For AdS$_3$, by the explicit calculation in section \ref{AdS}, we can rule out trajectories where $\gamma_L'(\la)$ and $\gamma_R'(\la)$ are both time-like and in the same quadrant, as shown in figure \ref{transit2x}b. However, for more generic backgrounds this type of trajectory may be possible. 

Hence we conclude that in AdS$_3$ for there to be a solution of $|v_\perp|=0$ the trajectory must begin and end in different quadrants. The latter then demands that either of the following inequalities is satisfied:
\be
(x_R'-t_R')(x_L'+t_L')>0 \label{constrA}
\ee
\begin{center} or\end{center}
\be
(x_R'+t_R')(x_L'-t_L')>0 \label{constrB}
\ee
These inequalities provide the generalization of the global constraint given previously in eq.~\reef{constrain88}  on the family of boundary intervals. Certainly one sees that both eqs.~\reef{constrA} and \reef{constrB}  reduce to $x_L'(\la)x_R'(\la)>0$, as appears in eq.~\reef{constrain88}, when $t'_R=0=t'_L$. In general, given a family of boundary intervals, 
it is possible for one, both, and neither of eqs.~\reef{constrA} or \reef{constrB} hold globally. If one holds, the generalized
hole-ographic construction will define a single bulk curve, while if both are satisfied globally, then our new construction defines two bulk curves for which the gravitational entropy equals the differential entropy.

This geometric formulation provides a wealth of possibilities for future research and exploration, and we further detail some features of this construction in \cite{Headrick2014}. As such, we turn to explicitly parameterizing the generic solution of this construction for AdS$_3$ to highlight some interesting features we hope to explore further in a more general setting. 

\pagebreak

\section{AdS$_3$ as a Case Study}
\label{AdS}

Next we try to build a better understanding of some of the generic properties of our
generalized hole-graphic construction by explicitly solving for the bulk curves for AdS$_3$ in Poincar\'e coordinates \reef{3metric}. Given a set of space-like boundary intervals with endpoints $\gamma_L(\la) = \coords{ x_L(\la), t_L(\la)}$ and $\gamma_R(\la) = \coords{ x_R(\la), t_R(\la)}$, first we change variables to a parameterization of the center, invariant length, and boost angle for each interval
\ban{ \notag &x_c(\la) = \frac 12 (x_L(\la)+x_R(\la))\\ 
\notag & t_c(\la)=\frac 12 (t_L(\la)+t_R(\la))\\ 
\notag & \Delta(\la) = \frac 12 \sqrt {(x_R(\la) - x_L(\la))^2-(t_R(\la)-t_L(\la))^2}\\ 
&\beta(\la)=\frac 12 \log \left[\frac{(x_R(\la)-x_L(\la))+(t_R(\la)-t_L(\la))}{(x_R(\la)-x_L(\la))-(t_R(\la)-t_L(\la))}\right]\label{param}} 
and we choose $x_R(\la) \geq x_L(\la)$. Note that we are only considering space-like intervals, \ie $|t_R(\la)-t_L(\la)|< x_R(\la)-x_L(\la)$ and hence the boost angle $\beta(\lambda)$ is everywhere finite and well-defined. For an interval at $\la$, with the parameterization $s\in[-1,1]$ the extremal curve has coordinates $\{Z,X,T\}$ given by 
\ban{
\Gamma(s;\la)= \coords{\sqrt{1-s^2} \Delta(\lambda), \, x_c(\lambda) + s\,  \Delta(\lambda) \cosh \beta(\lambda), \, t_c(\lambda) + s \, \Delta(\lambda)\sinh \beta(\lambda)}\label{geodB}
}

\noindent{\bf On a constant time slice:} First we consider boundary intervals which all lie on a constant time slice of AdS$_3$, characterized by $t_c(\la)=t_0$ and $\beta(\la)=0$. In this case the extremal curves are given by $\Gamma(s;\la)=\{ x_c(\la)+s\, \Delta(\la), \,\sqrt{1-s^2}\Delta(\la),\,t_0\}$. Let us note that in this case, the orthonormal basis introduced above becomes
\ban{
\hat u^\mu&=\frac{\Delta(\la)}{2L}\,\{ -s,\sqrt{1-s^2},0\}\,,\notag\\
\hat n_1^\mu&=\frac{\Delta(\la)}{2L}\,\{-\sqrt{1-s^2}, -s,0\}\,,\label{outside}\\
\hat n_2^\mu&=\frac{\Delta(\la)}{2L}\,\{ 0,0,1\}\,.\notag
}
For this case, we directly solve the intersection equation $\Gamma(s_+(\la);\la)=\Gamma(s_-(\la+d\la;\la+d\la)$, and we find exactly one solution given by 
\ban{s_+(\la)=-\frac{(x_c(\la)-x_c(\la+d \la))^2+\Delta(\la)^2-\Delta(\la+d \la)^2}{2\Delta(\la) (x_c(\la)-x_c(\la+d \la))}\label{intersection}}
In the continuum limit, this solution reduces to the result $s_+= -\Delta'(\la)/x'_c(\la)$. Imposing the condition $|s_+(\la)|<1$ in terms of $x_L(\la)$ and $x_R(\la)$ yields
\ban{
\left(\frac{x_L'(\la)-x_R'(\la)}{x_R'(\la)+x_L'(\la)}\right)^2<1
}
which can be rewritten as $x_L'(\la)x_R'(\la)>0$. That is, we have recovered the global constraint \reef{constrain88}
from this inequality. 
 
\noindent{\bf Generic families of intervals:} For the general case in AdS$_3$, the basis vectors become
 \ban{
 \hat u^\mu &= \frac{\Delta(\la)}{2L}\coords{-s, \, \sqrt{1-s^2} \cosh\beta(\la), \, \sqrt{1-s^2}  \sinh \beta(\la)} \notag\\
 \hat n_1^\mu &= \frac{\Delta(\la)}{2L}\coords{-\sqrt{1-s^2}, \, -s  \cosh\beta(\la), \, -s\,  \sinh \beta(\la)} \label{outside2} \\ 
 \hat n_2^\mu &=\frac{\Delta(\la)}{2L} \coords{0, \, \sinh \beta(\la), \,  \cosh\beta(\la)}\notag
 }
Now to determine when null vector alignment is achieved,  it is easiest to solve for when $|v_\perp(s;\la)|=0$. 
The projection of $\Gamma'(s;\la)$ into the transverse space is given by
\ban{v_\perp(s;\la) \propto &-(s\, x'_c(\la) \cosh \beta(\la) - s\, t'_c(\la) \sinh \beta(\la) + \Delta'(\la)) \hat n_1 \notag\\
&+(t'_c(\la) \cosh \beta(\la) - x'_c(\la) \sinh \beta(\la) + s\,\Delta(\la) \beta'(\la)) \hat n_2
}
We can explicitly solve for the parameters $s_\pm(\la)$
where $v^\mu_\perp$ is null,
\ban{s_\pm(\lambda)=-\frac{\Delta'(\la) \pm t'_c(\la) \cosh \beta(\la) \mp x'_c(\la) \sinh \beta(\la) }{\pm \Delta(\la) \beta'(\la)+x'_c(\la) \cosh\beta(\la)-t'_c(\la) \sinh\beta(\la) } \,. \label{nullcross}}
Note that this solution reduces to the constant time case, and we see an explicit confirmation that our intuition about intersection points was correct.

Next we turn to the constraints $|s_\pm(\la)|<1$. After some simplification, we see explicitly that $|s_+(\la)|<1$ corresponds to inequality \reef{constrA} and $|s_-(\la)|<1$ corresponds to inequality \reef{constrB}. Therefore the global
constraint \reef{constrA} ensures that a bulk curve exists corresponding to null vector alignment at $s_+(\la)$ while eq.~\reef{constrB} ensures the same at $s_-(\la)$. Furthermore,
we can interpret $s_+(\la)$ as the intersection with the null line $\hat n_1 + \hat n_2$ and $s_-(\la)$ as the intersection with the null line $\hat n_1 - \hat n_2$. These observations reveal that indeed for AdS$_3$, the previously mentioned trajectories in the transverse plane cross each light cone at most once. 

\noindent{\bf Equivalence classes of boundary intervals:}  
One of the lessons to learn thinking about intersections of entanglement wedges is that we did not need to limit ourselves to tangent vector projection in section \ref{time}. That is, given a curve in the bulk, we can build a family boundary intervals by projecting along a geodesic along $\bulkc_B'/|\bulkc_B'| +k$ where $k$ is any null vector such that $k\cdot \bulkc'_B=0$. These two conditions define a one-parameter family of possible vectors $k$, and therefore a one parameter set of boundary intervals for which the differential entropy equals the gravitational entropy of a given bulk curve. This defines a sort of gauge symmetry in the space of families of boundary intervals, and we can derive an explicit transformation between the parameters characterizing each family. However, even for AdS$_3$ this transformation doesn't reveal itself as an obvious symmetry of the boundary theory.\footnote{Note that lifting the restriction of planar symmetry will also expand the families of boundary regions which correspond to the same bulk curve.}

An alternative program might be to single out a particular `gauge' via some natural physical principle. One such gauge might be tangent vector alignment, where $s_+(\la)=s_-(\la)$. Solving this equation explicitly we get
\ban{
s_B(\la) &= \frac{\Delta'(\la)}{x'_c(\la) \cosh \beta(\la) - t'_c(\la) \sinh \beta(\la)}
}
subject to the condition
\ban{
{2 \Delta(\la) \Delta'(\la)}\beta'(\la)&= 2 x'_c(\la)t'_c(\la)\cosh(2\beta(\la))-(x'_c(\la)^2+t'_c(\la)^2)\sinh(2\beta(\la)) \label{TVAcond}
}
As usual we have the global condition $|s_B(\la)|<1$ which in our parameterization is $(x'_R{}^2-t'_R{}^2)(x'_L{}^2-t'_L{}^2)>0$, corresponding to a trajectory crossing both null lines. 

\begin{figure}[h!]
\begin{center}
\includegraphics[width=0.55\textwidth]{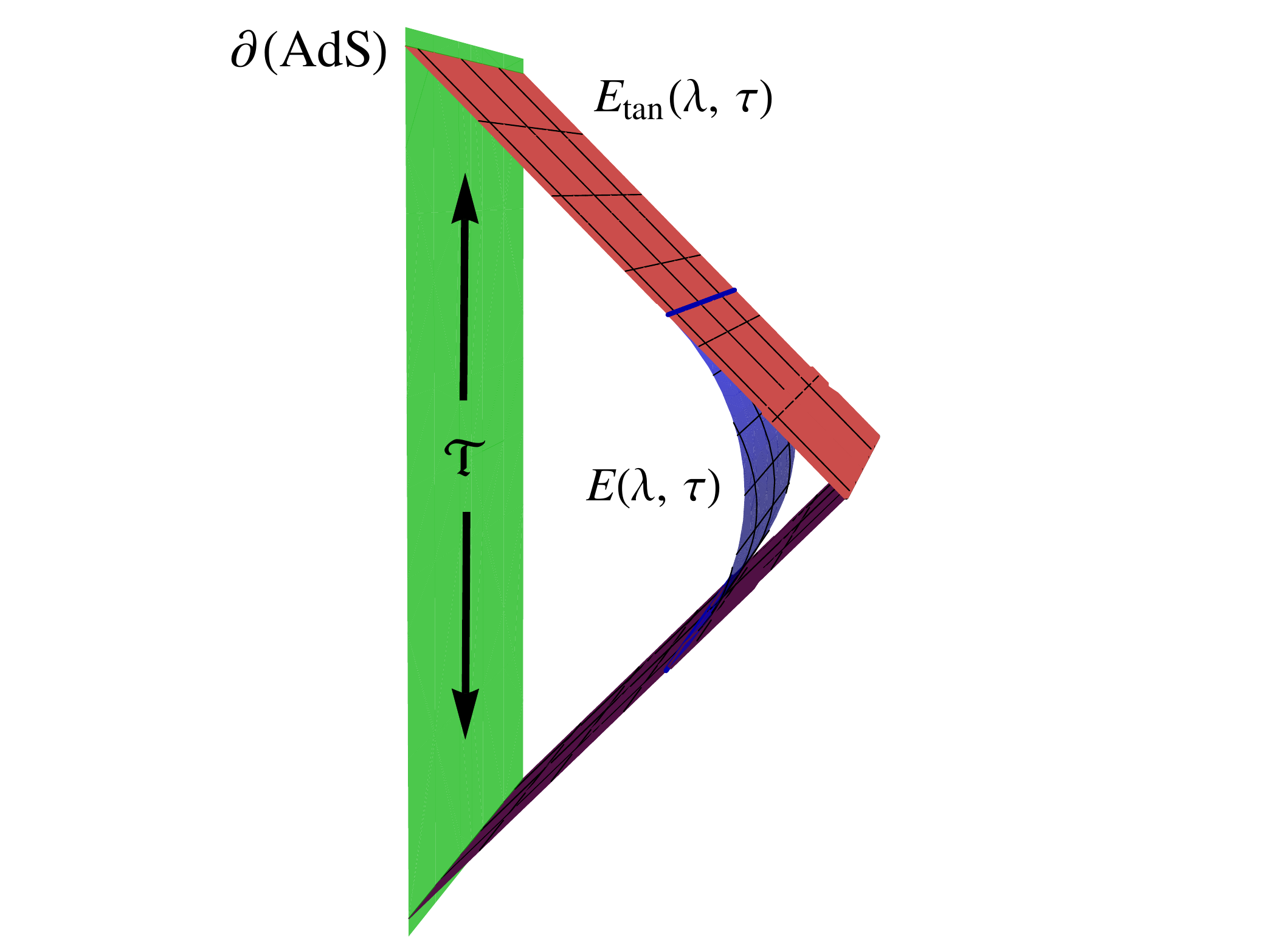}
\caption{We compare the surfaces $E_{\text{tan}}(\la,\tau)$, the boundary of the bulk region spatial to the time strip $\mathcal T$ for a family of boundary intervals, and the enveloping surface $E(\la,\tau)$. We see that generically $E_{\text{tan}}(\la,\tau)$ bounds the enveloping surface.}
\label{strip}
\end{center}
\end{figure}

Further insight comes from considering the enveloping surface shown in figure \ref{EWS}. 
If we compare the enveloping surface with surface $E_{\text{tan}}(\la,\tau)$ which is the boundary of the bulk region space-like to the union of causal diamonds of the family of boundary intervals $\mathcal T$, we find that generically the entangling surface lies inside of $E_{\text{tan}}(\la,\tau)$ as shown in figure \ref{strip}. Additionally, by construction in the case of tangent vector alignment the enveloping surface forms the boundary of this region. 
In \cite{Balasubramanian2014}, differential entropy was proposed as quantifying the residual uncertainty about the state after making measurements in the finite time strip $\mathcal T$. This observation suggests that the case of tangent vector alignment some how minimizes the differential entropy among the families of boundary intervals which share a time strip $\mathcal T$, as the bulk curve in this case extends the farthest into the bulk and has minimal area. 

Can we therefore understand tangent vector alignment in terms of minimizing the residual uncertainty of a family of intervals on the boundary? Stated more sharply, given two curves $\gamma_U(\la)$ and $\gamma_D(\la)$ which define the `upper' and `lower' boundaries of a time strip $\mathcal T$, does the condition of tangent vector alignment maximize the proper time experienced by a family of observers making measurements in $\mathcal T$? Generically, the answer is no, but in some special cases the answer is yes. 

Given two curves parameterized by 
\ban{
\gamma_{U/D}(\la)=\{x_c(\la)\pm\Delta(\la)\sinh \beta(\la), t_c(\la)\pm\Delta(\la)\cosh\beta(\la)\}
}
we derive a condition for when these parameters are a `maximal time protocol.' Note that these curves correspond to a family of intervals parameterized in the usual way. First, we unfix the relative parameterization between the upper and lower curves, writing $\gamma_U(\rho)$ and $\gamma_D(\la)$, and then we solve for the value of $\rho$ which maximizes the proper time of an observer starting at $\gamma_D(\la)$ and ending at $\gamma_U(\rho)$. If the given parameters are a maximal time protocol, then the solution should be $\rho=\la$ and we have the condition
\ban{
0=\left. \pd {\left|\gamma_U(\rho)-\gamma_D(\la)\right|}\rho\right|_{\rho=\la} = 2\Delta'(\la) +t'_c(\la)\cosh \beta(\la) -x'_c(\la)\sinh \beta(\la) \label{MTP}
}

\begin{figure}[h!]
\begin{center}
\includegraphics[width=0.5\textwidth]{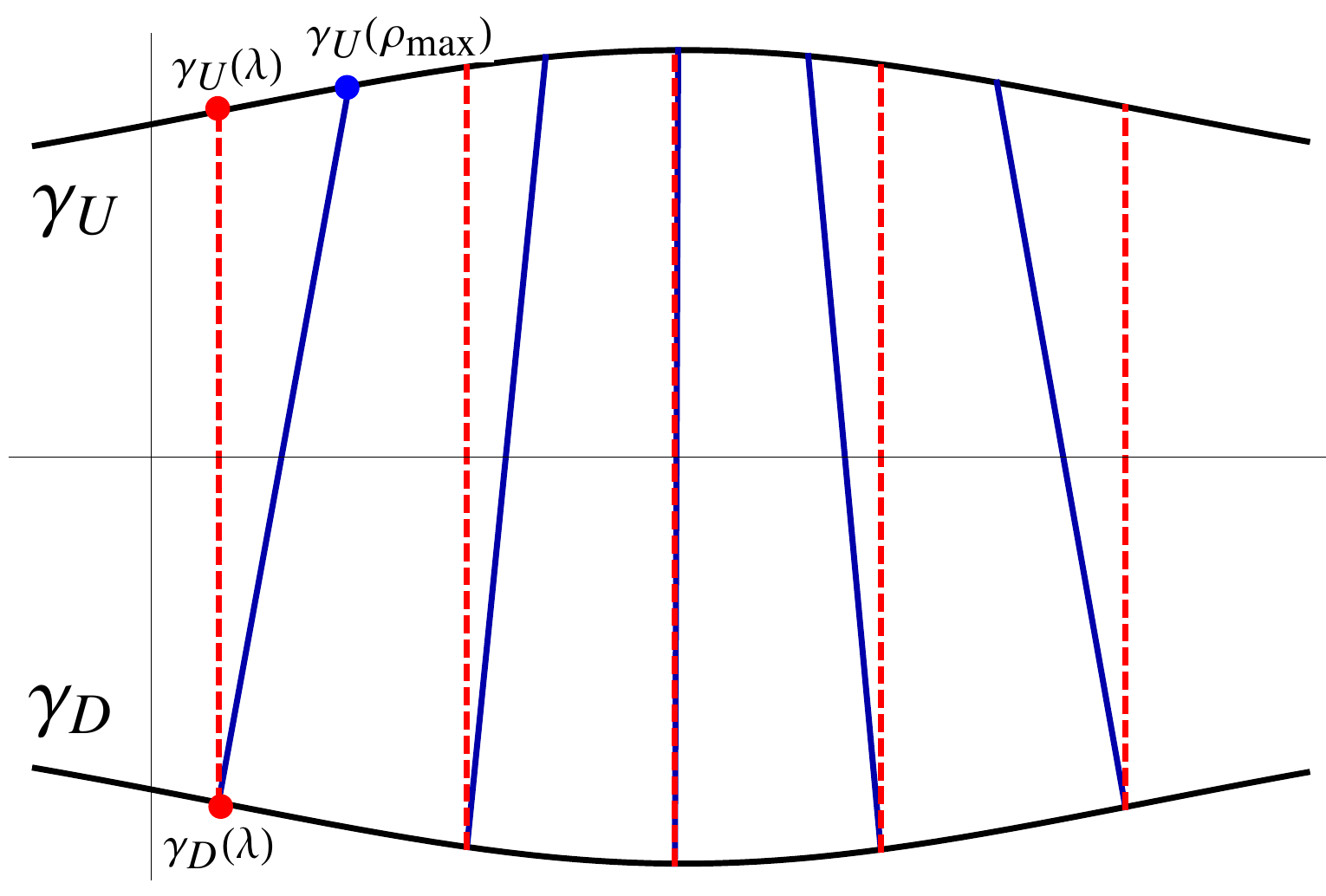}
\caption{For a given time strip, the path an observer takes under tangent vector alignment is in dashed red, and the path an observer takes under a maximal time protocol is in solid blue. If $\Delta'(\la)\neq 0$, the two conditions are not equivalent and $\gamma_U(\la)\neq \gamma_U(\rho_\text{max})$.}
\label{protocol}
\end{center}
\end{figure}

Comparing this condition to the condition for tangent vector alignment \reef{TVAcond}, we find that in general families of intervals which satisfy tangent vector alignment are not a maximal time protocol. See figure \ref{protocol} for an explicit example of the difference between the two approaches. However in the special case $\Delta'(\la)=0$ these conditions are indeed equivalent. In terms of our parameterization, this special case is when all of the boundary intervals have the same proper length. This result hints at a more direct interpretation of differential entropy, but more research is necessary to further understand this concept.

%

\chapter{Outlook}
\label{discuss}

In this essay, we showed that given an arbitrary closed space-like curve in the bulk of a three dimensional holographic spacetime, we are able to construct a family of boundary intervals whose differential entropy equals the gravitational entropy of the bulk curve. Additionally, we showed that the converse statement is also true.

For possible extensions, note that generic families of space-like boundary intervals produce bulk curves which may not be space-like everywhere via the construction from section \ref{generic}. Therefore, we infer it is possible to extend these constructions to surfaces which can also be time-like. In addition, an obvious extension is to lift the assumption of planar symmetry for higher dimensions. 

Additionally, note from the discussion at the end of section \ref{AdS} that there is a one-parameter set of boundary intervals corresponding to a given bulk curve. A new principle is therefore needed either to understand this equivalence in terms of a symmetry of the boundary theory or to pick out a preferred family of intervals for a given bulk curve. We showed that one of the simplest principles, \ie associating the condition of tangent vector alignment with a tiling of a given time strip that maximizes the proper time of observers, is in general not self-consistent. 

This observation might help shed light on answering if the differential entropy should be interpreted as an entropy, \ie counting a certain set of degrees of freedom in the boundary theory, or if it should be interpreted strictly as the directional derivative of entanglement entropy along a family of intervals. In either case, it is hoped that the holographic constructions outlined in this essay will lead to a better understanding of the interpretation of Bekenstein-Hawking entropy as the leading contribution to the entanglement entropy of an underlying quantum theory of gravity.

\appendix
\chapter*{APPENDICES}
\addcontentsline{toc}{chapter}{APPENDICES}

\chapter{Derivation of the Geometric Interpretation}
\label{geom}
In this appendix we show that the geometric interpretation of section \ref{new} satisfies the condition for the holographic correspondence established in section \ref{general} between the differential entropy evaluated on a family of boundary intervals and the gravitational entropy of a bulk curve. Given a family of intervals with extremal curves $\Gamma(s;\la)$, the bulk curve constructed via $\bulkc_B(\la)=\Gamma(s_B(\la);\la)$ must satisfy 
\ban{\left.\frac {\dot \bulkc_B(\la) \cdot \bulkc'_B(\la)}{|\dot \bulkc_B(\la)|} \right.= |\bulkc'_B(\la)| \,.\labell{aligned}}
As before we change variables to $\Gamma'(s;\la)$ and $\dot \Gamma(s;\la)$ with the relations in eq.~\reef{note99}, \ie
\be
 \bulkc'_B{}^{\!\!\mu}(\la)=\left. \Gamma'{}^{\mu}(s;\la)\right|_{s_B(\la)}+ \dot \Gamma^\mu(s;\la)|_{s_B(\la)}\ s_B'(\la)
\quad {\rm and}\quad 
\dot \bulkc_B^\mu(\la)=\dot \Gamma^\mu(s;\la)|_{s_B(\la)}\,.
\labell{note99A}
\ee
and then it is straightforward to show that eq.~\reef{aligned} becomes
\ban{\left. \frac{\dot \Gamma(s;\la) \cdot \Gamma'(s;\la)} { |\Gamma'(s;\la)|{|\dot \Gamma(s;\la)|} }\right|_{s=s_B(\la)}
=1\,. \labell{alignedG}}

Now recall the basis of orthonormal vectors established for each extremal curve in section \ref{new}. This basis consists of the tangent vector $\hat u(s;\la)=\dot \Gamma(s;\la)/|\dot \Gamma(s;\la)|$ and two orthogonal unit vectors $\hat n_1(s;\la)$ and $\hat n_2(s;\la)$. With this formalism, we defined $v^\mu_\perp(s;\la)$, the projection of $\Gamma'{}^\mu$ into the
subspace transverse to $\hat u(s;\la)$ in eq.~\reef{proj0}. The condition \reef{alignedG} for null vector alignment then
became $|v_\perp(s_B(\la);\la)|=0$.

\vskip 0.2cm

Now we begin by showing that when the boundary intervals all lie on a constant time slice, at the intersection between the curve $\Gamma(s;\la)$ and $\Gamma(s;\la+d\la)$ the condition of null vector alignment \reef{aligned} is satisfied in the continuum limit \ie as $d\la \to 0$. In fact since all the intervals are on a constant time slice, in this case we have tangent vector alignment. Let $s_\pm(\la)$ denote the intersection of the extremal curve $\Gamma(s;\la)$ with $\Gamma(s;\la \pm d\la)$.\footnote{As discussed in section \ref{new}, we expect neighboring curves to intersect at most once. However, in the situations where the curves are extremal but not minimal, it may be that they intersect more than once, as discussed in section \ref{discuss}. In this case, we can simply choose consecutive points such that eq. \reef{intersectp} holds.} By construction, the `right' intersection point for $\Gamma(s;\la)$ is equal to the `left' intersection point for $\Gamma(s;\la+d\la)$ so we have
\ban{
\Gamma(s_+(\la); \la) &= \Gamma(s_-(\la+d\la); \la+d\la)  \labell{intersectp}
}
We can expand this equation for $d\la \ll |\gamma_R(\la)-\gamma_L(\la)|$ to get
\ban{
\Gamma(s_+(\la);\la)=\Gamma(s_-(\la);\la)+\mathcal O(d\la)
}
And as we are assuming a bijective parameterization this equation implies that $s_+(\la)-s_-(\la)\sim \mathcal O(d\la)$. Therefore we can write
\ban{
s_+(\la) &= s_I(\la)+\delta s_+ (\la)d\la + \mathcal O(d\la^2)\\
s_-(\la) &= s_I(\la)+\delta s_- (\la)d\la + \mathcal O(d\la^2)
}
where we refer to $s_I(\la)$ as the `intersection point in the continuum limit.' Substituting these expressions into \reef{intersectp} we get
\ban{
&\Gamma(s_I(\la); \la)+\dot \Gamma(s_I(\la);\la) \delta s_+(\la) d\la + \mathcal O(d\la^2)\notag \\ &=\Gamma(s_I(\la);\la)+\dot \Gamma(s_I(\la);\la) \delta s_-(\la) d\la + \left( \Gamma'(s_I(\la); \la)+s'_I(\la)\dot \Gamma(s_I(\la);\la)\right)d\la+\mathcal O(d\la^2) \label{intersectdp}
}
\normalsize
And so we have that at the point $s_I(\la)$
\ban{
\alpha(\la) \dot \Gamma(s;\la)|_{s_I(\la)}= \Gamma'(s;\la)|_{s_I(\la)}
}
where $\alpha(\la)=\delta s_+(\la) - \delta s_-(\la)- s'_I(\la)$. In this way, at $s_I(\la)$ the curves satisfy tangent vector alignment, so the bulk curve can be thought of as being built from the intersection points between extremal curves and their neighbors, in the continuum limit. 

\vskip0.2cm

Next we repeat the above analysis for the general case. That is, we show that at the intersection of $\Gamma(s;\la)$ with the entanglement wedge boundary $W(s,\tau;\la\pm d\la)$, we have the desired relation $\Gamma'(s;\la)|_{s_\pm(\la)} \propto \dot \Gamma (s;\la)|_{s_\pm(\la)}+ k_\pm(\la)$. It is convenient to use the parameterization 
\ban{
W^\mu(s,\tau;\la)=\Gamma^\mu(s;\la)+  \tau\, k^\mu(s ;\la) \labell{Wparam}
}
where $k(s;\la)\cdot \dot \Gamma(s;\la)=0$, $|k(s;\la)|=0$, and $\tau \in [0,1]$. That is, the vector $k^\mu(s;\la)$ is the null separation between $\Gamma(s;\la)$ and the `cusp' of the entanglement wedge. Note that there are two such vectors which we denote $k_\uparrow$ and $k_\downarrow$, corresponding to the `upper' and `lower' parts of the entanglement wedge respectively. Hence they lie on two different light sheets and so they can not be smoothly deformed into one another while remaining null.

As above we denote the intersection of $\Gamma(s;\la)$ with $W(s,\tau;\la \pm d\la)$ by $s_\pm(\la)$. For concreteness we will assume that there exists one unique point for both $s_+(\la)$ and $s_-(\la)$, \ie the trajectory of $\Gamma'$ in the transverse plane can only cross each null line once, and we discuss the general case below. By construction we have
\ban{
\Gamma{}^\mu(s_+(\la);\la)&=\Gamma{}^\mu(s^*_+(\la+d\la); \la+d\la) + \tau^*_+(\la+d\la) k^\mu(s^*_+(\la+d\la); \la+d\la) \labell{EWintersect}
}
for some particular $\tau^*_+(\la+d\la)$ and $s^*_+(\la+d\la)$. Note that in general $s^*_+(\la)\neq s_-(\la)$ because the choice of $k_\uparrow$ or $k_\downarrow$ generically differs in the equation analogous to eq.~\reef{EWintersect} for $s_-(\la)$. We return to this point later in the discussion. The general setup and notation is illustrated in figure \ref{EWI}. Expanding eq. \reef{EWintersect} around $d\la$ we have
\ban{
\Gamma^\mu(s_+(\la);\la) = \Gamma^\mu(s^*_+(\la);\la) + \tau^*_+(\la)k^\mu(s^*_+(\la); \la) + \mathcal O(d\la)
}
This equation implies that the zeroth order separation between $\Gamma^\mu(s_+(\la);\la)$ and $\Gamma^\mu(s^*_+(\la);\la)$ is a null vector, but as the extremal curves are space-like this must vanish. Therefore we can write 
\ban{
s^*_+(\la)&=s_+(\la)+\delta s_+(\la)\, d\la + \mathcal O(d\la^2)\\
\tau^*_+(\la)&= \delta\tau_+(\la)\, d\la + \mathcal O(d\la^2)
}
Plugging this expansion into eq. \reef{EWintersect} and keeping all terms to first order we have
\ban{
\Gamma{}^\mu(s_+(\la);\la)=&\Gamma^\mu(s_+(\la);\la)+\dot \Gamma^\mu(s_+(\la);\la) \delta s_+(\la) d\la+ \dot \Gamma^\mu(s_+(\la);\la)s'_+(\la) d\la \notag\\
&+ \Gamma'{}^\mu(s_+(\la);\la) d\la + \delta\tau_+(\la) k^\mu(s_+(\la);\la)d\la 
}
and we see explicitly 
\ban{
 \Gamma'{}^\mu |_{s_+(\la)}=\alpha \left( \dot \Gamma|_{s_+(\la)}+\tilde k_+\right) \labell{NVA}
}
where $\alpha(\la)= -(\delta s_+(\la)+s'_+(\la))$ and $\tilde k_+(\la)= -\frac{\delta\tau_+(\la)}{\delta s_+(\la)+s'_+(\la)} k(s_+(\la);\la)$. By construction $\tilde k_+$ is null and $\tilde k_+ \cdot \dot \Gamma =0$, and therefore at this point the condition of null vector alignment is satisfied. We can then identify $s_+(\la)$ with $s_B(\la)$ to form the bulk curve as the continuum limit of these intersection points. 

Additionally, we can write down equations analogous to eqs.~\reef{EWintersect} and \reef{NVA} for the intersection point $s_-(\la)$. By repeating the above arguments, we can show that at $s_-(\la)$ the extremal curves satisfy null vector alignment: 
\ban{
 \Gamma'{}^\mu |_{s_-(\la)}=\alpha \left( \dot \Gamma|_{s_-(\la)}+\tilde k_-\right) \labell{NVAm}
}
where $\alpha(\la)= -(\delta s_-(\la)+s'_-(\la))$ and $\tilde k_-(\la)= -\frac{\delta\tau_-(\la)}{\delta s_-(\la)+s'_-(\la)} k(s_-(\la);\la)$. 

Further, note from the above definitions $\tilde k_\pm$ is proportional either to $k_{\uparrow}$ or $k_{\downarrow}$, and we see from the null vector alignment equations \reef{NVA} and \reef{NVAm} that $\tilde k_\pm$ is additionally proportional to the projection of $\Gamma'$ into the transverse plane. We also note that by definition the two null directions in the transverse plane are given exactly by the vectors $k_\uparrow$ and $k_\downarrow$, and so crossings of each null direction in the transverse plane are characterized by the null vector alignment equations \reef{NVA} and \reef{NVAm}. As we assume the trajectories in the transverse plane can cross each light cone only once, then in the continuum limit there can be at most one point on the extremal curve satisfying null vector alignment for each $k_\uparrow$ and $k_\downarrow$. Therefore, we have that if $\tilde k_+(\la) \propto \tilde k_-(\la)$ then $s_+(\la)-s_-(\la) \sim \mathcal O(d\la)$. However, note that this situation can only arise when the extremal curves $\Gamma(s;\la\pm d\la)$ are either both `above' or both `below' $\Gamma(s;\la)$, and so at $\la$ the time-like separation between extremal curves is either a maximum or a minimum. Therefore in the continuum limit, $\delta\tau_\pm(\la)$ vanishes and in this case we additionally have tangent vector alignment, corresponding to a trajectory crossing through the origin as in figure \ref{transit2}a.

\chapter{Characterization of Planar Symmetry}
\label{planarS}
For the proof in section \ref{general} to apply in higher dimensions, we are implicitly making some assumptions about the
relevant surfaces and the background geometry. In particular, given a general holographic $d+1$ dimensional
spacetime with coordinates $q^i=\{t, x, z\}$ and $y^a=\{y^1, \cdots, y^{d-2}\}$, we would like to consider a co-dimension two surface in the bulk parameterized by $\{\la, \sigma^a\}$ with a simple embedding which
factorizes as
\ban{
\gamma_B(\la,\sigma^b)=\lbrace q^i(\la, \sigma^b), y^a(\la, \sigma^b)\rbrace= \lbrace q^i(\la),\sigma^a\rbrace
 \label{planar}
}
Implicitly to describe the gravitational entropy of this bulk surface, it must be that the extremal surfaces appearing
in the holographic evaluation of the differential entropy have a similar simple description, \ie
\ban{
\Gamma(s,\sigma^b;\la)=\lbrace q^i(s(\la), \sigma^b(\la)),\ y^a(s(\la), \sigma^b(\la))\rbrace= \lbrace q^i(s(\la)),\ \sigma^a\rbrace
 \label{planar2}
}
However these extremal surfaces must be solutions to the equations of motion extremizing the given Lagrangian
and so this implicit property restricts the class of background spacetimes which we can consider.
If the surfaces admit the parameterization in eqs.~\reef{planar} and \reef{planar2}, we say that they have 
`planar symmetry' and we call $y^a$ the planar coordinates. Similarly, we say that the background geometry
has planar symmetry if the parameterization \reef{planar2} consistently applies for solutions of the equations
of motion determining the extremal surfaces. 

Towards identifying the class of backgrounds which admit solutions with planar symmetry, we restrict our attention
to the case of Einstein gravity in the bulk, for which appropriate entropy functional is simply the Bekenstein-Hawking entropy as in eq.~\reef{RT}. In the language of section \ref{general}, the Lagrangian is simply $\sqrt{h}/(4G_N)$, where
$h$ is the determinant of the induced metric on the bulk surface. Now, we
show that spacetimes for which we can `factor out' the $y^a$ coordinates admit planar symmetry. In particular, 
we consider spacetimes with a metric of the form
\ban{
ds^2 = g_{jk}(q^i)\, dq^j\, dq^k+ g_{bc}(q^i, y^a)\, dy^b\, dy^c \label{pmetric}
}
and where the determinant of $g_{bc}$ can be written as  $\det\!\[g_{bc}(q^i, y^a)\] = F(q^i)\,\Sigma(y^a)$. 
We now show that the ansatz \reef{planar2} indeed provides a solution of the corresponding
equations of motion for metrics of this form. 

First, the determinant of the induced metric can be written as 
\ban{
h = \varepsilon^{\alpha_0\cdots \alpha_{d-2}}& \left(g_{ij} \partial_s q^i \partial_{\alpha_0}q^j + g_{ab}\partial_s y^a \partial_{\alpha_0} y^b\right)\notag\\
&\times\left(g_{ij} \partial_{\sigma^1} q^i \partial_{\alpha_1}q^j + g_{ab}\partial_{\sigma^1} y^a \partial_{\alpha_1} y^b\right)\times\cdots \notag\\
&\times\left(g_{ij} \partial_{\sigma^{d-2}} q^i \partial_{\alpha^{d-2}}q^j + g_{ab}\partial_{\sigma_{d-2}} y^a \partial_{\alpha_{d-2}} y^b\right)
}
where $\varepsilon^{\alpha_0\cdots \alpha_{d-2}}$ is the totally antisymmetric symbol on the surface.
Next given ${\cal L}\propto\sqrt{h}$, the equations of motion can be written as 
\ban{
\pd h{\xi^\mu} -\partial_\alpha \pd{h}{(\partial_\alpha \xi^\mu)}+\frac{\partial_\alpha{h}}{2h} \, \pd{h}{(\partial_\alpha \xi^\mu)}=0 \label{eom99}
}
where $\xi^\mu=\{q^i, y^a\}$ and $\partial_\alpha = \lbrace\pd{}{s},\, \pd{}{\sigma^{\alpha_i}}\rbrace$. 
To simplify notation, we introduce $Q(s) = g_{ij}\,\partial_s q^i(s) \partial_s q^j(s)$. Next we evaluate each term in eq.~\reef{eom99} for the $q^i$ coordinates evaluated on the planar symmetry ansatz \reef{planar2}:
\ban{
\left. \pd{h}{q^i}\right|_{\Gamma_{\mathcal P}}&= \partial_s q^j(\la) \partial_s q^k(\la)  \Sigma(\sigma^a)\pd{}{q^i}{\left[F(q^i(s))\,g_{jk}(q^i(s))\right]}\notag\\
\left. -\partial_\alpha\pd h {(\partial_\alpha q^i)}\right|_{\Gamma_{\mathcal P}}&=-2\Sigma(\sigma^a) \pd{}\la \left[F(\la)g_{ij}(\la) \partial_\la x^j(\la)\right]
 \label{threep}\\
\left. \frac {\partial_\alpha h }{2h}\pd h {(\partial_\alpha q^i)}\right|_{\Gamma_{\mathcal P}}&=\frac {\Sigma(\sigma^a)}{Q(\la)} g_{ij}(\la)\partial_\la q^j (\la) \pd{}\la\left[Q(\la) F(\la)\right]  \notag
}
Summing the three above equations gives the equation of motion for $q^i$. Hence we see that all of the dependence on $\sigma^a$ is isolated in an overall factor of $\Sigma(\sigma^a)$. Hence, dividing out by this factor  (which we will assume
only vanishes at isolated points), all of $\sigma^a$ dependence drops out of these three equations of motion for $q^i$. We can additionally assume that our original spacetime is well enough behaved so that these resulting equations have a solution.

Next, we examine the equations of motion \reef{eom99} for $y^a$. Similarly we can write
\ban{
\left. \pd{h}{y^a}\right|_{\Gamma_{\mathcal P}}&=F(\la)Q(\la)\pd{\Sigma}{\sigma^a}\notag\\
\left. -\partial_\alpha\pd h {(\partial_\alpha y^i)}\right|_{\Gamma_{\mathcal P}}&=- 2F(\la) Q(\la)\pd{\Sigma}{\sigma^a}
 \label{gongshow}\\
\left. \frac {\partial_\alpha h }{2h}\pd h {(\partial_\alpha y^a)}\right|_{\Gamma_{\mathcal P}}&= F(\la) Q(\la)\pd{\Sigma}{\sigma^a} \notag}
and therefore we see that summing these three terms gives a vanishing result in eq.~\reef{eom99}. Hence we conclude that spacetimes with metrics of the form described by eq.~\reef{pmetric} have planar symmetry
and are accommodated by the construction in appendix \ref{AdSn}.

\chapter{Holographic Holes in AdS$_{d+1}$}
\label{AdSn}

We can extend the construction of chapter \ref{time} to arbitrary surfaces with planar symmetry in AdS$_{d+1}$ for $d>2$. Denote the directions with planar symmetry as $y^a$, and we can write the Poincar\'e metric as 
\ban{
ds^2=\frac {L^2}{z^2}\left( dz^2 -dt^2 +dx^2 + \sum_a dy_a^2\right) \label{Nmetric}
}
Additionally, we impose a periodicity $\ell_i$ in the spatial directions on the boundary as an infrared regulator of the area which we assume is much larger than the width of the strips constructed on the boundary. 

Given a surface parameterized by $\Sigma(\la, \sigma^a)=\coords{Z(\la), X(\la), T(\la), \sigma^1, \cdots, \sigma^{d-2}}$ with periodic boundary conditions $\Sigma(0, \sigma^a)=\Sigma(1,\sigma^a)$, we can construct the corresponding regions on the boundary by noting that the tangent extremal surface must respect the planar symmetry of the $y^a$ coordinates, as the vectors $\partial_{y_i}$ are Killing vectors of the induced metric. That is, we construct a family of co-dimension two strips on the boundary with a fixed proper width $\Delta(\la)$ in the $x$ direction and length $\ell_a$ in the $y^a$ directions. As in the case of AdS$_3$, each strip can have a nonzero boost angle and be centered on a different time coordinate. Note that by fixed proper width we mean $\Delta(\la)$ does not depend on the coordinates $y^a$, but in general it can vary between strips in the family, parameterized as usual by $\la$. 

The entanglement entropy of any such strip at $\la$ is given by \cite{Ryu2006}
\ban{
S(\la)=\frac{L^{d-1}}{4G_N}\frac{\ell_2\cdots \ell_{d-1}}{d-2} \left(\frac 2{\delta^{d-2}} - \frac {c_d ^{d-1}}{\Delta(\la)^{d-2}}\right) \label{EEn}
}
where the constant $c_d$ is the ratio of the strip width $\Delta(\la)$ with the maximum $z$ value $z_*$ achieved by the corresponding extremal surface. 
\ban{
c_d=\frac{\Delta(\la)}{z_*}=2\sqrt \pi \frac{\Gamma\left(\frac d{2d-2}\right)}{\Gamma\left(\frac 1{2d-2}\right)} \label{const}
}

The construction follows the same strategy as the case of AdS$_3$: at each point $\la$ on the bulk surface we find the tangent extremal surface and take its intersection with the boundary to define the strip at $\la$. First, we boost the coordinate system so that the tangent vector $\partial_\la \Sigma$ has no time-like component, where the problem reduces to the case of a surface on a constant time slice studied in \cite{Myers2014}. As $\partial_\la \Sigma(\la, \sigma^a)=\coords{Z'(\la), X'(\la), T'(\la), 0, \cdots, 0}$, we will drop the final vanishing components for simplicity.

In the coordinates boosted from the original coordinate system by the boost angle $\beta(\la)=\log \sqrt{\frac{X'(\la)+T'(\la)}{X'(\la)-T'(\la)}}$, the tangent vector is proportional to 
\ban{
u^*(\la) = \coords{\frac{Z'(\la)}{\sqrt{X'(\la)^2 - T'(\la)^2}}, 1,0}
}
In this way we can use the formulas computed in \cite{Myers2014} for a surface with a tangent vector $\coords{\tilde Z'(\la),1,0}$ where $\tilde Z'(\la)=\frac{Z'(\la)}{\sqrt{X'(\la)^2 - T'(\la)^2}}$. The parameters of the strip whose extremal surface is tangent to the bulk surface for this tangent vector are therefore given in these boosted coordinates by 
\ban{
x_c^*(\la)&= X(\la) + \frac 1{2{(d-1)}} Z(\la) \left(1+\frac{Z'(\la)^2}{X'(\la)^2-T'(\la)^2}\right)^{\frac 1{2{(d-1)}}} B\left[\left(1+\frac{Z'(\la)^2}{X'(\la)^2-T'(\la)^2}\right)^{-1}\right]\notag\\
\Delta(\la)&= c_{d} Z(\la)\left(1+\frac{Z'(\la)^2}{X'(\la)^2-T'(\la)^2}\right)^{\frac 1{2{(d-1)}}}
}
where $x_c^*(\la)$ denotes the $x$ coordinate of the center of the strip, $\Delta(\la)$ denotes the invariant width of the strip, and the function $B[\la]$ is given by 
\ban{
B[\la]=\int_\la^1 \frac 1{s^{\frac{d-2}{2(d-1)}}\sqrt{1-s}}ds
}
Applying the inverse boost, we recover the parameters of the strip in the original coordinate system. The invariant width remains the same, but the center of the strip is given by
\ban{
x_c(\la)&=X(\la)+(x_c^*(\la)-X(\la))\cosh \beta(\la)\notag\\
t_c(\la)&=T(\la)+(x_c^*(\la)-X(\la))\sinh \beta(\la)
}
With this parameterization, the ends of the strip are given by
\ban{\gamma_{R,L}(\la)=\left\{x_c(\la)\pm \Delta\cosh \beta(\la), t_c(\la)\pm \Delta \sinh \beta(\la)\right\}}

Given these curves, it is straightforward to compute the differential entropy of this family of strips. As in the case of AdS$_3$ we use the Lorentz symmetry of AdS$_{d+1}$ to write the entanglement entropy of a boosted strip as given by \reef{EEn} with $\Delta(\la)$ signifying the invariant width of the strip.  After much algebraic manipulation, the differential entropy can be written as 
\small
\ban{
&E=\frac{ L^{{(d-1)}} \ell_2 \cdots \ell_{{(d-1)}}}{4G_N}\int_0^1d\la\left[\frac{\sqrt{X'(\la)^2-T'(\la)^2+Z'(\la)^2}}{Z(\la)^{{(d-1)}}} -\frac{Z(\lambda )^{-{(d-1)}}
   Z'(\lambda )^2}{\sqrt{-T'(\lambda )^2+X'(\lambda )^2+Z'(\lambda )^2}} \right.\notag\\
&-\frac{Z(\lambda
   )^{-{(d-2)}} \left(Z''(\lambda ) \left(T'(\lambda )^2-X'(\lambda )^2\right)+Z'(\lambda )
   \left(X'(\lambda ) X''(\lambda )-T'(\lambda ) T''(\lambda )\right)\right)}{(d-2)
   \left(-T'(\lambda )^2+X'(\lambda )^2+Z'(\lambda )^2\right)^{3/2}}\notag\\
&-\frac{(d-2) B[f(\lambda)]Z'(\lambda ) \left(Z''(\lambda ) \left(T'(\lambda )^2-X'(\lambda)^2\right)+Z'(\lambda ) \left(X'(\lambda ) X''(\lambda )-T'(\lambda ) T''(\lambda)\right)\right)}{2 (d-2) {(d-1)}^2 \left(T'(\lambda )^2-X'(\lambda )^2\right) \left(T'(\lambda )^2-X'(\lambda )^2-Z'(\lambda )^2\right)}\notag\\
&\left.+\left(\frac 1{2n} B[f(\lambda)] Z(\lambda)^{-{(d-1)}} Z'(\lambda )  -\frac {1}{2 (d-2) {(d-1)}}Z(\lambda )^{-{(d-2)}} B'[f(\lambda )]\right) \left(1+\frac{Z'(\lambda )^2}{X'(\lambda )^2-T'(\lambda)^2}\right)^{\frac 12(d-2)}\right] \notag\\
\notag\\
&= \frac{ L^{{(d-1)}} \ell_2 \cdots \ell_{{(d-1)}}}{4G_N}\int_0^1d\la\left[\frac{\sqrt{X'(\la)^2-T'(\la)^2+Z'(\la)^2}}{Z(\la)^{{(d-1)}}} \right]\notag\\
&+ \frac{ L^{{(d-1)}} \ell_2 \cdots \ell_{{(d-1)}}}{4G_N({d-2})}\left[\frac{Z(\la)^{-{(d-2)}} Z'(\la)}{ \sqrt{-T'(\la)^2+X'(\la)^2+Z'(\la)^2}}-\frac{B[f(\la)]}{2 {(d-1)}} \left(Z(\la) \left(\frac{Z'(\la)^2}{X'(\la)^2-T'(\la)^2}+1\right)^{\frac{1}{2 {(d-1)}}}\right)^{-{(d-2)}}\right]_0^1
}\normalsize
The boundary term vanishes by the periodic boundary conditions we imposed on $\Sigma(\la,\sigma^a)$, and therefore we have
\ban{
E=  \frac{ L^{{(d-1)}} \ell_2 \cdots \ell_{{(d-1)}}}{4G_N}\int_0^1d\la\left[\frac{\sqrt{X'(\la)^2-T'(\la)^2+Z'(\la)^2}}{Z(\la)^{{(d-1)}}} \right]
}
Which we recognize as $E=S_{BH}=\frac {\mathcal A(\Sigma)}{4G_N}$. Indeed the tangent vector alignment construction extends to higher dimensions, albeit not quite as elegantly!

\chapter{Extension to Lovelock Gravity}
\label{lovelock}
An advantage of the general discussion in section \ref{general} is that it accommodates constructions similar to those of chapters \ref{time} and \ref{new} in higher curvature theories of gravity, provided that the Lagrangian is a function only of the canonical coordinates and their first derivatives. We show that for a time dependent holographic spacetime of a specific form, the entropy functional for Lovelock gravity contains only first derivatives. Therefore we can apply a construction similar to that of chapter \ref{time} to associate a family of boundary intervals with a given bulk co-dimension 2 surface. This proof is an extension of the one found in \cite{Myers2014}.

The entropy functional for Lovelock gravity is given in \cite{Jacobson1993} for a $d+1$ dimensional spacetime by
\ban{
S_{JM}= \frac{2\pi}{\ell_P^{d-1}} \int d^{d-1} x \sqrt{h} \left[ 1+ \sum _{p=2}^{\left\lfloor \frac{d+1}2\right\rfloor}p c_p L^{2p-2} \cL_{2p-2}(\mathcal R)\right]
}
where $h$ is the determinant of the induced metric on the horizon and $c_p$ are dimensionless coupling constants. The curvature dependence is given by
\ban{
\cL_{2p}(\mathcal R)= \frac 1{2^p} \delta ^{\nu_1 \cdots \nu_{2p}}_{\mu_1 \cdots \mu_{2p}}\mathcal R^{\mu_1\mu_2}{}_{\nu_1\nu_2}\cdots \mathcal R^{\mu_{2p-1}\mu_{2p}}{}_{\nu_{2p-1}\nu_{2p}}
}
We denote the intrinsic curvature tensor on the horizon as $\mathcal R_{\mu_1 \mu_2}{}_{\nu_1\nu_2}$, and the totally antisymmetric product of $n$ Kronecker delta symbols as $\delta^{\nu_1 \cdots \nu_{n}}_{\mu_1\cdots \mu_n}$. 

First, we consider a general holographic background with coordinates $\{t,z,x,y_i\}$, where $\{y_i\}$ denote the spatial coordinates with planar symmetry. Let $x^\mu$ denote the coordinates $\{t,z,x\}$. We write the metric in the form  
\ban{
ds^2=\tilde g_{\mu\nu}(t,z,x) dx^\mu dx^\nu +g_i(t,z,x)dy_i^2
}
where $\tilde g_{\mu\nu}$ denotes an arbitrary metric on the coordinates $\{x^\mu\}$ with one time-like direction. We consider a  $d-1$ dimensional surface with planar symmetry given by the embedding coordinates $\{t(\la), z(\la), x(\la), \sigma_i\}$. The induced metric is 
\ban{
ds^2_\text{ind}= Q(\la) d\la^2 + g_i(\la) d\sigma_i^2
}
where $Q(\la) = g_{\mu\nu} \partial_\la x^\mu \partial_\la x^\nu$. We would like to show explicitly how to remove all second derivatives of the coordinate functions, which in our notation are contained in any first derivatives $Q'(\la)$ or second derivatives $g_i''(\la)$. 

The induced Riemann tensor has non-vanishing components
\ban{
\mathcal R ^{\la\sigma_i}{}_{\la \sigma_i}&= \frac 1{2 g_i(\la) \sqrt{Q(\la)}}\left[ \frac{g_i'(\la)^2}{2 g_i(\la) \sqrt{Q(\la)}}- \left(\frac{g_i'(\la)}{\sqrt{Q(\la)}}\right)'\,\right] \notag \\
\mathcal R^{\sigma_k\sigma_l}{}_{\sigma_k \sigma_l}&= -\frac 14 \frac {g_k'(\la) g_l'(\la)}{g_k(\la) g_l(\la) Q(\la)}
}
The only terms we need to worry about are those involving the second term of the first line. We will manipulate these terms to write them as $F(g_i(\la)) \partial_\la {G(g_i(\la), g_i'(\la))}$ which we can integrate by parts to remove all second derivatives.

In general, the Lovelock entropy functional will contain a sum over many curvature terms, and we show that each term individually can be integrated by parts. All terms containing second derivatives will be of the form
\ban{
\sqrt{h} \mathcal L _{m+2} &\sim \sqrt{h} \mathcal R ^{\la\sigma_{i_1}}{}_{\la \sigma_{i_1}}\mathcal R^{\sigma_{i_2}\sigma_{i_3}}{}_{\sigma_{i_2}\sigma_{i_3}}\cdots \mathcal R^{\sigma_{i_{2m}}\sigma_{i_{2m+1}}}{}_{\sigma_{i_{2m}}\sigma_{i_{2m+1}}}\notag\\
&\sim\frac1{ F(g_i)} \left(\frac{g_{i_1}'(\la)}{\sqrt{Q(\la)}}\right)'  \frac 1 {Q(\la)^m} \prod _{k={i_2}}^{i_{2m+1}} g_k'(\la)+\cdots \label{badterms}
}
where the omitted terms in the second line depend only on first derivatives and $F(g_i)= g_{i_1}(\la)\cdots g_{i_{2m+1}}(\la)$. There are many such terms like the one in \reef{badterms}, each coming from part of the sum over the coordinates $\{\sigma_i\}$. To organize these terms, we first select an arbitrary $2m+1$ subset of these coordinates. Summing over only these coordinates (and expanding out the derivative term), we can write the contribution to the entropy functional from the terms like \reef{badterms} as 
\ban{
\frac1{F(g_i)}& \sum_{i=i_1}^{i_{2m+1}}\left[ \left( \frac {g_i''(\la)}{\sqrt{Q(\la)}}- \frac 12 \frac {Q'(\la) g_i'(\la)}{Q(\la)^{3/2}}\right) \frac 1{Q(\la)^m} \prod _{k \neq i} g_k'(\la)\right]\notag \\
\frac1{F(g_i)}& \left( \frac 1 {Q(\la)^{m+1/2}} \sum_{i=i_1}^{i_{2m+1}} \left[g_i''(\la) \left(\prod_{k \neq i} g_k'(\la)\right) \right]- \frac 12 (2m+1) \frac{Q'(\la) \prod_k g_k '(\la)}{Q(\la)^{m+3/2}}\right) \notag \\
\frac1{F(g_i)}& \left(\frac{\prod_k g_k'(\la)}{Q(\la)^{m+1/2}}\right)'
}
And so we can integrate this term by parts to get rid of all of the second derivatives. For every $2m+1$ subset of coordinates, we can apply the same trick, and so in this way we can write the entropy functional for Lovelock gravity completely in terms of the coordinate functions and their first derivatives. Therefore, the construction outlined in chapter \ref{time} for associating a family of boundary intervals with an extremal surface extends to higher curvature theories of gravity. 

Additionally, we would like to extend the construction of chapter \ref{new} to higher curvature theories of gravity. A major role in the construction was the observation that eq.~\ref{intersect} had a solution for more than just tangent vector alignment. We show how this feature works in general. 

In order for eq.~\ref{final} to hold, we required
\ban{
\pd{\cL}{\dot x^\mu_B} x'{}^\mu_B =\pd{\cL}{x'{}^\mu_B} x'{}^\mu_B \label{crucial}
}
If we impose the conditions
\ban{
\pd{\cL}{\dot x^\mu_B} =\pd{\cL}{x'{}^\mu_B}+k_\mu\\
x'{}^\mu_B k_\mu = 0 \label{ortho}
}
then we can write
\ban{
\pd{\cL}{\dot x^\mu_B} x'{}^\mu_B =\pd{\cL}{ x'{}^\mu_B}  x'{}^\mu_B = \cL(x_B, x'_B) \label{almost2}
}
and therefore the result of section \ref{general} is still valid. Note that for Einstein gravity we wrote $x'{}^\mu_B= \dot x ^\mu_B + k^\mu$ where $k_\mu x'{}^\mu_B=0$ and $|k|=0$, but as $\pd{\cL}{\dot x^\mu_B}=\frac{ \dot x^\mu_B}{|\dot x_B|}$ one can prove the two formulations are equivalent. 

\chapter[Scaled Differential Entropy]{Scaled Differential Entropy in Einstein Gravity}
\label{scaledE}

In this appendix we consider a family of boundary intervals and  an associated bulk curve which do not satisfy the conditions discussed in chapter \ref{new}. More explicitly, consider $\dot \bulkc_B(\la)=\left. \dot\Gamma (s;\la)\right|_{s_B(\la)}$ and $\bulkc'_B(\la)$ and define
\ban{
\alpha(\la) = \frac{ \dot \bulkc_B(\la)\cdot \bulkc'_B(\la)}{|\dot \bulkc_B (\la)||\bulkc'_B(\la)|}
}
In Einstein gravity $\cL(x, \dot x) = |\dot x|$, and therefore we can write eq.~\reef{almost} as 
\ban{
-\int_0^1 \frac{dS(\ga_L, \ga_R)}{d \ga_L^\mu}\deriv{\ga^\mu_L}{\la}d\la&=\int_0^1  \frac{\dot \bulkc_B \cdot \bulkc'_B}{|\dot \bulkc_B|} d\la \notag \\
&= \int_0^1 \alpha(\la) \cL(\bulkc_B, \bulkc'_B) \label{scaled}
}
Comparing this expression to eq.~\reef{final}, we see that when $\alpha(\la)=1$ the differential entropy of the boundary intervals equals the gravitational entropy of the bulk curve. Note that this case represents the constructions detailed in the main body of this essay. 

This calculation shows that for $\alpha(\la)\neq 1$, the relation between the differential entropy of the boundary intervals and the gravitational entropy of the bulk takes on a simple `scaled' form. In this way, we can think of the particular way of constructing the bulk curve from boundary intervals in chapter \ref{new} (or equivalently the way of constructing boundary intervals given a bulk curve in chapter \ref{time}) as a special case of the relation \reef{scaled}.

Note however that eq.~\reef{scaled} implies more general constructions would yield equality of differential entropy of a family of boundary intervals with the Bekenstein-Hawking of a bulk curve. For example, let $\alpha(\la) = 1+ \partial_\la \cL(\bulkc_B, \bulkc'_B)$, then eq.~\reef{final} still holds as $\cL \cL'$ is a total derivative. However, as mentioned in chapter \ref{new}, there is already a large class of families of boundary intervals corresponding to a given bulk curve, and therefore this additional freedom merely expands the proposed gauge-type symmetry in the space of families of boundary intervals. 

We can explicitly check the relation \reef{scaled} in AdS$_3$. This calculation follows the construction of section \reef{arbhole}, but projects along an arbitrary vector denoted $V(\la)=\{V_z(\la), V_x(\la), V_t(\la)\}$. We consider an arbitrary bulk curve $\bulkc_B(\la)=\coords{Z(\lambda),X(\lambda), T(\lambda)}$ with the condition that its tangent vector is space-like everywhere. Also, we impose the usual periodic boundary conditions on $\bulkc_B(\la)$ and $V(\la)$. 

To construct the geodesic along the vector $V(\la)$, we can follow the same method for contracting the tangent geodesic \reef{housefire} substituting $\bulkc'(\la)$ with $V(\la)$. We get 

\ban{
\notag  \G(s; \la)= &\left\{r^*(\la)  \sin s,X(\la)+ \frac{Z(\la) V_z(\la)V_x(\la) }{{V_x(\la)^2-V_t(\la)^2}}+\frac{V_x(\la)\, r^*(\la)}{\sqrt{V_x(\la)^2- V_t(\la)^2}}\cos s,\right.\\
&\hspace{0.6cm}\left.T(\la)+ \frac{Z(\la) V_z(\la)V_t(\la) }{{V_x(\la)^2- V_t(\la)^2}}+\frac{V_t(\la)\, r^*(\la)}{\sqrt{V_x(\la)^2- V_t(\la)^2}}\cos s\right\}
\label{ScaledGeod}}

Given this parameterization, it is straightforward to compute the differential entropy
\ban{
E=\frac L{4G_N} \int_0^1 d\la & \frac {\alpha(\la)}{Z(\la)}\sqrt{X'(\la)^2-T'(\la)^2+Z'(\la)^2}\notag\\
&+\frac{V_z(\la) (V_t(\la) V_t'(\la)-V_x(\la) V_x'(\la))+ V_z'(\la)(V_x(\la)^2-V_t(\la)^2)}{(V_x(\la)^2-V_t(\la)^2)\sqrt{V_x(\la)^2-V_t(\la)^2+V_z(\la)^2}}\notag\\
=\frac L{4G_N} \int_0^1 d\la & \frac {\alpha(\la)}{Z(\la)}\sqrt{X'(\la)^2-T'(\la)^2+Z'(\la)^2}+\left.\frac{L}{4G_N}{\sinh}^{-1}\left(\frac{V_z(\la)}{\sqrt{V_x(\la)^2-V_t(\la)^2}}\right)\right|_0^1\notag\\
=\frac L{4G_N} \int_0^1 d\la & \frac {\alpha(\la)}{Z(\la)}\sqrt{X'(\la)^2-T'(\la)^2+Z'(\la)^2}
}
where the total derivative vanishes due to the periodic boundary conditions. Indeed we have the claimed relation \reef{scaled}.


\bibliographystyle{apsrev_jacob}
	\cleardoublepage 
\phantomsection  
\renewcommand*{\bibname}{References}

\addcontentsline{toc}{chapter}{\textbf{References}}

\bibliography{references}

\begin{thebibliography}{10}
\expandafter\ifx\csname bibnamefont\endcsname\relax
  \def\bibnamefont#1{#1}\fi
\expandafter\ifx\csname bibfnamefont\endcsname\relax
  \def\bibfnamefont#1{#1}\fi
\expandafter\ifx\csname url\endcsname\relax
  \def\url#1{\texttt{#1}}\fi
\expandafter\ifx\csname urlprefix\endcsname\relax\def\urlprefix{URL }\fi
\providecommand{\bibinfo}[2]{#2}
\providecommand{\eprint}[2][]{\url{#2}}

\bibitem{Headrick2014}
\bibinfo{author}{\bibfnamefont{M.}~\bibnamefont{Headrick}},
  \bibinfo{author}{\bibfnamefont{R.}~\bibnamefont{Myers}}, \bibnamefont{and}
  \bibinfo{author}{\bibfnamefont{J.}~\bibnamefont{Wien}}, ``{Holographic Holes
  and Differential Entropy}'' , \eprint{arXiv:1408.4770}
  (\bibinfo{year}{2014}).

\bibitem{Bianchi2012}
\bibinfo{author}{\bibfnamefont{E.}~\bibnamefont{Bianchi}} \bibnamefont{and}
  \bibinfo{author}{\bibfnamefont{R.~C.} \bibnamefont{Myers}}, ``{On the
  Architecture of Spacetime Geometry}'' , \eprint{arXiv:1212.5183v1}
  (\bibinfo{year}{2012}).

\bibitem{Balasubramanian2014}
\bibinfo{author}{\bibfnamefont{V.}~\bibnamefont{Balasubramanian}},
  \bibinfo{author}{\bibfnamefont{B.~D.} \bibnamefont{Chowdhury}},
  \bibnamefont{and} \bibinfo{author}{\bibfnamefont{B.}~\bibnamefont{Czech}},
  ``{A Hole-ographic Spacetime}'', \bibinfo{journal}{Phys. Rev. D}
  \textbf{\bibinfo{volume}{89}}, \eprint{arXiv:1310.4204v1}
  (\bibinfo{year}{2014}).

\bibitem{Myers2014}
\bibinfo{author}{\bibfnamefont{R.~C.} \bibnamefont{Myers}},
  \bibinfo{author}{\bibfnamefont{J.}~\bibnamefont{Rao}}, \bibnamefont{and}
  \bibinfo{author}{\bibfnamefont{S.}~\bibnamefont{Sugishita}}, ``{Holographic
  Holes in Higher Dimensions}'' , \eprint{arXiv:1403.3416v1}
  (\bibinfo{year}{2014}).

\bibitem{Headrick2013}
\bibinfo{author}{\bibfnamefont{M.}~\bibnamefont{Headrick}}, ``{General
  Properties of Holographic Entanglement Entropy}'' ,
  \eprint{arXiv:1312.6717v1} (\bibinfo{year}{2013}).

\bibitem{Wall2012}
\bibinfo{author}{\bibfnamefont{A.~C.} \bibnamefont{Wall}}, ``{Maximin Surfaces
  , and the Strong Subadditivity of the Covariant Holographic Entanglement
  Entropy}'' , \eprint{arXiv:1211.3494v1} (\bibinfo{year}{2012}).

\bibitem{Bekenstein1972}
\bibinfo{author}{\bibfnamefont{J.~D.} \bibnamefont{Bekenstein}}, ``{Black Holes
  and the Second Law}'', \bibinfo{journal}{Lett. Nuovo Cim.}
  \textbf{\bibinfo{volume}{4}}, \bibinfo{pages}{737} (\bibinfo{year}{1972}).

\bibitem{Bekenstein1973}
\bibinfo{author}{\bibfnamefont{J.~D.} \bibnamefont{Bekenstein}}, ``{Black Holes
  and Entropy}'', \bibinfo{journal}{Phys. Rev. D} \textbf{\bibinfo{volume}{7}},
  \bibinfo{pages}{2333} (\bibinfo{year}{1973}).

\bibitem{Hawking1974}
\bibinfo{author}{\bibfnamefont{S.~W.} \bibnamefont{Hawking}}, ``{Black Hole
  Explosions}'', \bibinfo{journal}{Nature} \textbf{\bibinfo{volume}{248}},
  \bibinfo{pages}{30} (\bibinfo{year}{1974}).

\bibitem{Hawking1975}
\bibinfo{author}{\bibfnamefont{S.~W.} \bibnamefont{Hawking}}, ``{Particle
  Creation by Black Holes}'', \bibinfo{journal}{Communications In Mathematical
  Physics} \textbf{\bibinfo{volume}{43}}, \bibinfo{pages}{199}
  (\bibinfo{year}{1975}).

\bibitem{Wald1993}
\bibinfo{author}{\bibfnamefont{R.~M.} \bibnamefont{Wald}}, ``{Black Hole
  Entropy is Noether Charge}'', \bibinfo{journal}{Phys. Rev. D.}
  \textbf{\bibinfo{volume}{48}}, \bibinfo{pages}{3427},
  \eprint{arXiv:gr-qc/9307038v1} (\bibinfo{year}{1993}).

\bibitem{Jacobson1994}
\bibinfo{author}{\bibfnamefont{T.}~\bibnamefont{Jacobson}},
  \bibinfo{author}{\bibfnamefont{G.}~\bibnamefont{Kang}}, \bibnamefont{and}
  \bibinfo{author}{\bibfnamefont{R.~C.} \bibnamefont{Myers}}, ``{On Black Hole
  Entropy}'', \bibinfo{journal}{Phys. Rev. D.} \textbf{\bibinfo{volume}{49}},
  \bibinfo{pages}{6587}, \eprint{arXiv:gr-qc/9312023v2} (\bibinfo{year}{1994}).

\bibitem{Iyer1994}
\bibinfo{author}{\bibfnamefont{V.}~\bibnamefont{Iyer}} \bibnamefont{and}
  \bibinfo{author}{\bibfnamefont{R.~M.} \bibnamefont{Wald}}, ``{Some Properties
  of the Noether Charge and a Proposal for Dynamical Black Hole Entropy}'',
  \bibinfo{journal}{Physical Review D} \textbf{\bibinfo{volume}{50}},
  \bibinfo{pages}{846}, \eprint{arXiv:gr-qc/9403028} (\bibinfo{year}{1994}).

\bibitem{Sorkin1983}
\bibinfo{author}{\bibfnamefont{R.~D.} \bibnamefont{Sorkin}}, ``{On the Entropy
  of the Vacuum outside a Horizon}'', \bibinfo{journal}{General Relativity and
  Gravitation} \textbf{\bibinfo{volume}{1}}, \bibinfo{pages}{734},
  \eprint{arXiv:1402.3589} (\bibinfo{year}{1983}).

\bibitem{Bombelli1986}
\bibinfo{author}{\bibfnamefont{L.}~\bibnamefont{Bombelli}},
  \bibinfo{author}{\bibfnamefont{R.}~\bibnamefont{Koul}},
  \bibinfo{author}{\bibfnamefont{J.}~\bibnamefont{Lee}}, \bibnamefont{and}
  \bibinfo{author}{\bibfnamefont{R.}~\bibnamefont{Sorkin}}, ``{Quantum Source
  of Entropy for Black Holes}'', \bibinfo{journal}{Physical Review D}
  \textbf{\bibinfo{volume}{34}}, \bibinfo{pages}{373} (\bibinfo{year}{1986}).

\bibitem{Srednicki1993}
\bibinfo{author}{\bibfnamefont{M.}~\bibnamefont{Srednicki}}, ``{Entropy and
  Area}'', \bibinfo{journal}{Physical Review Letters}
  \textbf{\bibinfo{volume}{71}}, \bibinfo{pages}{666},
  \eprint{arXiv:hep-th/9303048} (\bibinfo{year}{1993}).

\bibitem{Frolov1993}
\bibinfo{author}{\bibfnamefont{V.}~\bibnamefont{Frolov}} \bibnamefont{and}
  \bibinfo{author}{\bibfnamefont{I.}~\bibnamefont{Novikov}}, ``{Dynamical
  Origin of the Entropy of a Black Hole}'', \bibinfo{journal}{Physical Review
  D} \textbf{\bibinfo{volume}{48}}, \bibinfo{pages}{4545},
  \eprint{arXiv:gr-qc/9309001} (\bibinfo{year}{1993}).

\bibitem{Mcgreevy2009}
\bibinfo{author}{\bibfnamefont{J.}~\bibnamefont{Mcgreevy}}, ``{Holographic
  Duality with a View Toward Many-Body Physics}'' , \eprint{arXiv:0909.0518v3}
  (\bibinfo{year}{2009}).

\bibitem{Horowitz2006}
\bibinfo{author}{\bibfnamefont{G.~T.} \bibnamefont{Horowitz}} \bibnamefont{and}
  \bibinfo{author}{\bibfnamefont{J.}~\bibnamefont{Polchinski}}, ``{Gauge /
  Gravity Duality}'' , \eprint{arXiv:gr-qc/0602037v3} (\bibinfo{year}{2006}).

\bibitem{Hooft1993}
\bibinfo{author}{\bibfnamefont{G.}~\bibnamefont{`t~Hooft}}, ``{Dimensional
  Reduction in Quantum Gravity}'' , \eprint{arXiv:gr-qc/9310026}
  (\bibinfo{year}{1993}).

\bibitem{Susskind1995}
\bibinfo{author}{\bibfnamefont{L.}~\bibnamefont{Susskind}}, ``{The World as a
  Hologram}'', \bibinfo{journal}{Journal of Mathematical Physics}
  \textbf{\bibinfo{volume}{36}}, \bibinfo{pages}{6377},
  \eprint{arXiv:hep-th/9409089} (\bibinfo{year}{1995}).

\bibitem{Maldacena1998}
\bibinfo{author}{\bibfnamefont{J.}~\bibnamefont{Maldacena}}, ``{The Large N
  Limit of Superconformal Field Theories and Supergravity}'',
  \bibinfo{journal}{Adv. Theor. Math. Phys.} \textbf{\bibinfo{volume}{2}},
  \bibinfo{pages}{231}, \eprint{arXiv:hep-th/9711200v3} (\bibinfo{year}{1998}).

\bibitem{Gubser1998}
\bibinfo{author}{\bibfnamefont{S.}~\bibnamefont{Gubser}},
  \bibinfo{author}{\bibfnamefont{I.}~\bibnamefont{Klebanov}}, \bibnamefont{and}
  \bibinfo{author}{\bibfnamefont{A.}~\bibnamefont{Polyakov}}, ``{Gauge Theory
  Correlators from Non-Critical String Theory}'', \bibinfo{journal}{Physics
  Letters B} \textbf{\bibinfo{volume}{428}}, \bibinfo{pages}{105},
  \eprint{arXiv:hep-th/9802109} (\bibinfo{year}{1998}).

\bibitem{Witten1998}
\bibinfo{author}{\bibfnamefont{E.}~\bibnamefont{Witten}}, ``{Anti De Sitter
  Space And Holography}'' , \eprint{arXiv:hep-th/9802150}
  (\bibinfo{year}{1998}).

\bibitem{Ryu2006a}
\bibinfo{author}{\bibfnamefont{S.}~\bibnamefont{Ryu}} \bibnamefont{and}
  \bibinfo{author}{\bibfnamefont{T.}~\bibnamefont{Takayanagi}}, ``{Holographic
  Derivation of Entanglement Entropy from AdS/CFT}'', \bibinfo{journal}{Phys.
  Rev. Lett.} \textbf{\bibinfo{volume}{96}}, \eprint{arXiv:hep-th/0603001v2}
  (\bibinfo{year}{2006}).

\bibitem{Nishioka2009}
\bibinfo{author}{\bibfnamefont{T.}~\bibnamefont{Nishioka}},
  \bibinfo{author}{\bibfnamefont{S.}~\bibnamefont{Ryu}}, \bibnamefont{and}
  \bibinfo{author}{\bibfnamefont{T.}~\bibnamefont{Takayanagi}}, ``{Holographic
  Entanglement Entropy: An Overview}'', \bibinfo{journal}{J. Phys. A}
  \textbf{\bibinfo{volume}{42}}, \eprint{arXiv:0905.0932v2}
  (\bibinfo{year}{2009}).

\bibitem{Takayanagi2012}
\bibinfo{author}{\bibfnamefont{T.}~\bibnamefont{Takayanagi}}, ``{Entanglement
  Entropy from a Holographic Viewpoint}'', \bibinfo{journal}{Class. Quant.
  Grav.} \textbf{\bibinfo{volume}{29}}, \eprint{arXiv:1204.2450v2}
  (\bibinfo{year}{2012}).

\bibitem{Ryu2006}
\bibinfo{author}{\bibfnamefont{S.}~\bibnamefont{Ryu}} \bibnamefont{and}
  \bibinfo{author}{\bibfnamefont{T.}~\bibnamefont{Takayanagi}}, ``{Aspects of
  Holographic Entanglement Entropy}'', \bibinfo{journal}{JHEP}
  \textbf{\bibinfo{volume}{0608}}, \eprint{arXiv:hep-th/0605073v3}
  (\bibinfo{year}{2006}).

\bibitem{Lewkowycz2013}
\bibinfo{author}{\bibfnamefont{A.}~\bibnamefont{Lewkowycz}} \bibnamefont{and}
  \bibinfo{author}{\bibfnamefont{J.}~\bibnamefont{Maldacena}}, ``{Generalized
  Gravitational Entropy}'', \bibinfo{journal}{JHEP}
  \textbf{\bibinfo{volume}{1308}}, \eprint{arXiv:1304.4926v2}
  (\bibinfo{year}{2013}).

\bibitem{Hubeny2007}
\bibinfo{author}{\bibfnamefont{V.~E.} \bibnamefont{Hubeny}},
  \bibinfo{author}{\bibfnamefont{M.}~\bibnamefont{Rangamani}},
  \bibnamefont{and}
  \bibinfo{author}{\bibfnamefont{T.}~\bibnamefont{Takayanagi}}, ``{A Covariant
  Holographic Entanglement Entropy Proposal}'', \bibinfo{journal}{JHEP}
  \textbf{\bibinfo{volume}{0707}}, \eprint{arXiv:0705.0016v3}
  (\bibinfo{year}{2007}).

\bibitem{mattEW}
\bibinfo{author}{\bibfnamefont{M.}~\bibnamefont{Headrick}},
  \bibinfo{author}{\bibfnamefont{V.~E.} \bibnamefont{Hubeny}},
  \bibinfo{author}{\bibfnamefont{A.}~\bibnamefont{Lawrence}}, \bibnamefont{and}
  \bibinfo{author}{\bibfnamefont{M.}~\bibnamefont{Rangamani}}, ``{Causality and
  Holographic Entanglement Entropy}'' , \bibinfo{note}{in preparation}.

\bibitem{Jacobson1993}
\bibinfo{author}{\bibfnamefont{T.}~\bibnamefont{Jacobson}} \bibnamefont{and}
  \bibinfo{author}{\bibfnamefont{R.}~\bibnamefont{Myers}}, ``{Black Hole
  Entropy and Higher Curvature Interactions}'', \bibinfo{journal}{Physical
  Review Letters} \textbf{\bibinfo{volume}{70}}, \bibinfo{pages}{3684},
  \eprint{arXiv:hep-th/9305016} (\bibinfo{year}{1993}).

\end{thebibliography}

\end{document}